\documentclass[12pt]{book}
\usepackage{beton}
\usepackage{euler}
\usepackage{epsfig}

\usepackage{fancyheadings}
\font\Bbbfont=msbm10
\def\Bbb#1{\hbox{\Bbbfont#1}}

\date{\vskip 40pt
Academic year 1997/98}

\author{\Large {\bf Candidate}\\
Gabriele Gionti
\and
\Large {\bf Supervisors}\\
Prof. Mauro Carfora\\
Prof. Alessandro D'Adda}

\title{\vskip 200pt
{\Large \bf INTERNATIONAL SCHOOL\\ FOR\\ ADVANCED STUDIES\\ 
(SISSA-ISAS) - TRIESTE}
\vskip 50pt
{\Large \bf Thesis submitted for the degree of Doctor Philosophi\ae}
\vskip 50pt 
{\Huge \bf Discrete Approaches Towards the Definition of a Quantum 
Theory of Gravity}}
\begin{document}
\pagestyle{empty}
\maketitle

\pagestyle{fancyplain}\pagenumbering{arabic}
  \section*{\Huge Acknowledgments}
{\it I am grateful to Prof. Mauro Carfora for introducing me to the subject of
Dynamical Triangulations, for collaboration and discussions. I am indebted 
also with Prof. Alessandro D'Adda for collaboration on Regge calculus and 
friendship toward me. Many thanks to Dr. Ruth M. Williams for 
having read this thesis and for remarks and questions which 
have improved it a lot. It is a pleasure to acknowledgment Davide Gabrielli 
for daily discussions and collaborations. I have shared with him my experience
in Sissa. I want to single out all my friends, now spread in the world, with
whom I began my experience in Sissa: Carlo Acerbi, Francesca Colaiori, Paolo De
Los Rios, 
Alessandro Flammini, Beppe Turco. In particular Stefano Panzeri and Loredana
Fantauzzi and their baby (not "universe") Matteo 
(also if up to now I have not seen him), my friendship
with them is one of the best thing in these years. I cannot forget Azer Akhmedov
(Azeglio)
and his moustaches which gave him the  appearance of an Islamic fundamentalist. 
It is still surprising for me how he is able to understand mathematical concepts
and misunderstand every other thing. 
A special mention is for Antonio Liguori, outstanding
scientist, with whom I have had many scientific discussions,   
and whose friendship I enjoyed this last year in Sissa. He has been the only
person who has given attention (and I think that this is something which could
be used against him)  to my tentative to interpret our world in a Platonic sense.
I think that in order to verify these interpretations one time  he conducted me
in the
most dangerous streets of Naples. 

\noindent I want to remember Fabrizio Melchionna and his nostalgia of Ragusa.
The two
astronomers Michele Maris and Cesario Lia and our metaphysical discussions in
which 
each one of us remains always on his own position.  
Last 
but not least are Marco Bertola, Stefano Bianchini, 
and Davide Guzzetti with whom, recently, 
I have daily scientific and not "strictly" scientific discussions.} 
\vskip 0.1cm
\noindent { \large   This thesis and my scientific work in Sissa is dedicated to
my mother 
Angela, my brother Oreste and my uncle Attilio whose material life has been
recently 
destroyed by the cancer}

\tableofcontents

\newtheorem{theorem}{\bf Theorem}[section]
\newtheorem{lemma}{\bf Lemma}[section]
\newtheorem{corollary}{\bf Corollary}[section]
\newtheorem{definition}{\bf Definition}[section]
\chapter{Introduction}

In spite of many recent developments, in particular in string theory, the 
problem of quantization of General Relativity is still an open one 
both from mathematical and physical point of view.

In this thesis we discuss some contributions to the lattice approach  
to Euclidean Quantum Gravity, namely to Regge Calculus and Dynamical
Triangulations that offer the most
 natural  discretization of 
General Relativity. This  discretization consists in going from a Riemannian 
manifolds to triangulations of Piecewise-Linear manifolds. In both cases the
partition 
function is a sum over the triangulations of Piecewise-Linear manifolds. Each
triangulation 
is weighted by a factor equal to the exponential of minus the discretized
version of the 
Einstein-Hilbert action (Regge-Einstein action). Moreover the diff-invariant
continuum measure on the Riemannian structures of a manifold $M$ is
replaced, in general, by a DeWitt-like measure for the edge lengths in Regge
Calculus and 
by a micro-canonical measure in the dynamical triangulations. 

The bulk of this thesis is based on two original contributions which are 
reported in chapter three and four, while
in chapter two we introduce very briefly the main notions
about the model of dynamical triangulations. All the notions that we give in
this chapter 
are well known and established in the literature
and we introduce them since they will be used in chapter three.

\noindent Along the main stream of the connection between Euclidean quantum
field theory and classical statistical mechanics we introduce the notions
of micro-canonical, canonical, and grand-canonical partition functions. In 
particular in two dimensions we mention different analytical ways in which
the number of combinatorial inequivalent triangulations was calculated. Recent
results on the estimate of micro-canonical partition function in three and
four dimensions are summarized. This estimate will be extensively used for 
the classification of the elongated phase of dynamical triangulations in four 
dimensions in chapter three.

\noindent Successively
we give the definition of the two point Green function in the context of
dynamical
triangulations. As a consequence of the definition of the Green function, we
give 
the definition of the susceptibility function. From its behavior near the
critical line we
define the string susceptibility exponent $\gamma_{str}$.

\noindent  
In chapter three we shall study the elongated phase of dynamical triangulations
in four 
dimensions. We begin by introducing Walkup's theorem which characterize the
triangulations
with the topology of the sphere in four dimensions. We review the
kinematical bounds
which are fixed by this theorem. Furthermore by using the expression of the
estimated
canonical partition
function, we stress that the average curvature is saturated in
correspondence to the
kinematical bound of Walkup's theorem. This means that for values of $k_2$, the
inverse of 
gravitational constant, greater than the value $k_{2}^{c}$ for which  the Walkup
bound is reached, there is in the statistical ensemble 
of equilateral triangulations of the sphere $S^{4}$
a prevalence of particular triangulations called "Staked Spheres". These
configurations are
the only triangulations for which the Walkup bound is realized. They have a
simple
tree-like structure
that can be mapped into branched polymers structures. Anyway the map is not one
to one
in the sense that combinatorial inequivalent stacked spheres can be mapped into
the same 
branched polymer. We recognize that a stacked sphere fits with the model of a
network of baby
universes which has been formulated from the analysis of the results of the
Monte Carlo 
simulations in four dimensions.

\noindent We construct two distinct models of branched polymers and we put them 
in correspondence with the stacked spheres by the dual map. This analysis shows
that 
the string susceptibility of the stacked spheres is less than $1$.

\noindent At the end we analyse a model taken from the theory of random surfaces
and adapted
to the stacked spheres. The aim of this analysis is to show some evidences on
the analogy 
between the stacked spheres and self-energy Feynman graphs relative to matrix
models of two
dimensional triangulations. What we learn from this analysis and from all
previous
considerations is that there is a strong evidence that the stacked spheres
correspond to a
mean field phase in which the string susceptibility exponents is
$\gamma_s={1\over 2}$, so that
any attempt of performing a continuum limit in this phase will give, even if
we assume the
convergence of the Schwinger functions, a Gaussian measure.

Recent numerical evidence of a first order phase transition of the 
 model of dynamical triangulations in four dimensions and the previous strong
 evidence of a trivial elongated phase
suggest that more efforts towards new discrete models for simplicial quantum
gravity might be required.

Along this line and following the work of various authors we study
 Regge 
calculus as a local theory of Euclidean Poincar\'e group
in the first order formalism. The reason for a first
order 
formalism is both theoretical and technical. With respect to traditional Regge
Calculus the novelty lies both in its formulation as a gauge theory and in the 
first order formalism. The gauge theory approach results mainly in the deficit
angle
being replaced by its sine. The first order formalism has others effect of
smoothing
out some pathological configurations, like "spikes", which might prevent the
theory from
having a smooth continuum limit. These configurations are in fact in the region
of large
deficit angles where the first order formalism and the second order formalism
are not 
equivalent on a lattice. 

\noindent We first review and improve some definitions of a previous work on
this subject.
In particular it is stressed how a group theoretical formulation of Regge
calculus
allows to write an action on the dual Vorono{\"\i} complex of the original 
simplicial complex which is quite similar to a gauge theory and, more precisely,

looks like the Wilson action for lattice gauge theory.

\noindent We prove that this action does not depend from the orientation of the 
Vorono{\"\i} plaquette.

We formulate a first order principle in which we have two sets of independent
variables:
 the normals to the $n-1$-faces and the connection matrices. The normals are
 considered 
 as the analogous of the $n$-bein in the continuum theory, and the connection
 matrices as 
 the connection one form in General Relativity. The main result of this chapter 
 is that we prove in the case of 
 "small deficit angles" that Regge calculus is a solution of the first order
 formalism. 
 This result is not obvious if we vary indipendently the two sets of variables
 above. 

\noindent Then we derive the general field equations for the connection matrices
and for the
normals. We use the method of Lagrange multipliers to take in account the
constraints of the
theory. We propose a method for the calculation of the Lagrange multipliers by
using the one
to one correspondence among the normals to the faces of the $n$ simplices and
the
circumcentric coordinates of the vertices.

\noindent A measure for the  path integral for this simplicial theory of gravity
is introduced
and it is shown 
that it is locally invariant under $SO(n)$. As a last step we propose a coupling
of this
lattice theory of gravity with fermionic matter. This coupling is entirely
performed by
following the general prescription of the continuum theory. In other formulation
of discrete
gravity (Regge calculus and dynamical triangulations) the coupling with
fermionic matter 
is usually introduced "ad hoc". In this approach the coupling with fermionic
matter
is given by considering spinorial representation of $SO(n)$.

\chapter{Dynamical Triangulations}
\label{dyna}

\section{Introduction}
\label{int1}

\noindent In this chapter we shall introduce the basic tools for 
the simplicial approach \cite{will1} \cite{will2} \cite{hamber} 
to Euclidean Quantum Gravity (see \cite{hawking1} for 
the main articles on the subject and also reference \cite{giampi}) 
{\it via} the 
theory of Dynamical Triangulations \cite{mauro} \cite{1Ambjorn} 
\cite{LJ} \cite{david}. A precursor of Dynamical Triangulations has been
Weingarten \cite{wein}.
R{\"o}mer and Z{\"a}hringer \cite{romer} proposed for the first time this model
as a gauge fixing of Regge calculus.
 In section one we begin by considering 
the class of equilateral triangulations of Picewise-Linear (PL) 
manifolds (for a review on PL-manifods see 
\cite{rourke} \cite{thurstonb} \cite{seifert})
which are used in dynamical triangulations. Successively 
we define the action for dynamical triangulations as a restriction of the 
Regge-Einstein action \cite{regge}
to the equilateral triangulations of PL-manifolds. 
The partition function for dynamical triangulations 
is defined over the  ensemble of equilateral triangulations of PL-manifolds. 

\noindent In this framework we address the counting problem 
of the number of combinatorial inequivalent triangulations
and we briefly review the different methods
of enumeration in the two dimensional case. 
In three and four dimensions some recent analytical results are
illustrated \cite{mauro} \cite{mauro1} \cite{mauro2} \cite{mauro3}. 

In section \ref{verde} following
\cite{1Ambjorn} we give the definition of the Green 
functions in the 
framework of dynamical triangulations. The exposition always follows  
the connection with classical 
statistical mechanics and at the end we give the definition of susceptibility
function as a direct consequence of the definition of the 
grand-canonical Green function.

All this chapter has to be considered as an introduction to the main 
concepts of dynamical triangulations which will be used in the study
of the elongated phase in 4-D dynamical triangulations.

\noindent Anyway  it is important in our opinion to remark that 
the whole approach of dynamical triangulations must be considered in the 
general framework of Euclidean Lattice field theory as explained 
in \cite{sokal} chap. 15. It is crucial in this approach 
to determine the Green functions (the Schwinger functions) and to look
for a second order phase transition in the parameter space. If we have
a second order phase transition then the quantum theory of gravity
may be defined at the critical point by looking for the scaling
limit of the Green functions. If this limit exists we ask if it fits
with the Osterwalder-Schrader axioms \cite{osterwalder1} \cite{osterwalder2}.
In the affermative case the Riemannian Green functions are the Wick rotated 
version of the Lorentian Green Function (the Wightman functions) 
\cite{wightman} \cite{strocchi}.  
A first
implementation of the above ideas of Euclidean lattice 
field theory to simplicial quantum gravity has been given by Rocek and Williams
\cite{will1}
\cite{will2}. 

It can happen 
(as in the case of $\lambda \phi^{4}$) that the continuum 
limit gives a gaussian measure (free theory), then the theory is trivial.
We want to stress that the final goal of these theories is to find a 
non trivial continnum limit. 

\section{The Model of the Dynamical Triangulations}
\label{mode}

The standard rule in Dynamical Triangulations (for a review on the recent
results
see \cite{loll} and also \cite{rovelli})  is 
to consider all the triangulations of PL-manifolds 
made by equilateral
simplices of fixed edge lengths, say  $a$. 
This implies that the geometrical structures are even more
rigid with respect to Regge 
Calculus \cite{regge} (for a recent review on
Regge calculus see \cite{ruth1}). 
The set of piecewise-linear maps on these simplicial complexes depend
only by their 
combinatorial structures. So that two triangulations $T_{a}$ and ${T'}_{a}$ 
are equivalent if there is a piecewise 
linear map $\phi$ between them such that  
it maps one to one the vertices of $T_{a}$ into the vertices of ${T'}_{a}$ in
such 
a way that $(\phi(v_{i}),\phi(v_{j}))$ is an edge of ${T'}_{a}$ if and only if 
$(v_{i},v_{j})$ is an edge of $T_{a}$ and so on for every simplex of any
dimension.

\noindent The $n-2$ simplices are called {\it bones} $B$. 
The dihedral angle \cite{regge} of a $n$
simplex $\sigma^{n}$ 
on a bone is $cos^{-1}{1\over n}$. If we indicate by $q(B)$ the number of
$n$-dimensional 
simplices which share the bone $B$, the Regge curvature \cite{regge} on the bone
$B$ is

\begin{equation}
K(B)=2\left(2\pi - q(B)cos^{-1}{1\over n}\right)V(B)
\label{dcurv}
\end{equation}

\noindent where $V(B)$ is the $n-2$-dimensional volume of the bone. By standard
formula we know that
the volume of a $n$-dimensional equilateral simplex $V(\sigma^{n})$ is 

\begin{equation}
V(\sigma^{n})={a^{n}\sqrt{n+1} \over (n)!\sqrt{2^{n}}}\;\;\;\; .
\label{sivol}
\end{equation}

The Regge-Einstein action with cosmological term for dynamical triangulations
$T_{a}$ without boundary
can be seen as a functional of the form

\begin{equation}
S_{R}(\Lambda,G,T_{a})\equiv {\Lambda \over 8\pi G}
\sum_{\sigma^{n}}V(\sigma^{n})
-{1\over 16\pi G}\sum_{B}K(B)\;\;\;\; ,
\label{cosm}
\end{equation}

\noindent where the first sum is over all $n$-dimensional simplices of the
triangulation $T_{a}$.

We define this two bare coupling constants

\begin{eqnarray}
k_{n}&\equiv& {\Lambda \over 8\pi G}V(\sigma^{n}) 
+n(n+1){cos^{-1}{1\over n} \over 16\pi G}V(B) \nonumber \\
k_{n-2}&\equiv&{V(B) \over 4G}\;\;\;\; ,
\label{running}
\end{eqnarray}

\noindent then the action on a triangulation $T_{a}$ takes the standard form

\begin{equation}
S(k_{n},k_{n-2},T_{a})=k_n N_{n} -k_{n-2}N_{n-2}
\label{zuppa}
\end{equation}

\noindent where $N_{n}$ and $N_{n-2}$ are the numbers, respectively, of the $n$
and $n-2$ simplices
of the triangulation $T_{a}$.

The partition function on the triangulation $T_{a}$
of a PL-manifolds
${\mathcal PL}$ is defined as

\begin{equation}
Z({\mathcal PL},k_n,k_{n-2})\equiv \sum_{T_{a}}e^{-k_n N_{n}
+k_{n-2}N_{n-2}}\;\;\;\; .
\label{partita}
\end{equation}

\noindent Notice that all the triangulations of the PL-manifold ${\mathcal PL}$,
 have the same weight in the
path-integral (for a discussion about the measure of dynamical triangulations
see \cite{marinari}) . 

\noindent The partition function \ref{partita} can be rewritten in the following
way

\begin{equation}
Z(k_{n},k_{n-2})\equiv
\sum_{{\mathcal PL}}\sum_{N_{n}}e^{-k_{n}N_{n}}
\sum_{N_{n-2}}e^{k_{n-2}N_{n-2}}
\sum_{\# T_{a}({\mathcal PL},N_{n},N_{n-2})}1
\label{frittata}
\end{equation}

\noindent where $\sum_{{\mathcal PL}}$ means the sum over all piecewise-flat
topologies and 
$\sum_{\# T_{a}({\mathcal PL},N_{n},N_{n-2})}$ is 
the number of combinatorial inequivalent equilateral
triangulations
$T_{a}({\mathcal PL},N_{n},N_{n-2})$ of a fixed PL-topology, fixed number
$N_{n}$ of $n$-simplices  
and fixed number $N_{n-2}$ of $n-2$-simplices. Now we can consider 

\begin{equation}
\rho_{a}({\mathcal PL},N_{n},N_{n-2})\equiv 
\sum_{\# T_{a}({\mathcal PL},N_{n},N_{n-2})}1
\label{micocanonica}
\end{equation}

\noindent as the micro-canonical partition function over the ensemble of the
equilateral triangulations
of edge length $a$ with fixed PL-topology. The following formula 

\begin{equation}
Z({\mathcal PL},N_{n-2},k_{n-2})\equiv \sum_{N_{n-2}}e^{k_{n-2}N_{n-2}}
\rho_{a}({\mathcal PL},N_{n},N_{n-2}) 
\label{canonical}
\end{equation}

\noindent is the canonical partition function, in which $\sum_{N_{n-2}}
e^{k_{n-2}N_{n-2}}$ plays 
the role of the 
Gibbs measure. Finally 

\begin{equation}
Z(k_{n},k_{n-2})\equiv
\sum_{{\mathcal PL}}\sum_{N_{n}}e^{-k_{n}N_{n}}
Z(N_{n},k_{n-2})
\label{grancanonica}
\end{equation}

\noindent is the grand-canonical partition function in which $e^{-k_{n}N_{n}}$
can be considered as
the equivalent of the chemical potential. Anyway since even in two dimensions
the sum over the 
PL-topologies is divergent ( see ref. \cite{witten} for a brief account), 
many times we will restrict to 
the only topology of the sphere in every dimension.

Let's start to examine the two dimensional case. The Dehn-Sommerville equations
(see \cite{kunel} p. 62  and also \cite{Walkup} p. 80) are:

\begin{eqnarray}
N_{2}-N_{1}+N_{0}&=&\chi\left(T_{a}({\mathcal PL})\right)\\ \nonumber 
2N_{1}&=&3N_{2}\;\;\;\; ,
\label{2d}
\end{eqnarray}

\noindent where $\chi\left(T_{a}({\mathcal PL})\right)$ is the Euler-Poincar\'e
characteristic of
the PL-manifold ${\mathcal PL}$ whose triangulation is $T_{a}({\mathcal PL})$.
As is well known in
two dimensions, topological manifolds and PL-manifolds are equivalent and the
topology, for compact, connected and orientable two-dimensional manifolds , is
completely 
classified by the genus $g$ of the manifold \cite{dubrovin}. 
From the Dhen-Sommerville equations \ref{2d} we can express
all the components of the $f$-vector (see \cite{Walkup} p. 78) as function of
$N_{2}$ 
and $\chi\left(T_{a}({\mathcal
PL})\right)$, that is to say

\begin{eqnarray}
N_{0}&=&{N_{2} \over 2} +\chi\left(T_{a}({\mathcal
PL})\right)\\ \nonumber
N_{1}&=&{3 \over 2}N_{2}
\label{eguaglio} \;\;\;\;.
\end{eqnarray}

The asymptotic number of combinatorial inequivalent triangulations of the sphere
$S^{2}$ 
was calculated for the first 
time by the mathematician Tutte \cite{tutte}. The underlying idea has been to
map by the
stereographic projection the triangulations of the sphere in planar
triangulations and to
enumerate them by the techniques of generating series built up by using the
geometrical 
properties of the planar triangulations. The result is the following

\begin{equation}
\rho_{a}(S^{2},N_{2}) \asymp N_{2}^{-{7 \over 2}}e^{k^{c}_{2}N_{2}}\;\;\;\;,
\label{prima}
\end{equation}

\noindent where $k^{c}_{2}$ is a numerical constant. It is important that the
growth of the
number of the triangulations is at most exponential. In fact in the opposite
case the divergences 
make it 
impossible to define a statistical theory like \ref{frittata}. The result
\ref{prima} 
can be obtained again 
by using the quantum field theory techniques of 
the matrix models \cite{bessis}. This techniques
is based on the  use of the generating functional

\begin{equation}
Z_{N}(g)=\int dM\; exp \left(-{1\over 2}tr(M^{2}) 
-{\lambda \over \sqrt{N}}tr(M^{3})\right) 
\label{matmod}
\end{equation}

\noindent in which $M$ is an $N\times N$ Hermitian matrix and $\lambda$ the
coupling constant,
$dM$ is the Haar-measure on this matrix group. The path-integral \ref{matmod}
will generate a
perturbative series whose Feynman diagrams can be represented as double line one
for each
index of the matrix $M_{ij}$ (see figure 2.1). 

\begin{figure}
\begin{center}
\epsfig{figure= 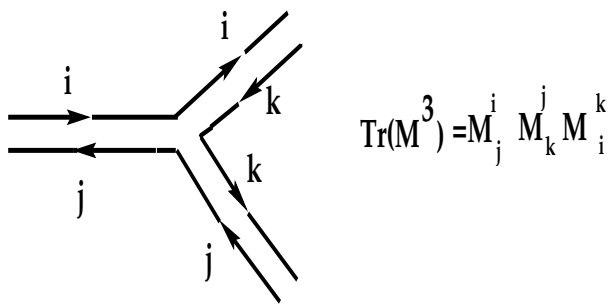,height=4.0truecm,width=8.0truecm}
\caption{ Feynman graph for the matrix model relative to the triangulations. 
Each line has two entries 
one for the row and other for the column of the matrix $M$. In each vertex there
are always
three double lines}
\end{center}
\label{fig:matrix1}
\end{figure}

\begin{figure}
\begin{center}
\epsfig{figure= 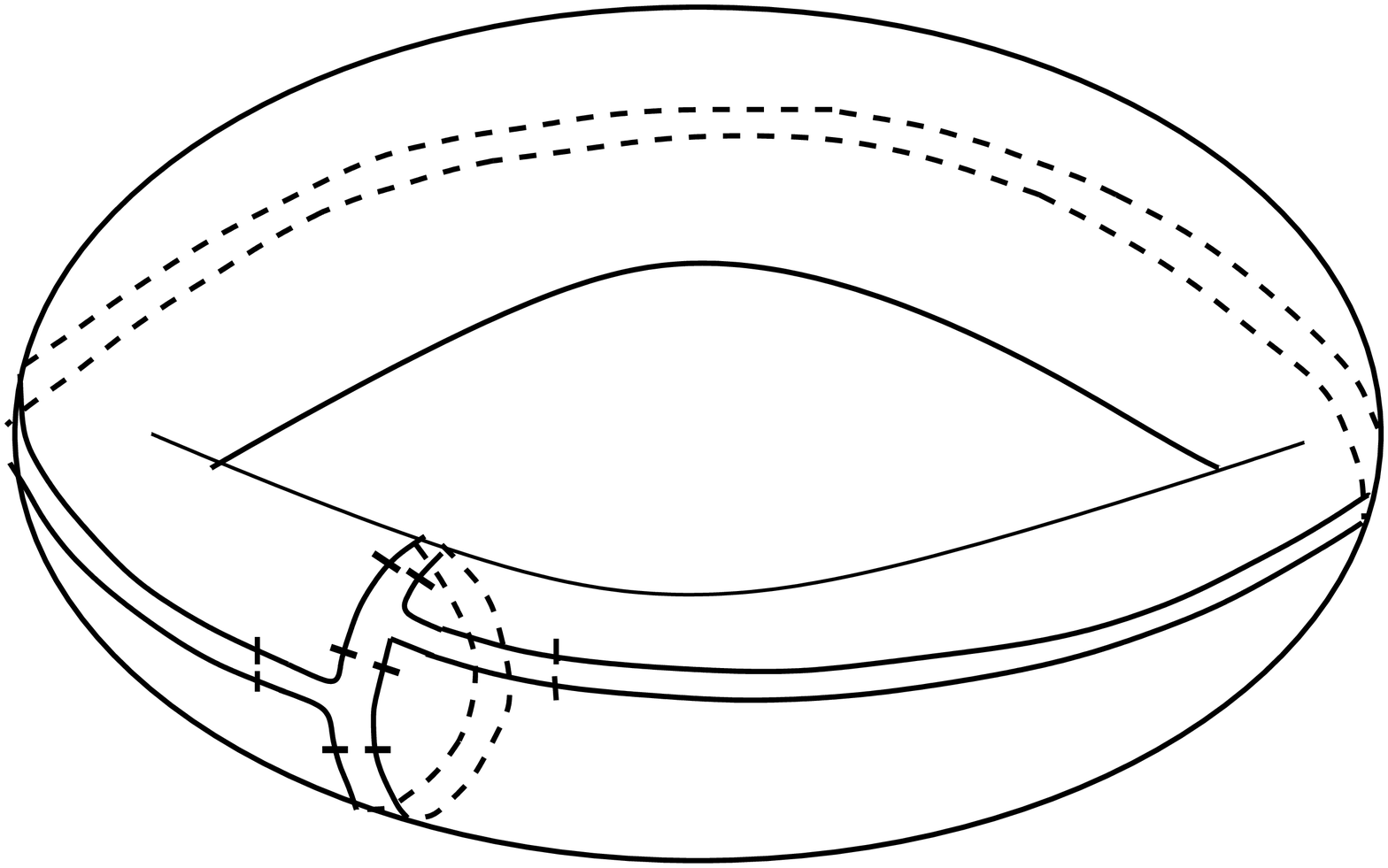,height=4.0truecm,width=8.0truecm}
\caption{ An example of a second order perturbation term which is 
rappresented by two double graphs of $Tr(M^{3})$ which can be closed
on a two dimensional torus }
\end{center}
\label{fig:toro1}
\end{figure}

Since in the equation \ref{matmod} there is a
trace of the product of 
matrices, the Feynman diagrams have to be closed graphs. 
The fact that there are two lines and the possibility
of all combinations of the indices means that in general 
the graphs can be closed only on Riemann surfaces (see figure \ref{fig:toro1}).
Furthermore each Feynman graph corresponds to a dual triangulation.

\noindent It
can be show that the perturbative series in $\lambda$ 
relative to connected Feynman double graphs generated by 
\ref{matmod} can be arranged as power serie in $N$
of the genus $g$ of
the Riemann surfaces. It follows that the coefficients of this series give 
the generating functions for the number of 
combinatorial inequivalent dual triangulations of a fixed topology.
Since there is a one to one 
correspondence between any triangulation and its (topological or metrical) dual
we can obtain the generating functions 
of the combinatorial inequivalent triangulations for any topology (see
\cite{bessis} 
and also \cite{frasca}).

Another derivation of the formula \ref{prima} has been 
recently found \cite{mauro} and is based on the enumeration of 
the curvature assignment of the bones of the triangulation. This estimate can be

extended to all two-dimensional topologies and 
to higher dimensions.

It follows that the two dimensional grandcanonical partition function can be
written 
as 

\begin{equation}
Z(k_{2},k_{0})=\sum_{g}e^{k_{0}\chi(g)}
\sum_{N_2}N_{2}^{\gamma_{str}-3}
e^{-(k_2 -{k_0\over 2}-k^{c}_{2})N_{2}}
\label{diverge}
\end{equation}

\noindent where $\chi(g)=2-2g$. $k_2 -{k_0\over 2}-k^{c}_{2}$ is called the
critical line since 
for $k_{2}>{k_0\over 2}+k^{c}_{2}$ the partition function is convergent and for
$k_{2}<{k_0\over
2}+k^{c}_{2}$ the partition function is divergent. $\gamma_{str}$ is the 
exponent of the string
susceptibility. This exponent, that in two dimensions is
equal to
$-{1 \over 2}$, controls the divergent part of the 
partition function. More precisely if we fix the genus $g$ and consider the
limit 
 $k_{2} \mapsto \left({k_0\over 2}+k^{c}_{2}\right)^{+}$,
we have that from \ref{diverge}

\begin{equation}
\lim_{k_{2}\mapsto \left({k_0\over 2}+k^{c}_{2}\right)^{+}}Z_{g}(k_{2},k_{0})=
Z_{g}^{reg}+(k_{2}-{k_0\over 2}-k^{c}_{2})^{2-\gamma_{str}}
\label{rossa}
\end{equation}

\noindent with $Z_{g}^{reg}$ the finite part (if any) of the partition function
near the critical line.
It is now clear that the discontinuity points of the partition function are on
this line, so
an eventual critical point is on this line. 

\noindent As regard the sum over the genus $g$ of the surface 
it is divergent and even not Borel
summable. There is a (non-rigorous) way to define a renormalizzation for this
series known 
as double scaling limit (see \cite{witten} for a brief account). Anyway, in
general, we 
will restrict to a given topology, for example the sphere $S^{2}$.

In three and four dimensions in the case of the sphere topology the estimate
\cite{mauro} gives, for the microcanonical partition function, an exponential 
bound. In particular by posing $N_{n-2}=N_{n}\eta$ we have that \cite{mauro}

\begin{equation}
\rho_{a}(S^{n},N_{n},N_{n-2})=g(N_{n},\eta)e^{N_{n}q(\eta)},\;\; n=3,4\;\;\;\;\;
,
\label{micropuz}
\end{equation}

\noindent where $g(N_{n},\eta)$ has a subleading growing in $N_{n}$  respect to
the exponential factor $e^{N_{n}q(\eta)}$. 
As we will see in the 
following geometrical arguments fix the range of $\eta$ in the closed interval 
$[\eta_{1},\eta_{2}]$. So that using the Euler-Maclaurin summation formula
\cite{estrada} we 
can approximate the sum by the integral so that the canonical partition function
is

\begin{equation}
Z(S^{n},N_{n},k_{n-2})\asymp
\int_{\eta_{1}}^{\eta_{2}}g(N_{n},\eta)
e^{N_{n}q_{1}(\eta,k_{n-2})}d\eta
\label{canepuzzu}
\end{equation}

\noindent where $q_{1}(\eta,k_{n-2})=q(\eta)+\eta k_{n-2}$. We can use the
Laplace method 
for giving an asymptotic estimate of the integral \ref{canepuzzu}. We have to
compute the 
point of absolute maximum $\eta^{*}$ of the function $q_{1}(\eta,k_{n-2})$ 
in the interval of integration. In
general this point will be a function of $k_{n-2}$, 
that is $\eta^{*}=\eta^{*}(k_{n-2})$. So we have

\begin{equation}
Z(S^{n},N^{n},k_{n-2})\asymp g(N_{n},\eta^{*}(k_{n-2})) 
e^{N_{n}q_{1}(\eta^{*}(k_{n-2}),k_{2})}\;\;\;\;.
\label{laplace}
\end{equation}

\noindent We define $k^{c}_{n}(k_{n-2})\equiv q_{1}(\eta^{*}(k_{n-2}),k_{2})$,
so that
the 
grand-canonical partition function is

\begin{equation}
Z(S^{n},k_{n},k_{n-2})\asymp
\sum_{N_{n}}g(N_{n},\eta^{*}(k_{n-2})) 
e^{-N_{n}(k_n - k^{c}_{n}(k_{n-2}))}\;\;\;\; .
\label{granpuzzu}
\end{equation}

\noindent It is clear that now the critical line is $k_{n}=k_{n}^{c}(k_{n-2})$
with the same
meaning as in the two dimensional case.

\section{Green Functions in Dynamical Triangulations}
\label{verde}

Since the geometry in General Relativity has a dynamical role, the definition of
the Green 
function in Quantum Gravity is different from ordinary quantum field theory. 

\noindent The formal continuum definition of the unnormalized two point function
is the following

\begin{eqnarray}
&&G_{2}(\Lambda,G,r)\equiv \sum_{Top(M)}
\int {\mathcal D}(g)e^{-S_{E-H}(M,g)} \nonumber \\
&&\int_{(M,g)}d^{n}y\sqrt{det\;g(y)}
\int_{(M,g)}d^{n}x\sqrt{det\;g(x)}
\delta(d_{g}(x,y)-r)
\label{verde}
\end{eqnarray}

\noindent where $\Lambda$ and $G$ are respectively the cosmological and
gravitational 
constant, the sum is over the topological structures of the differentiable
manifold
$M$, the measure ${\mathcal D}(g)$ is over the Riemannian structure allowed by
$M$, 
$S_{E-H}$ is the Einstein-Hilbert action and $d_{g}(x,y)$ is the geodesic
distance
between 
the points $x$ and $y$ on the Riemannian manifold $(M,g)$. To normalize
$G_{2}(\Lambda,G,r)$
we have to divide it by the partition function. So in this definition
we have to
perform a sort of average over all points of the Riemannian manifold $(M,g)$
that are at 
distance $r$. In the discrete we cannot transfer {\it verbatim} the above
definition due to the 
coordinate invariance of the dynamical triangulations. It is necessary to define
the two point
function in such a way that the analogous quantities, in the discrete,  
of $x$ and $y$ in the continuum  have a dependence from the cut-off $a$. 

\begin{figure}
\begin{center}
\epsfig{figure= 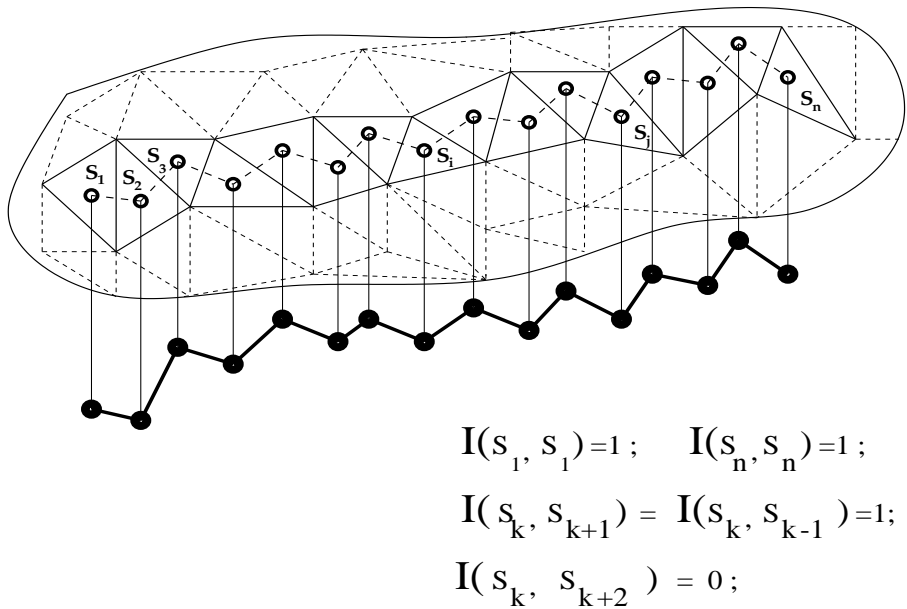,height=8.0truecm,width=12.0truecm}
\caption{A simplicial path between the two simplices $S_1$ and $S_n$ of a
triangulation.
The simplicial path is simple and its dual is obtained by joining the
baricenters or the 
circumcenters of the path's simplices. On this path the properties of the 
incidence matrix $I$ are illustrated }
\end{center}
\label{fig:walk}
\end{figure}

A path in a triangulation $T_{a}$ of a n-dimensional PL-manifolds is a sequence 
$\left\{S_{j}\right\}_{j=1}^{l}$ of 
$l$ $n$-dimensional simplices with the property that $S_{j}$ and $S_{j+1}$
$(S_{j+1}\neq S_{j})$  
have a common face, that is to say $I(S_{j},S_{j+1})=1$ (see \cite{frohlich} p.
526). 
Suppose that the path
does not intersect 
itself, that is to say $I(S_{p},S_{q})=1$ if and only if or $q=p-1$ or $q=p+1$, 
we call it a simple path. 
In this case we can define the 
length of the path $\left\{S_{j}\right\}_{j=1}^{l}$ as the number of the faces
that the $l$ simplices 
have in common, that are $l-1$. This definition comes from graph theory, in fact
if we 
think to the dual of the path $\left\{S_{j}\right\}_{j=1}^{l}$, every simplex is
mapped into a vertex, 
and the face between  two simplices is mapped into a edge joining the two
vertices. So the length 
of the path is equal to the number of the dual edge of the path (see fig.
\ref{fig:walk} ). 

\noindent We define the distance between two simplices $A$ and $B$ of the
triangulation $T_{a}$ as the 
length 
of the simple path which has the minimal length among all simple paths 
$\left\{S_{j}\right\}_{j=1}^{l_q}$ 
such that $S_{1}=A$ and $S_{l_q}=B$.   

Let's now give the definition of unnormalized micro-canonical Green function.
The
unnormalized 
micro-canonical Green function is defined as \cite{1Ambjorn}

\begin{equation}
G({\mathcal PL},r,N_{n},N_{n-2})\equiv \sum_{\# T_{a}({\mathcal
PL},r,N_{n},N_{n-2})}1
\label{green1}
\end{equation}

\noindent where $\sum_{\# T_{a}({\mathcal PL},r,N_{n},N_{n-2})}$  is the number
of inequivalent equilateral triangulations $T_{a}({\mathcal
PL},r,N_{n},N_{n-2})$ of a PL-manifold
${\mathcal PL}$ 
with two labelled $n$-simplices at fixed distance $r$ and with fixed number
$N_{n}$ and $N_{n-2}$ 
of respectively $n$ and $n-2$ simplices.
The normalized partition function 
is obtained by dividing \ref{green1}
by the microcanonical partition function. So on we have that the canonical Green
function is
 
\begin{equation}
G({\mathcal PL},r,N_{n},k_{n-2})\equiv \sum_{N_{n-2}}e^{-k_{n-2}N_{n-2}}
G({\mathcal PL},r,N_{n},N_{n-2})    
\label{green2}
\end{equation}

\noindent and finally the grand-canonical Green function

\begin{equation}
G({\mathcal PL},r,k_{n},k_{n-2})\equiv \sum_{N_{n}}e^{k_{n}N_{n}}
G({\mathcal PL},r,N_{n},k_{n-2})
\label{green3}\;\;\;\; .
\end{equation}

From the grand-canonical Green Function we can define another quantity often
used
in the literature
the {\it susceptibility} $\chi({\mathcal PL},k_{n},k_{n-2})$ which is defined in
the following 
way

\begin{equation}
\chi({\mathcal PL},k_{n},k_{n-2})\equiv \sum_{r=0}^{\infty}
G({\mathcal PL},r,k_{n},k_{n-2})\;\;\;\; .
\label{susceptibility}
\end{equation}

\noindent In the previous definition we can exchange the sum over $r$ with all
the sums 
in the definition of the grand-canonical Green function \ref{green3} up the
micro-canonical Green function 
\ref{green1}. So that we will have

\begin{equation}
\chi({\mathcal PL},k_{n},k_{n-2})=\sum_{N_{n}}e^{k_{n}N_{n}} 
\sum_{N_{n-2}}e^{-k_{n-2}N_{n-2}}
\sum_{r}\sum_{\# T_{a}({\mathcal PL},r,N_{n},N_{n-2})}1\;\;\;\; .
\label{retti}
\end{equation}

\noindent The last two sums in the left hand side of the previous equation are
the number of 
inequivalent triangulations with two labelled n-dimensional simplices. For large
value of 
$N_{n}$ we will expect that asymmetrical triangulations will prevail in the
micro-canonical ensemble 
(this is, for example, true for two-dimensional convex polyhedra \cite{bender},
for same 
particular planar triangulations \cite{tutte2} and  planar maps \cite{richmond}
and it is conjectured
for all planar maps. In four 
dimensions this is verified, so to say, indirectly because once we do this {\it
antsaz} 
and derive the formula for the number of {\it baby universes}(see appendix
\ref{ababy}) 
it is in good
agreement with
the numerical simulations \cite{baby4} \cite{4dynamical} \cite{catteral}). In
this hypothesis
for each labelling of a $n$-simplex we will produce a number of combinatorially
inequivalent 
triangulation that asymptotically is $N_{n}\rho_{a}({\mathcal
PL},N_{n},N_{n-2})$. Finally the leading asymptotic estimate of the number 
of combinatorially inequivalent triangulations with two 
labelled simplices is

\begin{equation}
\sum_{r}\sum_{\# T_{a}({\mathcal PL},r,N_{n},N_{n-2})} 
\asymp N_{n}^{2}\rho_{a}({\mathcal PL},N_{n},N_{n-2})  
\label{semivera}
\end{equation}

\noindent Then the susceptibility \ref{susceptibility} is equal to 

\begin{equation}
\chi({\mathcal PL},k_{n},k_{n-2}) \asymp \sum_{N_{n-2}}N_{n}^{2}
e^{-k_{n}N_{n}}Z({\mathcal PL},N_{n},k_{n-2}) 
\label{rela}
\end{equation}

\noindent so that

\begin{equation}
\chi({\mathcal PL},k_{n},k_{n-2}) \asymp
{\partial^{2} \over \partial k_{n-2}^{2}} Z(k_{n},k_{n-2},{\mathcal PL})\;\;\;\;
.
\label{cuccu}
\end{equation}

\noindent Near the critical line $k_{n}\mapsto (k^{c}_{n}(k_{n-2}))^{+}$ we will
have the following asymptotic behaviour of the susceptibility

\begin{equation}
\lim_{k_{n}\to (k^{c}_{n}(k_{n-2}))^{+}}
\chi({\mathcal PL},k_{n},k_{n-2}) \asymp \chi^{reg}+ 
{1 \over (k_{n}-k^{c}_{n}(k_{n-2}))^{\gamma_{str}}}
\label{scuscia}
\end{equation}

\noindent where $\chi^{reg}$ is the finite part (if any) of the susceptibility.

\chapter{The Geometry of the Elongated Phase of 4-D Simplicial Quantum Gravity}
\label{geo}

\section{Introduction}
\label{int2}

In this chapter we are going to discuss in detail the elongated phase 
of 4-D dynamical triangulations. The early numerical simulations in four
dimensions
 have produced some evidence of the presence of
two distinct phases in the ensemble of four
dimensional triangulations: 
the crumpled phase and the elongated phase \cite{mig1}
\cite{mig2} \cite{ambo} 
(it is interesting to remark that this picture emerges also 
in the simulations of 4D quantum Regge calculus \cite{hamber4} 
\cite{hambe1} \cite{hambe2}). The 
crumpled phase is characterized by the presence of few bones and many
four simplices incident on them. There are very few baby universes
(see appendix \ref{ababy}). The elongated phase is characterized by 
the presence of many bones and of many baby universes of smallest size
(blips). These facts have encouraged people to use the relative 
abundance of baby 
universes as an order parameter to distinguish these two phases.

\noindent Anyway it was still unclear why in the elongated phase there 
is an upper kinematical bound on the values of $\eta={N_2 \over N_{4}}$ 
and a prevalence of simple tree-graphs (branched polymers) which appear
as a model of proliferating baby universes \cite{DeBakkerPh.D.}. 
Established in \cite{mauro} 
in section 
\ref{Wa} we explain that the upper kinematical 
bound on the values of $\eta$ is due to the
Walkup's theorem for the triangulations of the four sphere. In particular 
this theorem fixes the upper value of $\eta$ in the large-$N_4$ limit  
$N_4 \mapsto \infty$. In section \ref{cano} we introduce a recent 
\cite{mauro} \cite{me4} analytical estimate for the canonical partition 
functions for dynamical triangulations in four dimensions relative to the
triangulations of the sphere $S^{4}$. This estimate has the right 
behaviour in the limit $N_4 \mapsto \infty$, because the average curvature
\ref{vmagnetr} in the elongated phase remains constant in perfect agreement 
with  Walkup's theorem. Furthermore this theorem says also that the 
dominating configurations in this phase are "Stacked Spheres". In section 
\ref{elong} we discuss how the stacked spheres have a simple tree 
structure and can be mapped into branched polymers structures (see 
appendix \ref{abranch}). However the problem arises that the map is not a one to
one, 
since there are more stacked spheres than branched polymers. So in section 
\ref{stacco} we consider the action of dynamical triangulations restricted 
to stacked spheres and show that the partition function over the ensemble
of the stacked spheres is equal or greater than the partition 
function of branched polymers which are the immages of the above map, and 
equal or less than a particular model of branched polymers.  
This argument
shows that the susceptibility exponent $\gamma_s$ is less than one. To 
show that $\gamma_s={1\over 2}$ we use an argument taken from the physical 
literature \cite{1Ambjorn} \cite{LJ}. This argument
follows the line of a model introduced 
for the first time in the theory of random surfaces
\cite{jonsson}.

\noindent We take into account some particular kind of non standard
triangulations   
and put it in correspondence with the model of stacked spheres considered 
as a model of proliferating baby universes. Physical arguments ensure that 
the two models belong to the same universality class and we show that 
the susceptibility exponent is $\gamma_s={1 \over 2}$ as for the branched 
polymers.

\noindent These types of analytical results were already well known from
Monte Carlo simulations. An early analytical attempt, in the direction of
interpreting
the elongated phase as a stacked spheres phase, has been done in reference
\cite{semi} in which a map from a sort of stacked sphere triangulations to
branched polymers is considered. Anyway the deep reason for 
polymerization mechanism and with the kinematical 
bound on the values of $\eta$ is given only by Walkup
theorem.

All these facts point out that the elongated phase of 4-D dynamical
triangulations is a mean field phase, so that we expect that if we take the
continuum limit
in the 
framework of Euclidean lattice field theory (c.f. \cite{sokal} chap. 15), in 
this phase, we will obtain a probabilistic Feynman measure which is Gaussian,
or,
in other words, a trivial theory 

\section{Walkup's Theorem}
\label{Wa}

As anticipated in section \ref{mode} of chapter \ref{dyna} due to geometrical
constraints the 
variable $\eta$ in four dimensions can vary in a finite interval
$[\eta_{1},\eta_{2}]$.
The limits $\eta_{1}$ and $\eta_{2}$ of this interval are fixed respectively by
the following 
Walkup theorem and Dehn-Sommerville equations \cite{Walkup}. 

Let us label the bones \cite{me} by an index $\alpha$ and denote by $q(\alpha)$ 
the number of four simplexes
incident on it; the average number of simplexes incident on a bone is
\begin{eqnarray}
b\equiv \frac{1}{N_{2}}\sum_{\alpha}q(\alpha)=
10\left(\frac{N_4(T_{a})}{N_{2}(T_{a})} \right)
\label{avnumber}
\end{eqnarray}
since each simplex is incident on $10$ different bones. b ranges 
between two kinematical
bounds which follow from Dehn-Sommerville and a theorem
by Walkup \cite{Walkup}. This latter theorem is also relevant in
classifying the "elongated phase" of four-dimensional simplicial quantum
gravity as we shall see below 

\begin{theorem}
 If T is a triangulation of a closed, connected
four-dimensional manifold then

\begin{equation}
N_1(T)\geq 5N_0(T) -{15\over 2}\chi(T)
\label{importante}
\end{equation}

\noindent
 Moreover, equality holds if and only if $T\in H^{4}(1-{1\over 2}\chi(T))$, 
where the class of triangulations
 $H^{4}(n)$ is defined inductively according to: 
(a) The boundary complex of any abstract five-simplex ($Bd\sigma$)
is a member of $H^{4}(0)$. (b) If $K$ is in $H^{4}(0)$ and $\sigma$ is a
four-simplex of $K$, then $K'$ is in $H^{4}(0)$, where $K'$ is any complex
obtained from $K$ by deleting $\sigma$ and adding the join of the boundary
complex $Bd\sigma$ and a new vertex distinct from the vertices of $K$. (c) 
If $K$ is in $H^{d}(n)$, then $K'$ is in $H^{d}(n+1)$ if there exist two 
four-simplexes $\sigma_1$ and $\sigma_2$ with no common vertices and a 
dimension preserving simplicial
map $\phi$ from $K - {\sigma_1} - {\sigma_2}$ onto $K'$ which identifies
$Bd\sigma_1$ with $Bd\sigma_2$ but otherwise is one to one.
\end{theorem}

\noindent 
In other words $H^{4}(0)$ is built up by gluing together
five-dimensional simplexes through their four dimensional faces and
considering only the boundary of this resulting complex. $H^{4}(n)$
differs from $H^{4}(0)$ by the fact that it has n handles. This way of
constructing a triangulation of a four-sphere has a natural connection with the
definition of a baby universe (see appendix \ref{ababy}).  

\noindent A baby universe is associated with a triangulation in which 
we can distinguish two pieces. A piece that contains the majority of 
the simplices of the triangulation that is called the "mother", and a 
small part called the "baby". In the "minbus" (minimun neck baby
universes) the two parts are glued together along the boundary of a 
four dimensional simplex (in four dimensions) that is the "neck" of
the baby universes. Thus the "stacked
spheres" can be considered as a network of minbus in which the
mother is disappeared and  the baby universes have a minimal volume, that
is the boundary of a five simplices minus the simplices of the necks through
which they are glued to the others . 
We will exploit this parallel in section \ref{elong} to
give an estimate of the number of distinct stacked spheres. 

\noindent In four-dimensions Dehn-Sommerville equations \cite{Walkup} are

\begin{eqnarray}
N_{0}-N_{1}+N_{2}-N_{3}+N_{4}&=&\chi \\ \nonumber
2N_{1}-3N_{2}+4N_{3}-5N_{4}&=&0 \\ \nonumber
2N_{3}=5N_{4} \;\;\;\; ,
\label{esplico}
\end{eqnarray}

\noindent substituting the third equation in the second we have

\begin{equation}
N_{1}={3\over 2}N_{2} -{5\over 2}N_{4}\;\;\;\; ,
\label{miscela}
\end{equation}

\noindent and substituting these last equation in the first equation of
\ref{esplico} we get

\begin{equation}
N_{0}-{1\over 2}N_{2}+N_{4}=\chi
\label{rimiscela}\;\;\;\; .
\end{equation}

\noindent This imply that

\begin{equation}
b\leq 5+{10\chi \over N_2}\;\;\;\; ,
\label{sup}
\end{equation}

\noindent and substituting \ref{miscela} and \ref{rimiscela} in \ref{importante}
we obtain

\begin{equation}
b\geq 4-{10\chi \over N_2}\;\;\;\; .
\label{inf}
\end{equation} 

\noindent Thus in the limit of large $N_2$, $4\leq b\leq 5$.

\section{Canonical Partition Function}
\label{cano}

Now we introduce more extensively the results in four dimensions for the
canonical 
partition function which has been rapidly summarized in section \ref{mode} of
chapter
\ref{dyna}. 

\noindent Recently it has been analytically shown \cite{mauro} \cite{me4} 
that the dynamical
triangulations in four dimensions are characterized by two phases: a
strong coupling phase in the region $k_2^{inf}=log 9/8 < k_2 <
k_2^{crit}$ and a weak coupling phase for  $ k_2 > k_2^{crit}$.
$k_2^{crit}$ is the value of $k_2$ for which in the infinite volume
limit the theory has a phase transition from the strong to the weak
phase (for a detailed analysis see \cite{mauro}). The 
transition between these two phases is characterized by the fact that
the sub dominant asymptotic of the number of distinct
triangulations passes from an exponential to polynomial behaviour 
(c.f.\cite{mauro}). 
Presently the precise value of $k_2^{crit}$ has not been established
yet, it is just known that it is close but distinct from $k_2^{max}=log4$,
recent
numerical simulations suggest that $k^{crit}_{2}=1.24$.

\noindent In the strong coupling phase the leading term of the 
asymptotic expansion of the canonical partition function is (c.f.
\cite{mauro}) 

\begin{eqnarray}
Z(N_4,k_2)=c_4 \left({A(k_2)+2 \over 3A(k_2)}\right)^{-4}\;\;\;
N{_4}^{-5}exp[-m(\eta^{*})N^{1\over n_H}]\nonumber\\
exp\left[[10log{{A(k_2)+2}\over 3}]N_4\right]
\label{strong}
\end{eqnarray}

where for notational convenience we have set
\begin{eqnarray}
 A(k_{2})&\equiv&
 { \left [\frac{27}{2}e^{k_{2}}+1+
\sqrt{(\frac{27}{2}e^{k_{2}}+1)^2-1} \right ]}^{1/3}\\ \nonumber
&+&
{ \left [\frac{27}{2}e^{k_{2}}+1-
\sqrt{(\frac{27}{2}e^{k_{2}}+1)^2-1} \right ]}^{1/3}-1
\end{eqnarray}
and 
\begin{eqnarray}
\eta^*(k_{2})=\frac{1}{3}(1-\frac{1}{A(k_{2})})
\;\;\;\; .
\end{eqnarray}

The explicit form of 
$m(\eta^{*}(k_2))$ and $n_H$, the Hausdorff dimension, are 
at present unknown. 

In the weak phase $ k_2 > k_2^{crit}$ the
number of distinct dynamical triangulations with equal curvature assignment have
a power law behaviour in $N_4$. The phase is characterized by two distinct
asymptotic regimes
of the leading term of the canonical partition function: the critical coupling
regime and
the weak coupling regime. In the critical coupling regime 
$k_2^{crit}<k_2<k_2^{max}$ we have  

\begin{eqnarray}
Z(N_4,k_2)\asymp c_4 \left({A(k_2)+2 \over 3A(k_2)}\right)^{-4}\;\;\;
N{_4}^{\tau(\eta^{*})-5}\nonumber\\
exp\left[[10log{{A(k_2)+2}\over 3}]N_4\right]
\label{crit}\;\;\;\; ,
\end{eqnarray}
  
\noindent in which $\tau(\eta^{*})$ is not explicitly known.

Instead in the weak coupling regime $k_2>k_2^{max}$ 

\begin{eqnarray} 
Z(N_4, k_{2})&\asymp&\frac{c_4e}{\sqrt{2\pi}}
\frac{\eta_{max}^{-1/2}(1-2\eta_{max})^{-4}}{\sqrt{(1-3\eta_{max})(1-
2\eta_{max})}}
(\widehat{N_4+1})^{\tau -11/2}\nonumber \\
&&\cdot\frac{e^{(\widehat {N_4+1})f(\eta_{max},k_2)}}{k^{sup}_{2}-k_{2}}
\label{lower}
\end{eqnarray}

in which $\eta^{max}={1\over 4}$ for the sphere
$S^{4}$,$\widehat{N_4+1}=10(N_4+1)$  and 

\begin{equation}
f(\eta,k_{2})\equiv
-\eta\log\eta+(1-2\eta)\log(1-2\eta)-
(1-3\eta)\log(1-3\eta)+ k_{2}\eta
\label{effe}\;\;\;\; .
\end{equation}

\noindent From this form of the canonical partition function it follows that in
the 
case of the sphere $S^{4}$ and in the infinite volume limit the average
value of $b$, see equation (\ref{avnumber}), is a decreasing function of
$k_2$ in the critical coupling regime and it is constantly equal to $4$ in the
weak coupling regime, that is to say 

\begin{eqnarray}
\lim_{N_4 \mapsto \infty}<b>_{N_4}={1\over\eta(k_2)},\;\;\;\; k_2^{crit}\leq 
k_2\leq k_2^{max} \nonumber\\
\lim_{N_4 \to \infty}<b>_{N_4}=4,\;\;\;\; k_2 \geq k_{2}^{max}
\label{curv}\;\;\;\; .
\end{eqnarray}

If we look at the average curvature we have

\begin{equation}
\lim_{N_4 \to \infty} {1 \over N_4}<\sum_{B}K(B)vol_4(B)>
= \pi a^2 \sqrt{3}\left(10\eta^{*}(k_2) - 
{5\over \pi}cos^{-1}{1\over4}\right)
\;\;\;\;\;\;\;\;\;\; k_{2}^{crit}< k_2\leq k_{2}^{max} 
\label{vmagnetl}
\end{equation}

and

\begin{equation}
\lim_{N_4\to\infty}{1 \over N_4}<\sum_{B}K(B)vol_4(B)>
= \pi a^2 \sqrt{3} \left({5\over2} - 
{5\over \pi}cos^{-1}{1\over4}\right)\;\;\;\;\;\;\;\;\;\; 
k_2 > k_2^{max}\;\;\;\; ,
\label{vmagnetr}
\end{equation}

in which $vol_4(B)$ is the volume of the four simplexes incident on the bone
$B$, and it is
$a^{4}{\sqrt{5}\over 2^{4}\cdot 6}q(B)$. As we have already said the value of
the average
curvature is 
saturated. 

\noindent This result was already known in numerical simulations
\cite{DeBakkerPh.D.} and, as we will see in detail in the next section,
it is due to the prevalence of stacked spheres in the weak coupling 
regime. 

\section{Elongated Phase}   
\label{elong}

We have seen that in the case of triangulations of the sphere $S^{4}$
for $k_2 \mapsto log\;4$, in the infinite volume limit $<b>_{N_4}
\mapsto 4$. Walkup's theorem \cite{Walkup} implies that the minimum
value of $b$ is reached on triangulations $K$ of the sphere $S^4$ that
belong to $H^{4}(0)$ (stacked spheres). Then for $k_2>
log\;4$  one has $<b>_{N_4}=4$,
that means that in this region of $k_2$ the statistical
ensemble of quantum gravity is strongly dominated by stacked spheres. 

\noindent As explicitly shown in \cite{Walkup},  
the elements of $H^{4}(0)$ can be put in correspondence with a
tree structure. Let us recall that a $d-$dimensional simplicial
complex $T$, $d\geq 1$, is called a simple $d-$tree if it is the closure
of its $d-$simplexes $\sigma_1,...,\sigma_t$ and these $d-$simplexes
can be ordered in such a way that: 

\begin{equation}
Cl\;\sigma_j\bigcap\left\{ \bigcup_{i=1}^{j-1} Cl\;\sigma_i \right\}= 
Cl\;\tau_j
\end{equation}

\noindent for some $(d-1)$-face $\tau_j$ of $\sigma_j$, $j\geq 2$, and
where the $\tau_j$ are all distinct. This ordering of the simplexes of $T$
induces a natural ordering of its vertices in $ v_1,...,v_{t+d}$, where
$v_{i+d}$ is the vertex of $\sigma_i$ not in $Cl\;\tau_i$. Note that the
interior
part of $T$ contains the simplexes $\sigma_i$ and faces $\tau_i$. The
boundary of $T$, $Bd\;T$, consists of the boundary of the $\sigma_i$
minus the $\tau_i$, and is topologically equivalent to $S^{d-1}$. 

\noindent It can be shown\cite{Walkup}, and it is very easy to check, that any
element of $K\in H^d(0)$ is the boundary of a simple $(d+1)$-tree $T$, and
that $K$ determines uniquely the simple $(d+1)$-tree $T$ for $d\geq 2$.

Note that any stacked sphere $K\in H^{4}(0)$ can be mapped into a tree graph 
(see also the reference \cite{Davide}).
This mapping is defined in the following
way, let's consider the (unique) simple five-simple tree $T$ associated
with $K$, every five-simplex is mapped into a vertex and every
four-dimensional face in common with two five simplexes is mapped into an
edge which has endpoints at the two vertices which represent the two
five simplexes (see fig. 3.1). 

\begin{figure}
\begin{center}
\epsfig{figure=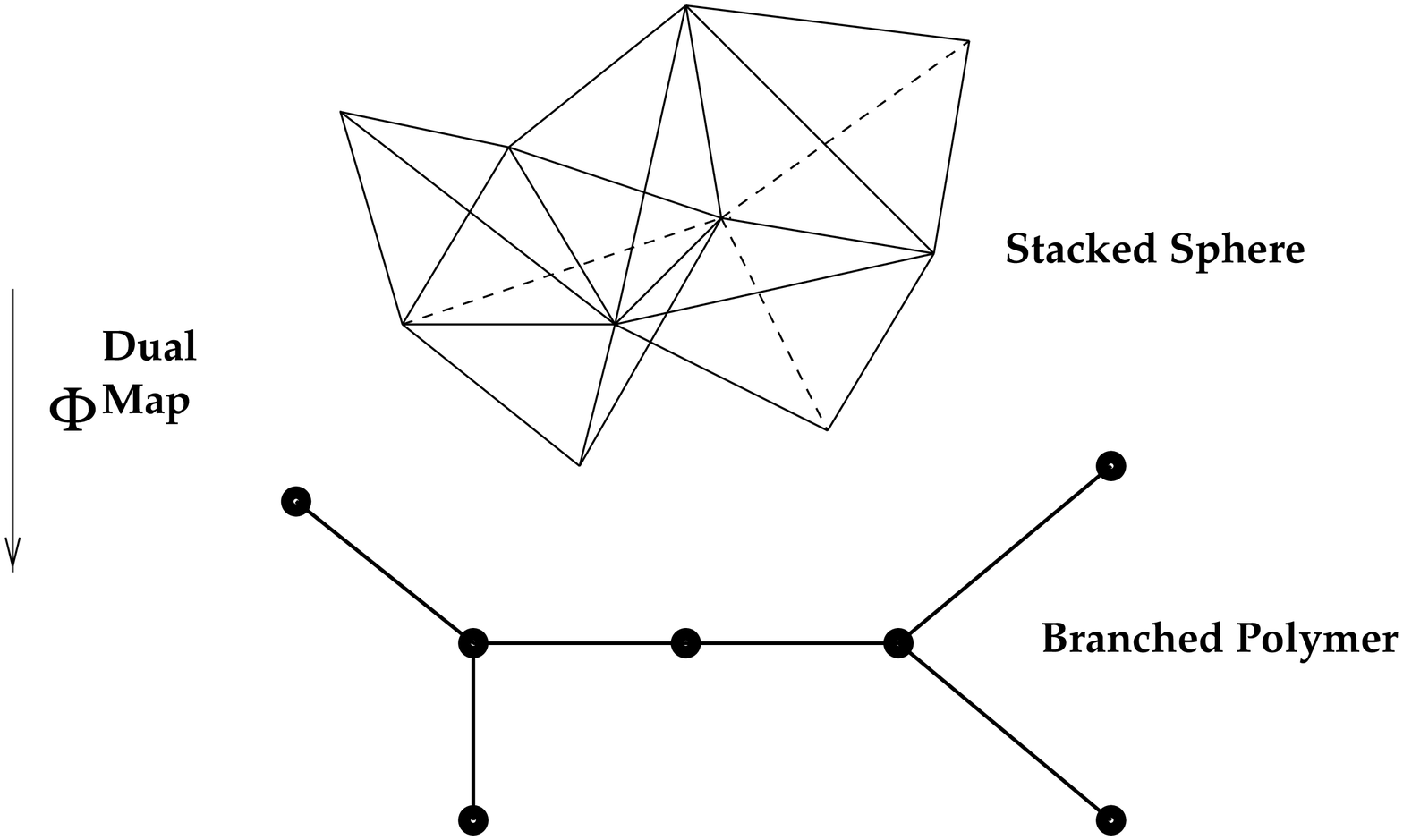,height=8.0truecm,width=12.0truecm}
\caption{ Two dimensional stacked sphere that is mapped into a branched polymer
by the dual map
$\Phi$}
\end{center}
\label{fig:stack}
\end{figure}

\noindent Since the map between $K$ and $T$ is one to one, we have
a map from a stacked-sphere into a tree-graph whose number of links at
every vertex can be at most $6$ ( since a five simplex has $six$ faces).
This map, from the stacked spheres $H^{4}(0)$ to the simple tree-graphs,
is not one to one. The mathematical reason is that the previous
construction maps every $T$ to a tree-graph by an application that is
the dual map restricted to the five and four dimensional simplexes of
$T$ in its domain and whose image is the tree graphs that are the
$1$-dimensional skeleton of the dual complex. It is well known that the
dual map is a one to one correspondence only if we take in account the
simplexes of $T$ of any dimension. It follows that our map is such that
a simple tree-graph may correspond to many stacked spheres. To
illustrate better this point we shall use a picture closer to the
physical intuition. Let us consider a stacked sphere $K$ and its
simple tree $T$. Consider on $T$ one of its ordered simplexes $\sigma_j$
and let $\tau_j$ be the face that it shares with one of the simplexes
${\sigma_1,..., \sigma_t}$ introduced previously. Repeating the same
arguments used in the calculations of the inequivalent triangulations
for  baby universes \cite{jain} (see also appendix \ref{ababy}), we can cut $T$
in $\tau_j$ in two
parts, such that each part has two copies of $\tau_j$ as a part of its
boundary. Since $\tau_j$ is a four dimensional simplex it is easy to
admit that we can glue this two different copies in $5\;\cdot 4\;\cdot
3$ ways to rebuild again a simple five tree $T'$. In general all the
possible gluings will generate distinct triangulations  and
consequently distinct stacked spheres (if the two parts are highly 
symmetric triangulations some of the $60$
ways of joining will not be distinct but, since for large $N_4$ asymmetrical
triangulations will dominate, the number of case in which this will happen will
be
negligible). The corresponding tree graph associated with these configurations
$T'$
will always be the same since this operation has not modified the one
skeleton of the tree $T$. 

In statistical mechanics simple-tree
structures correspond to "branched-polymers" (see appendix \ref{abranch}). So we
have
that for $k_2\geq k_2^{max}$ the dominant configurations for the
triangulations of $S^4$ are branched polymers. This fact was already
observed in numerical simulations in four dimensions. 
In \cite{Ambjorn2}a network of
baby universes of minimum neck (minbu) and of minimum size (blips) 
without a mother universe was obtained. These have been
interpreted as branched polymer like structures. 

\section{Stacked Spheres and Branched Polymers}
\label{stacco}

In this section we will establish a parallel between the standard mean field 
theory of branched polymers and  the stacked spheres, in the sense
that we will use the counting techniques of branched polymers to give an upper
and
lower estimate of the number of inequivalent stacked spheres.  
As we have stressed in the previous section there exists a map between
the stacked spheres and tree graphs and this map is {\it not} one to one
in the sense that there are more stacked spheres than tree graphs. 
 This
means that the number of inequivalent stacked spheres with a fixed
number $N_5$ of five-simplexes is bounded below by the corresponding
number of tree graphs. Now we will study the statistical behaviour of
the tree graphs  using the measure of simplicial quantum gravity 
and restricting it
to the stacked spheres. In this analysis we shall follow the theory
of branched polymers as explained in appendix \ref{abranch}.
The reader may refer to it 
for the details. A more mathematical analysis on the
enumeration of inequivalent tree graphs is contained in \cite{Otter}. 

First of all we notice that the boundary of a five-simplex has six 
four-simplexes and every edge of a tree graph correspond to a cancellation of
two four-simplexes in the corresponding boundary of the stacked sphere.
Since in a tree graph with $N_5$ vertices there are $N_5 - 1$ edges, we 
have that the boundary of a stacked sphere, whose corresponding tree
graph has $N_5$ points, is made by a number $N_4$ of four-simplexes

\begin{equation}
N_4=4N_5 + 2
\;\;\;\; .
\end{equation}  

\noindent From the condition for stacked spheres (\ref{importante}), we
have that the Einstein-Hilbert action, for dynamical triangulations,
restricted to the stacked spheres is 

\begin{equation}
S=N_4(k_4 - {5\over 2}k_2) - 5k_2
\label{res}\;\;\;\; . 
\end{equation}

It follows that the Gibbs factor for the ensemble of tree graphs
(branched polymers) corresponding to stacked spheres is

\begin{equation}
exp\left(-(4N_5 + 2)(k_4 - {5\over 2}k_2) +5k_2 \right)
\label{gibbs}\;\;\;\; .
\end{equation}

Following the same notation as in \ref{root} , let
$r^{6}(N_5)$ be the number of inequivalent 
rooted tree graphs with $N_5$ vertices and
with at most six incident edges on each vertex, with one rooted vertex
with one incident edge. 
$\xi^{6}(N_5)$ is the number of inequivalent tree graphs
with $N_5$ vertices. By \ref{iden1} we have

\begin{equation}
\xi^{6}(N_5)={1\over N_5}r^{6}(N_5 +1)
\label{asy}
\end{equation}

\noindent Now let $R^{6}(k_4,k_2)$ be the partition function for rooted
tree graphs with statistical weight (\ref{gibbs}), and $Z^{6}(k_4,k_2)$ be
the partition function for unrooted tree graphs. We have 

\begin{eqnarray}
R^{6}(k_4,k_2)&\equiv& \sum_{N_5=2}^{\infty} 
e^{-\left(4(N_5 -1)+2\right)(k_4-{5\over 2}k_2) +5k_2}r^{6}(N_5)\\
Z^{6}(k_4,k_2)&\equiv& \sum_{N_5=1}^{\infty} 
e^{-(4N_5 +2)(k_4-{5\over 2}k_2) +5k_2}\xi^{6}(N_5 )
\label{part}
\end{eqnarray}

\noindent Let's define

\begin{eqnarray}
R^{*6}(\triangle k_4)&\equiv&\left(e^{-2(k_4-5)}R^{6}(k_4,k_2)\right)
=\sum^{\infty}_{N_5=2}e^{-4(N_5-1)\triangle k_4}r^{6}(N_5)\\
Z^{*6}(\triangle k_4)&\equiv&\left(e^{-2(k_4-5)}Z^{6}(k_4,k_2)\right)
=\sum^{\infty}_{N_5=1}e^{-4N_5\triangle k_4}\xi^{6}(N_5)
\label{ric}\;\;\;\; ,
\end{eqnarray}

\noindent where $\triangle k_4\equiv k_4 - {5\over 2}k_2$.
 It is easy to see, from \ref{asy}, that 
 
 \begin{equation}
 R^{*6}(\triangle k_4)=-{1\over 4}{d\over d\triangle k_4}Z^{*6}(\triangle k_4)
 \label{pappy}\;\;\;\;.
 \end{equation}
 
 \noindent From \ref{rivoto} we know that the asymptotic behavior of the
susceptibility $\chi(k_4,k_2)$  is given by 
 
 \begin {equation}
 \chi(k_4,k_2)\asymp {\partial^{2}\over \partial k_4^{2}} Z(k_4,k_2)
 \label{scuccu}
 \end{equation}
 
 \noindent Then
 
 \begin{equation}
 \chi(\triangle k_4)\asymp {d^{2}\over d(\triangle k_4)^2}Z^{*6}(\triangle k_4) 
 \asymp {d\over d(\triangle k_4)} R^{*6}(\triangle k_4) 
 \label{sugg}
 \end{equation}
 
Now we shall study the critical behaviour of this system in order to
obtain some information about the critical behaviour of the stacked spheres. 

\noindent From the identity \ref{recurro} we have that 

\begin{equation}
R^{*6}(\triangle k_4)=e^{-4\triangle k_4}
\left(\sum_{\gamma=0}^{5}{1\over \gamma !}
[R^{*6}(\triangle k_4)]^{\gamma}\right)
\label{rec}\;\;\;\; .
\end{equation} 

\noindent This last equation suggest to define a function $F(R^{*6},\triangle
k_4)$ of
$R^{*6}$ and $\triangle k_4$ and to apply the implicit function theorem to it in

order to obtain information on $\triangle k_4$ as a 
function of $R^{*6}$. 

\noindent More precisely the following function $F(R^{*6},\triangle k_4)$ of
$R^{*6}$ and $\triangle k_4$

\begin{equation}
F(R^{*6},\triangle k_4)\equiv R^{*6}-e^{-4\triangle k_4}
\left(\sum_{\gamma=0}^{5}{1\over \gamma !}
[R^{*6}]^{\gamma}\right)
\label{suso}
\end{equation}

\noindent at the point $\left((R^{*6})^{0},(\triangle k_4)^{0}\right)$ where 

\begin{equation}
F((R^{*6})^{0},(\triangle k_4)^{0})=0\;\;\;\; ,
\label{dini}
\end{equation}

\noindent if
it is true the following condition

\begin{equation}
\left.{\partial F \over 
\partial \triangle k_4}\right|_{\left((R^{*6})^{0},
(\triangle k_4)^{0}\right)}\;\;\;\; ,
\label{dervf}
\end{equation}

\noindent  defines by the implicit function theorem, locally, $\triangle k_4$ as

function of $R^{*6}$.

\noindent Differentiating 
equation \ref{rec} we get a differential equation for $R^{*6}(\triangle k_4)$

\begin{equation}
{d\over d\triangle k_4}R^{*6}(\triangle k_4)=- 4
R^{*6}(\triangle k_4)\left[1+{e^{-4\triangle k_4}\over
5!}\big(R^{*6}(\triangle k_4)\big)^{5}
-R^{*6}(\triangle k_4)\right]^{-1}
\label{diffe}
\end{equation}

\noindent So $R^{*6}(\triangle k_4)$ shows a singularity when
\begin{equation}
R^{*6}\left((\triangle k_4)^{c}\right)=1+{e^{-4(\triangle k_4)^{c}}\over
5!}\left(R^{*6}\big((\triangle k_4)^{c}\big)\right)^{5}
\label{crite}
\end{equation}

\noindent This equation has only one positive solution 
and since ${d\over d\triangle k_4}R^{*6}|_{(\triangle
k_4)^{c}}$ diverges, the inverse function tends to zero 

\begin{equation}
\left.{d(\triangle k_4)(R^{*6})\over dR^{*6}}\right|_{(\triangle
k_4)^{c}}\mapsto 0
\label{inver}
\end{equation}

\noindent Observe that $F(R^{*6},\triangle k_4)$ is an analytic function, 
and in the point $\left((R^{*6})^{c},(\triangle k_4)^{c}\right)$ fixed 
by equation \ref{crite}, we have 

\begin{equation}
\left.{\partial F \over 
\partial \triangle k_4}\right|_{\left((R^{*6})^{c},(\triangle k_4)^{c}\right)}
=4e^{-4(\triangle k_4)^{c}}
\left(\sum_{\gamma=0}^{5}{1\over \gamma !}
[(R^{*6})^{c}]^{\gamma}\right)>0
\label{dderi}
\end{equation}

\noindent and pairwise

\begin{equation}
\left.{\partial F \over 
\partial R^{*6} }\right|_{\left((R^{*6})^{c},(\triangle k_4)^{c}\right)}
=1-e^{-4(\triangle k_4)^{c}}
\left(\sum_{\gamma=0}^{5}{1\over \gamma !}
[(R^{*6})^{c}]^{\gamma}\right)=0\;\;\;\; ,
\label{comeprima}
\end{equation}

\noindent this last equation explains why we cannot express in the neighbourhood
of
$\left((R^{*6})^{c},(\triangle k_4)^{c}\right)$  
$R^{*6}$ as function of $\triangle k_{4}$
by applying the implicit function theorem to \ref{suso}. These arguments
are enough to say that in the neighbourhood of 
$\left((R^{*6})^{c},(\triangle k_4)^{c}\right)$ $(\triangle k_4)(R^{*6})$ is an
analytic 
function of $R^{*6}$. Let us fix the following notation 

\begin{eqnarray}
F_{R^c}\equiv \left.{\partial F \over 
\partial R^{*6} }\right|_{\left((R^{*6})^{c},(\triangle
k_4)^{c}\right)}\;\;&&\;\; 
F_{\triangle k_{4}^{c}}\equiv \left.{\partial F \over 
\partial \triangle k_4}\right|_{\left((R^{*6})^{c},(\triangle
k_4)^{c}\right)}\;\;\;\; 
\\ \nonumber 
F_{R^c R^c}&\equiv&\left.{\partial^{2} F \over 
\partial (R^{*6})^{2} }\right|_{\left((R^{*6})^{c},(\triangle k_4)^{c}\right)}
\label{defio}
\end{eqnarray}

\noindent and so on. It is straightforward to calculate that

\begin{equation}
F_{R^c R^c}=-e^{-4(\triangle k_4)^{c}}
\left(\sum_{\gamma=0}^{3}{1\over \gamma !}
[(R^{*6})^{c}]^{\gamma}\right)< 0 \;\;\;\; .
\label{storia}
\end{equation}

\noindent Applying again the implicit function theorem we obtain

\begin{equation}  
\left.{d^{2}(\triangle k_4)(R^{*6})\over d(R^{*6})^{2}}\right|_{(\triangle
k_4)^{c}}
=-{F_{R^c R^c} \over F_{\triangle k_{4}^{c}}}> 0 \;\;\;\; .
\label{beauty}
\end{equation}

All these facts allow to write the following  expansion

\begin{eqnarray}
\triangle k_4(R^{*6}) - {\triangle k_4}^{c}&=&
{1\over 2}\left.{d^{2}(\triangle k_4)(R^{*6})
\over d(R^{*6})^{2}}\right|_{R^{*6}\left((\triangle k_4)^{c}\right)} 
\left(R^{*6}(\triangle k_4) - R^{*6}({\triangle k_4}^{c})\right)^{2}\nonumber\\
& &+ o\left(\left(R^{*6}(\triangle k_4) - R^{*6}({\triangle
k_4}^{c})\right)^{2}\right)
\label{exp}\;\;\;\; .
\end{eqnarray}

\noindent So near $(\triangle k_4)^{c}$ we can write

\begin{equation}
R^{*6}(\triangle k_4)\asymp R^{*6}\big((\triangle k_4)^{c}\big)
+ C\left((\triangle k_4) - (\triangle k_4)^{c}\right)^{1\over 2}
\label{svilu}
\end{equation}

\noindent From (\ref{sugg}) we get

\begin{equation}
\chi^{6}(\triangle k_4)\asymp
\left((\triangle k_4) - (\triangle k_4)^{c}\right)^{-{1\over 2}}
\label{scippu}
\end{equation}

\noindent This means that the critical exponent of the susceptibility
for the system of these tree graphs is $\gamma={1\over 2}$.

The partition function that has been studied is a lower estimate of the
partition function of the stacked spheres. Consider, now, a stacked
sphere $K$ and a face $\tau_j$ through which two five-simplexes are
glued together (a link on the corresponding tree graph). We can glue 
two four face of a stacked sphere in $5{\cdot} 4{\cdot} 3$ different
ways, for large number of simplexes this will generate distinct
configurations of stacked spheres whose associated tree-graph is always
the same. Repeating the same argument for every $j$, $j=1,...,N_5-1$,
(that is to say for every link of the dual tree graph) we obtain a
factor, $(5{\cdot} 4{\cdot} 3)^{N_5-1}$, that multiplied by
$r^{6}(N_5)$ and $\xi^{6}(N_5)$, gives an upper bound on the number of ,
respectively, rooted and unrooted inequivalent stacked spheres. In other
words 

\begin{equation}
Z_{tree}(k_4,k_2)\leq Z_{s.s.}(k_4,k_2) \leq {\tilde Z}_{tree}(k_4,k_2)
\label{dise}
\end{equation} 

\noindent where $Z_{tree}(k_4,k_2)$ is the partition function for tree
graphs studied above, $Z_{s.s.}(k_4,k_2)$ is the partition function for
stacked spheres and ${\tilde Z}_{tree}(k_4,k_2)$ is the partition
function for tree graphs with the additional weight defined above.

\noindent A similar analysis as for $Z_{tree}(k_4,k_2)$ shows that the
critical line of ${\tilde Z}_{tree}(k_4,k_2)$ is, of course, a straight
line parallel and above respect to the $k_4$ axis to $Z_{tree}(k_4,k_2)$
one, and the susceptibility exponent is again $\gamma ={1\over 2}$. 

\noindent Obviously the estimates (\ref{dise}) are true for the canonical
partition functions too, 

\begin{equation}
Z_{tree}(N_4,k_2)\leq Z_{s.s.}(N_4,k_2) \leq {\tilde Z}_{tree}(N_4,k_2)
\label{estimo}\;\;\;\; ,
\end{equation}

\noindent and the previous calculations show that

\begin{equation}
Z_{tree}(N_4,k_2)\asymp N_{4}^{-{5\over 2}}
e^{-N_4(k_4 -{5\over 2}k_2 -t_{4})}
\label{estimo1}
\end{equation}

and

\begin{equation}
{\tilde Z}_{tree}(N_4,k_2)\asymp N_{4}^{-{5\over 2}}
e^{-N_4(k_4 -{5\over 2}k_2 +t_{4}-{1\over 4}log\;60)}
\label{estimo2}\;\;\;\; ,
\end{equation}

\noindent where $t_4$ is a constant that may be calculated by the 
equation \ref{crite}. More easily from the table in reference \cite{Otter}
we get that $t_4 \approx {1\over 4}log\;0.34$. 

\noindent The circumstance that in the weak coupling region 
the partition function of
quantum gravity, as found in an analytically way in \cite{mauro}, is
strongly dominated  by stacked spheres, allows us to write 

\begin{equation}
Z_{s.s.}\asymp N_4^{\gamma_s - 3}e^{N_4 k_4^{c}(k_2)}
\label{funct}\;\;\;\; ,
\end{equation}

\noindent where $\gamma_s$ is the susceptibility exponent \cite{4dynamical}.

\noindent 
Thus the critical line of the stacked spheres is a straight line parallel and 
among the critical lines of the systems of the two branched polymers. This 
implies

\begin{equation}
 k_4 -{5\over 2}k_2 +t_{4} \leq k_4^{c}(k_2) \leq 
 k_4 -{5\over 2}k_2 +t_{4}-{1\over 4}log\;60
\label{estimo3}
\end{equation}

Moreover from the equations \ref{estimo1} and \ref{estimo2} 
the one loop green functions \cite{1Ambjorn} 
of the two model
of 
branched polymers have respectively , near their critical lines ,  
the asymptotic behaviour

\begin{equation}
G_{tree}(\triangle k_4)\asymp cost_{1} + (\triangle k_4 - {\triangle
k_4}^{c})^{1\over 2}
\;\;\;\;\;\;\;
{\tilde G}_{tree}(\tilde{\triangle k_4})\asymp
cost_{2} + (\tilde{\triangle k_4} - {\tilde{\triangle k_4}}^{c})^{1\over 2}
\label{1green}
\end{equation}

This last equation together the equations \ref{estimo} and \ref{funct}
prove that the one loop function of the stacked spheres $G_{s.s.}(k_4,k_2)$
near the critical line has the asymptotic behaviour

\begin{equation}
G_{s.s.}(k_4,k_2)\asymp cost_{3} +(k_4 -k_4^{c}(k_2))^{1-\gamma_s}
\label{sgreen}\;\;\;\; ,
\end{equation}

\noindent with $\gamma_s < 1$ (the value $\gamma_s=1$ is not allowed because in
this case 
the one loop green function of stacked spheres at the critical line would have a
behaviour
like $\sum_{N_4=5}^{\infty}1/N_4$ that is divergent and then incompatible with
the
upper bound 
given by the second equation of \ref{1green}).

Motivated by physical considerations, we can use a well known argument
\cite{1Ambjorn} \cite{LJ} in favor of the fact that the susceptibility exponent
of the stacked spheres is $\gamma_s={1\over 2}$. 
More precisely we will show that a model of proliferating baby
universes, with the measure of quantum gravity restricted to stacked
spheres, can be put in correspondence with the statistical system of
stacked spheres. 

\noindent Let us consider four dimensional triangulations that are
(boundary of the) stacked spheres in which there can be loops made by
two three-dimensional simplexes. This is possible whenever the stacked spheres
are pinched on a three simplex creating a bottle neck loop of two
three-simplexes. These loops could be either  the loops of a Green
function or the minimum bottle neck of a baby universes. This class of
triangulations, following the notation in literature, is called
${\bf T}_{2}$. The other class of triangulations is the stacked spheres
in which the minimum loop length can be made by the boundary of a
four-simplex that are five three-simplexes. We call this last class ${\bf T}_5$.

In the two dimensional theory, the introduction in ${\bf T}_{2}$ of
two-link loops (the two dimensional analogous of two three-simplex loop)
corresponds in the matrix model $\phi^{3}$ to consider Feynman diagrams
with self-energy (c.f. \cite{1Ambjorn} \cite{LJ}). 

\noindent Since the minima loops of ${\bf T}_2$ and ${\bf T}_5$, for
which they differ, are of the order of lattice spacing we will expect
that the two classes of triangulations, as a statistical
mechanics system, coincide in the scaling
limit, that is to say they belong to the same universality class. 
  
\noindent Let's consider the minimum neck one loop function $G(\triangle
k_4)$ in ${\bf T}_2$. In every triangulation of ${\bf T}_2$ we can cut
out the maximal size baby universe of minimum neck and close the two
three-simplex loop. We will obtain again a triangulation that belongs to
${\bf T}_2$. In this way we will obtain all the triangulations of ${\bf
T}_2$ from the triangulations of the stacked spheres ${\bf T}_5$
considering that for each three-simplex either leave them in their
actual form or we can open the triangulation to create a two
three-simplex loop and gluing on it a whole one loop universe
$G(\triangle k_4)$.  We note that in the triangulations of ${\bf T}_5$
$\overline {N}_3=5/2\overline {N}_4$ (Dehn-Sommerville).  Calling the
one loop function  of ${\bf T}_5$ $\overline{G}(\overline{\triangle
k_4})$, the above considerations lead to the identity 
 
 \begin{equation}
 G(\triangle k_4)=\sum_{{\bf T}\in {\bf T}_5}
 e^{-\overline{N}_4\triangle k_4 }
 (1+G(\triangle k_4))^{\overline{N}_3}=
 \sum_{{\bf T}\in {\bf T}_5}e^{-\overline{N}_4\overline{\triangle k_4}}
 =\overline{G}(\overline{\triangle k_4})
 \label{selfenerg}\;\;\;\; ,
 \end{equation}
 
\noindent in which $\triangle k_4=k_4 - 5/2 k_2$ comes out from restricting the 
 action of quantum gravity to stacked spheres \ref{res}, and where we have 
 defined
 
 \begin{equation} 
 \overline{\triangle k_4}=\triangle k_4 - 
 {5\over 2}log\left(1+G(\triangle k_4)\right)
 \label{ini}\;\;\;\; .
 \end{equation}

\noindent By \ref{selfenerg} we can also write last equation as  

\begin{equation} 
 {\triangle k_4}=\overline{\triangle k_4} + 
 {5\over 2}log\left(1+\overline{G}(\overline{\triangle k_4})\right)
 \label{idi}\;\;\;\; .
 \end{equation}

\noindent By universality and the estimates (\ref{estimo}) it follows that near
the 
critical point we have that $G(\triangle k_4)\asymp cost+(\triangle k_4 -
\triangle k^{c}_4)^{1-\gamma_s}$ with $\gamma_s < 1$. 

\noindent The susceptibility functions of ${\bf T_2}$ and ${\bf T}_5$ by
\ref{selfenerg}
are
\begin{equation}
\chi(\triangle k_4)\asymp -{d \over d\triangle k_4}G(\triangle k_4)\;\;\;\; 
\overline{\chi}(\overline{\triangle k_4})\asymp -{d \over d\overline{\triangle
k_4}}
\overline {G}(\overline{\triangle k_4})
\end{equation}

\noindent From \ref{idi} we have

\begin{equation}
{d(\triangle k_4)\over d(\overline{\triangle k_4})}
=1-{5\over 2}
{\overline{\chi}(\overline{\triangle k_4})\over
(1+\overline{G}(\overline{\triangle k_4})}
\label{travo}\;\;\;\; .
\end{equation}

\noindent If we calculate di derivative with respect to $\triangle k_4$
of the one loop function $G(\triangle k_4)$ and use the previous
equation we get 

\begin{equation}
\chi(\triangle k_4)={\overline{\chi}(\overline{\triangle k_4})
\over {1-{5\over 2}
{\overline{\chi}(\overline{\triangle k_4})\over
(1+\overline{G}(\overline{\triangle k_4})}}}
\label{rela}\;\;\;\; .
\end{equation}

Now it is clear that $\chi(\triangle k_4) \mapsto +\infty$ for $\triangle k_4
\mapsto 
(\triangle k_4)^{c}$ and with the same critical exponent 
$\overline{\chi}(\overline{\triangle k_4)} \mapsto +\infty$ for 
$\overline{\triangle k_4} \mapsto (\overline{\triangle k_4})^{c}$. 
From equation (\ref{rela}) when  $\chi(\triangle k_4) \mapsto +\infty$ we have
that 
$\overline{\chi}\left(\overline{\triangle k_4}\right)\mapsto
 {2\over 5}\Bigl(1+\overline{G}\left(\overline{\triangle k_4}
({\triangle k_4}^{c})\right)\Bigr) < +\infty$, 
then the system ${\bf T}_5$ is above his critical line, i.e.
$\overline{\triangle k_4}({\triangle k_4}^c)>\overline{\triangle k_4}^{c}$. 
These facts  
imply that at $\overline{\triangle k_4}({\triangle k_4}^{c})$ 
$d(\triangle k_4)/d(\overline{\triangle k_4})=0$ and around it 
$\overline{\chi}(\overline{\triangle k_4})/
(1+\overline{G}(\overline{\triangle k_4}))$ 
is a decreasing monotonic
function  
by equation \ref{travo} because $\chi(\triangle k_4) \mapsto +\infty$ 
, we can expand equation \ref{idi} and \ref{rela} 
around $\overline{\triangle k_4}({\triangle k_4}^{c})$ 

\begin{equation}
{\triangle k_4} -{\triangle k_4}^{c}= 
cost\left(\overline{\triangle k_4} -\overline{\triangle k_4}({\triangle
k_4}^{c})\right)^{2}
\label{solita}\;\;\;\; .
\end{equation}

\begin{equation}
\chi(\triangle k_4)\asymp
{1\over {\overline{\triangle k_4} -\overline{\triangle k_4}({\triangle
k_4}^{c}})}
\asymp {1\over \sqrt{\triangle k_4 - {\triangle k_4}^{c}}}
\label{final}\;\;\;\; .
\end{equation}

\noindent The  second asymptotic equality of the last equation implies

\begin{equation}
\gamma_s={1\over 2}
\label{provola}
\end{equation}

The dominance of stacked spheres in the weak phase allows us 
to fix the parameter $\tau$ in the partition function 
of quantum 
gravity in the weak coupling regime.

\begin{equation}
\tau -{11\over 2}=-{5\over 2} \Longrightarrow \tau=3
\label{crippu}
\end{equation}     

These considerations put evidence that the branched polymer phase of 4-D
Dynamical 
Triangulations is a mean field phase.

\chapter{Lattice Gauge Theory of Gravity}
\label{latte}

\section{Introduction}
\label{int3}

The model of dynamical triangulations, that we have analysed up to now, 
has been very popular in recent years. 
The reason of this is due to the excellent results in two dimensions 
where the critical exponents coincide perfectly with the critical exponents of 
two dimensional continuum quantum gravity. This has encouraged people to explore

the model in four dimensions. The main aim has been to explore
the dynamics of this discrete theory in order to define a four-dimensional
quantum theory of gravity at the critical point (if any) of the parameters
space. At the beginnings a
second order phase transition has been observed by numerical 
simulations \cite{4dynamical}. Successive and more accurate simulations
\cite{deBakker} \cite{Bialas} have shown that the phase transition
is of first order.

\noindent At the moment the situation is not completely clear and further
investigations seem necessary, but the general agreement of the scientific
community about the first order nature of the phase transition points out 
that it is necessary to find a new lattice theory of gravity which fits with
the general requirements of Euclidean lattice field theory. 
The following model is an attempt in
this direction and might 
 be an answer to this request. It has been proposed originally and 
independently by different authors \cite{frohlich} \cite{Ale} \cite{nielsen}, 
and the original part of the content of this
chapter, based on the work \cite{first},  is a continuation and an 
improvement of these attempts, in particular of  \cite{Ale}
(connected to this work are \cite{Ale1} \cite{Ale2}). We
begin in section \ref{ciube} by reviewing some concepts contained 
in \cite{Ale} and in particular we show how to associate to each dual 
Vorono{\"\i} edge (see appendix \ref{voronoi} and references \cite{nobel1}
\cite{nobel2} and \cite{ruth} p. 395)
of the original $n$-dimensional simplicial complex a Poincar\`e transformation
by fixing an orthonormal reference frame in each $n$-dimensional 
simplex. Then we recall the group theoretical action which is a 
functional of the sine of the deficit angles. Respect to the original
work \cite{Ale} it is shown that the action is independent from the orientation
of the hinges and from the starting $n$-simplex in which it is written .

In section \ref{dudu} following \cite{Ale} we write the group theoretical action

completely defined on the dual Vorono{\"\i} complex of the original simplicial 
complex. Two sets of variables play an important role: the connection matrices
$U^{a}_{b}(\alpha,\beta)$ and the normals $b^{a}_{\alpha\beta}(\alpha)$ to the 
$n-1$ faces of each $n$-simplex $\alpha$. In the second order formalism the 
connection matrices $U^{a}_{b}(\alpha,\beta)$ are functions of the 
$b^{a}_{\alpha\beta}(\alpha)$. As anticipated in \cite{Ale} we follow a 
first order formalism in which we consider the $U^{a}_{b}(\alpha,\beta)$ 
and the $b^{a}_{\alpha\beta}(\alpha)$ as independent variables. A first
order formalism for Regge calculus has been done by J. Barrett \cite{barrett}.

\noindent  We impose 
that the first order field variables satisfy 
only the two constraints \ref{relatto} and \ref{chiusura}. 
Finally, respect to \cite{Ale}, in the first order formalism we define a 
modified action in such a way that it is dependent only from each plaquette
of the dual Vorono{\"\i} complex .

In section \ref{ore} it is proved that the modified action  
is independent from the orientation of each plaquette. 

In section \ref{fieno} the field equations for the connection matrices 
$U^{a}_{b}(\alpha,\beta)$ are derived in the first order formalism in 
the approximation of "small deficit angles". It is proved that Regge 
calculus is a solution for these equations. This result is no longer 
obvious in the first order formalism when we consider the connection 
matrices independent from the $b^{a}_{\alpha\beta}(\alpha)$.

\noindent The equivalence of the first order formalism to
the (second order) Regge calculus is the main new result in this 
chapter.

In section \ref{pupu} we obtain the general field equations on lattice.
These equations are divided in two sets. The set of equations for the 
connection matrices $U^{a}_{b}(\alpha,\beta)$ obtained by requiring 
that the action is stationary under the variation of the
$U^{a}_{b}(\alpha,\beta)$, 
and the set of equations for $b^{a}_{\alpha\beta}(\alpha)$. 
These two sets of 
equations are obtained by using the method of the Lagrange multipliers through
which we take into account the two sets of constraints, one set 
on the linear dependence among the $b^{a}_{\alpha\beta}(\alpha)$ of the same
simplex $\alpha$ and the second set on the condition that the  
$b^{a}_{\alpha\beta}(\alpha)$ and $b^{b}_{\beta\alpha}(\beta)$,
as the expressions of the same normal to the
$n-1$ face seen, respectively, in the orthonormal reference frame of $\alpha$
and $\beta$,
are connected by the matrix $U^{a}_{b}(\alpha,\beta)$.  
At the end of section \ref{pupu} we introduce a method to calculate the Lagrange
multipliers too.

In section \ref{quanta} we define a path-integral for this classical theory
of gravity on lattice and we prove that this quantum measure is invariant 
under local  transformation of $SO(n)$.

Finally in section \ref{copy} we show as it is possible to define in this
framework
a coupling with fermionic matter following, as usual in lattice field theory,
the 
example of the continuum theory.

The  main feature of this model is that it is more
general than  Regge calculus and dynamical triangulations. 
The introduction of a reference frame in each 
$n$-dimensional simplex is the key concept for which this version of discrete
gravity is different from the previous one. This is a way to translate on 
lattice the general principle of local invariance under a general Lie 
group which implies a structure , in the continuum, of a Principal
Fiber Bundle with the fibers isomorphic to a Lie group (gauge theories) or 
to orthonormal reference frames connected among them by matrices of 
$SO(n)$ (General Relativity in the Riemmanian version). In this scheme we can
naturally write on lattice the coupling of gravity with other fundamental forces
and consider extensions to more general theories than gravity like, for example,
supergravity (see for example \cite{Ale3} and \cite{ale4}).

\section{Group Action for Simplicial Gravity}
\label{ciube}

In this section we shall introduce a group action for simplicial gravity
\cite{Ale} on the dual 
Vorono{\"\i} complex \cite{miller} (see appendix \ref{voronoi} for a definition
of 
Vorono{\"\i} cells ) of the original simplicial complex. In particular we will
see
that it is possible to associate to each dual edge $(\alpha,\alpha+1)$ 
a "Poincar\'e" transformation 
$U(\alpha,\alpha+1)$ as in lattice gauge theories. Of course by a "Poincar\'e"
transformation from now on
we mean in general a $SO(n)$ rotation and a translation. 

We are now going to introduce a reference frame in each $n$-dimensional simplex
 and we shall see how these reference frames can be connected by using
coordinates of the vertices of the common $n-1$-dimensional faces and a notion
of Levi-Civita connection on simplicial manifolds. Consider an hinge $h$ and let
$\{P_1,...,P_{n-1}\}$ its vertices. 
Suppose that this 
hinge is shared by $N$ $n$-simplices $\{S_1,...,S_N\}$ whose vertices are
labelled in this way 

\begin{equation}
S_{\alpha}\equiv \{P_1,...,P_{n-1},
Q_{\alpha-1,\alpha},Q_{\alpha,\alpha+1}\}\;\;\;\; (\alpha=1,...,N)\;\;\;\;.
\label{vertices}
\end{equation}

In each simplex $S_{\alpha}$  we can fix an
origin and a reference frame. In this frame the vertices of the simplex
$S_{\alpha}$ have 
the following coordinates

\begin{eqnarray}
P_{i}&\equiv& \{y_{i}^{a}(\alpha)\} \;\;\; a=1,...,n\;,\; i=1,...,n-1
\nonumber\\
Q_{\alpha-1, \alpha}&\equiv& \{z^{a}_{\alpha-1,\alpha}(\alpha)\} \nonumber \\
Q_{\alpha, \alpha+1}&\equiv& \{z^{a}_{\alpha,\alpha+1}(\alpha)\}
\label{alpha}
\end{eqnarray}

Let's indicate by $D_{\alpha}$ the circumcenter of the simplex $\alpha$ 
and let $x^{a}(\alpha)$ their coordinates.

\noindent Now it is possible to associate uniquely to a dual edge  an element 
of the Poincar\'e group $U(\alpha,\alpha+1)\equiv 
\{ U^{a}_{b}(\alpha,\alpha+1), U^{a}(\alpha,\alpha+1) \}$ by requiring that

\begin{eqnarray}
U^{a}_{b}(\alpha,\alpha+1)y^{b}_{i}(\alpha+1)
+U^{a}(\alpha,\alpha+1)&=&y^{a}_{i}(\alpha)\nonumber \\
U^{a}_{b}(\alpha,\alpha+1)z^{b}_{\alpha,\alpha+1}(\alpha+1)
+U^{a}(\alpha,\alpha+1)&=&z^{a}_{\alpha,\alpha+1}(\alpha)\;\;\;\; ,
\label{connec}
\end{eqnarray}

\noindent in other words we are embedding $\alpha$ and $\alpha+1$ in ${\Bbb
R}^{n}$ and adopting 
 the standard notion of parallel displacement in ${\Bbb R}^{n}$ 
 and we transport the origin of the reference frame of $\alpha+1$ from
 $\alpha+1$ to the
 origin of the orthonormal reference frame of $\alpha$ \cite{frohlich}
 \cite{Ale}.
 This operation 
 makes the position vectors in $\alpha+1$ of the vertices of the common face
  $S_{\alpha}\cap S_{\alpha+1}$ coincident with 
  the position vectors of the same vertices in $\alpha$. 
 It follows that the matrix 
 $U^{a}_{b}(\alpha,\alpha+1)$ is an orthogonal matrix which describes the change
 from the reference frame of  $\alpha+1$ to $\alpha$, 
 considered now as two different reference frame
 of the same vector space (\cite{frohlich}). 
 Since the simplicial manifold is assumed orientable \cite{hawking} we can
 choose the reference frame 
 in $\alpha$ and $\alpha+1$ in such a way that $U^{a}_{b}(\alpha,\alpha+1)$ are 
 
 $SO(n)$ matrices.
 
 The family of all these matrices  
 determine a unique connection that is the {\it Levi Civita or Regge
 connection}.

The arbitrariness of the choice of the reference frame in each simplex $\alpha$
implies a sort of local  gauge invariance under Poincar\'e 
transformations
$\Lambda(\alpha)=\{\Lambda^{a}_{b}(\alpha),\Lambda^{a}(\alpha)\}$:

\begin{equation}
U(\alpha,\alpha+1)\mapsto
\Lambda(\alpha)U(\alpha,\alpha+1){\Lambda}^{-1}(\alpha+1)\;\;\;\;,
\label{gauge}
\end{equation}

\noindent in particular the coordinates of the dual vertices (the circumcenters
of the simplices
$S_{\alpha}$) transform as $x^{a}(\alpha)\mapsto
\Lambda^{a}_{b}(\alpha)x^{b}(\alpha)+\Lambda^{a}(\alpha)$.
Anyway from now on we decide to put the origin of the reference frames  in the
circumcenters, in such a way
we always have $x^{a}(\alpha)=0$.  

To the hinge $h$ it corresponds the dual two-dimensional plaquette that we still
label by $h$. 
Now consider 
the following plaquette variable 

\begin{equation}
U_{\alpha\alpha}^{(h)}=U(\alpha,\alpha+1)
U(\alpha+1,\alpha+2)...U(\alpha-1,\alpha)\;\;\;\; .
\label{ruota}
\end{equation}

\noindent It is evident that $U_{\alpha\alpha}^{(h)}$ leaves the coordinates of
the vertices 
of the hinge $h$ unchanged and its translational part is zero. So it is a
rotation of 
an angle $\theta(h)$
in the two-dimensional plain orthogonal to the hinge. It is easy to see
\cite{sorkin1} that the
angle $\theta(h)$ is the deficit angle  \cite{regge}

\begin{equation}
\theta(h)=2\pi -\sum_{\alpha=1}^{N}\theta(\alpha,h)\;\;\;\; ,
\label{semo}
\end{equation}

\noindent where $\theta(\alpha,h)$ are the dihedral angle  of the 
simplices $S_{\alpha}$ with the hinge $h$.

Let's choose $n-2$ linear independent edge vectors
$E^{a}_{1}=y^{a}_{1}-y^{a}_{n-1},... 
E^{a}_{n-2}=y^{a}_{n-2}-y^{a}_{n-1}$ belonging to $h$ , and consider  the
oriented
volume ${\mathcal V}^{(h)}{\;}^{ab}(\alpha)$ of $h$ 

\begin{equation}
{\mathcal V}^{(h)}{\;}^{ab}(\alpha)=\equiv {1\over (n-2)!}
\epsilon^{ab}_{c_{1}...c_{n-2}}E_{1}^{c_1}(\alpha)...E_{n-2}^{c_{n-2}}(\alpha)\;\;\;\;
.
\label{orio}
\end{equation}

 At this point it seems natural to propose the following
gravitational 
action

\begin{equation} 
I=-{1\over 2}\sum_{h}\Bigl( U_{\alpha\alpha}^{a_{1}a_{2}}{\;}^{(h)}{\mathcal
V}^{(h)}{\;}_{a_1a_2}(\alpha) \Bigr)
\label{haction}
\end{equation}

\noindent where  $U_{\alpha\alpha}^{a_{1}a_{2}}{\;}^{(h)}$ are the elements of
the orthogonal matrix
\ref{ruota}. 
If we choose a reference frame in which the first two axis, $1$ and $2$, are on
the 
two-dimensional plane orthogonal to the hinge $h$ the matrix
$U_{\alpha\alpha}^{a_{1}a_{2}}{\;}^{(h)}$
will be diagonal, with the diagonal elements equal to $1$, 
except in the intersection of the first two rows and columns where
it is

\begin{equation}
\pmatrix{{cos\;\theta(h)} & {-sin\;\theta(h)} \cr
          {sin\;\theta(h)} & {cos\;\theta(h)}} \;\;\;\; .
\label{lesto}
\end{equation}           

\noindent Pairwise in this reference frame the only non-zero components of
the antisymmetric tensor ${\mathcal V}^{(h)}{\;}^{ab}(\alpha)$ are 
${\mathcal V}^{(h)}{\;}^{12}(\alpha)$ and ${\mathcal V}^{(h)}{\;}^{21}(\alpha)$,
because in 
this reference frame the $n-2$ independent vectors of $h$ have zero components
on
the first two axes. 
In particular the component ${\mathcal V}^{(h)}{\;}^{12}(\alpha)$ of the
antisymmetric tensor
is equal to $V(h)$ by the definition of the volume $V(h)$ of the hinge $h$.
Since 
the trace in \ref{haction} is independent from the orthogonal reference frame
chosen in the $n$-simplex
$\alpha$, 
the action \ref{haction} is equal to
 
\begin{equation}
I= \sum_{h}sin\;\theta(h)\;V(h)\;\;\;\; ,
\label{calc}
\end{equation}

\noindent that for small deficit angles $\theta(h)$ reduces to Regge action
\cite{regge}.

\noindent Notice that in the definition \ref{ruota} we have chosen the sequence 
of simplices $\alpha,\alpha+1,...,\alpha-1,\alpha$. This sequence induces a
direction 
of rotation in the two-dimensional plane orthogonal to the hinge $h$ determined
by the rotation
matrix \ref{ruota}. We have implicitly assumed that this direction is the same
direction 
generated by the ordered sequence of axes $\{1,2\}$ in the rigid rotation from
the 
axis $1$ to the axis $2$.  

\noindent Since the simplicial manifolds under consideration 
are assumed orientable \cite{hawking}, we can choose the same orientation in
each $n$-dimensional 
simplex \cite{seifert}. 
This orientation induces an orientation on each face but no orientation on the
hinge
because the two faces of the simplex, that share the hinge, induce  opposite
orientations on it.
It follows that in principle there is no preferred orientation on the
two-dimensional plane orthogonal
to the hinge. So  we can pairwise choose the sequence
$\alpha,\alpha-1,...,\alpha+1,\alpha$,
the rotation matrix will be the transposed of \ref{ruota} and the direction of
rotation is now from 
the axis $2$ to the axis $1$, that is to say the opposite of the previous one.
In other words
respect to the first orientation, now 
the axis $1'$ coincides with $2$ and $2'$ with $1$. It follows that this change 
of orientation implies that
$\epsilon^{{a'}_1{a'}_2...{a'}_n}=-\epsilon^{a_1a_2...a_n}$ so that  
${\mathcal V}^{(h)}{\;}^{1'2'}(\alpha)={\mathcal V}^{(h)}{\;}^{21}(\alpha)$ .   

Bearing in mind that the analogous of \ref{lesto} with this new orientation is
equal to its transposed 
, we have that 
the action now \ref{haction} is equal again to \ref{calc}. In this way we have
proved that
the action \ref{haction} is independent from the chosen orientation.  

\noindent The action \ref{haction} by construction is invariant under local 
Poincar\'e transformations (gauge invariance) and is independent from the
starting simplex $\alpha$. 
This last remark may be done more clear by the following lemma

\begin{lemma}
\label{liber}
\noindent The action (\ref{haction}) does not depend from the 
reference frame where it is written, in the sense that if $\alpha$ and
$\delta_{i}$
are two simplices which have in common the same hinge $h$ then

\begin{equation}
U_{\alpha\alpha}^{a_{1}a_{2}}{\;}^{(h)}
{\mathcal V}^{(h)}{\;}_{a_1a_2}(\alpha)
=U_{\delta_{i}\delta_{i}}^{a_{1}a_{2}}{\;}^{(h)} 
 {\mathcal V}^{(h)}{\;}_{a_1a_2}(\delta_i)
\label{invectiva}
\end{equation}

\end{lemma}

\noindent {\bf Dim :} Let $\left\{\alpha...\delta_{i} \right\}$ the simplicial 
path from the simplex $\alpha$ to $\delta_{i}$ such that the simplices of the 
path share the hinge $h$. We have that from the definition of Levi-Civita
connection

\begin{equation}
y_{j}^{a}(\alpha)-y_{n-1}^{a}(\alpha)=\left(U(\alpha,\alpha+1)...
U(\delta_{i-1},\delta_{i})\right)^{a}_{b}
\left(y_{j}^{b}(\delta_{i})-y_{n-1}^{b}(\delta_{i})\right)
\label{trasform}
\end{equation}

\noindent So for $SO(n)$ connection matrices is true the following

\begin{eqnarray}
{\mathcal V}^{(h)}{\;}_{a_1a_2}(\alpha)&=& 
 \left(U(\alpha,\alpha+1)...
U(\delta_{i-1},\delta_{i})\right)^{a'_1}_{a_{1}}\nonumber\\
&&\times {\mathcal V}^{(h)}{\;}_{a'_1a'_2}(\delta_i)
\left(U(\delta_{i},\delta_{i-1})...
U(\alpha+1,\alpha)\right)^{a'_2}_{a_2}
\label{eguaglio}
\end{eqnarray}

\noindent By substituting this last equation in

\begin{equation}
 U_{\alpha\alpha}^{a_{1}a_{2}}{\;}^{(h)}
 {\mathcal V}^{(h)}{\;}_{a_1a_2}(\alpha)
\label{lastessa}
\end{equation}

\noindent we get the equality (\ref{invectiva}).

 It is interesting to note that this action is similar to Wilson action
 \cite{wilson} for
lattice gauge theory (for basic notions of lattice field theory
see \cite{drouffe} chap.6). Anyway in the form \ref{haction} the action is in a
second
order formalism since the plaquette variables $W_{\alpha}^{a_{1}a_{2}}(h)$ are
complicated functions by \ref{connec} of the coordinates of the vertices.
Moreover 
the field variables belong both to the original triangulation and its
Vorono{\"\i} 
dual. We are now going to introduce a formalism that is written completely on
the 
Vorono{\"\i} complex and is a first order formalism like the Palatini version
\cite{palatini} of
general relativity.

\section{Action on the Dual Vorono{\"\i} Complex}
\label{dudu}

Let's consider the $n-1$-dimensional face $f_{\alpha\beta}\equiv\{P_1,...,P_n\}$
between the $n$-dimensional simplices $\alpha$ and $\beta$. We 
label the coordinates of the vertices of the face $f_{\alpha\beta}$ by
$y^{a}_{i}(\alpha)$ in the reference frame $\alpha$. We define the vector
$b^{a}_{\alpha\beta}$

\begin{eqnarray}
b_{\alpha\beta}^{a}(\alpha)&=&
\epsilon_{ab_{1}...b_{n-1}}(y_{1}}(\alpha)-y_{n}(\alpha))^{b_{1}}...
(y_{n-1}(\alpha)-y_{n}(\alpha))^{b_{n-1}\nonumber\\
&=&
\epsilon_{ab_{1}...b_{n-1}}E^{b_{1}}_{1}...E^{b_{n-1}}_{n-1}\;\;\;\;.
\label{1tetra}
\end{eqnarray}

\noindent The analogous vector $b^{a}_{\beta\alpha}(\beta)$ can be calculated in
the 
reference frame of $\beta$. The two vectors are related by the formula

\begin{equation}
b^{a}_{\alpha\beta}(\alpha)=U^{a}_{b}(\alpha,\beta)b^{b}_{\beta\alpha}(\beta)
\label{relatto}\;\;\;\;
\end{equation}

\noindent By construction these vectors are orthogonal to the face
$f_{\alpha\beta}$ so that, from a 
mathematical point of view, they live on the dual Vorono{\"\i} edge
$(\alpha,\beta)$ (see appendix \ref{voronoi}). We choose the order
of  the the edges vectors $E^{a}_{i}$ in formula \ref{1tetra}   
in such a way that $b^{a}_{\alpha\beta}(\alpha)$
points toward the outside of the simplex $\alpha$. Among the vectors
$b^{a}_{\alpha\beta}(\alpha)$
of the same $n$-dimensional simplex there is the following identity

\begin{equation}
\sum_{\beta=1}^{n+1}b_{\alpha\beta}^{a}(\alpha)=0\;\;\;\; .
\label{chiusura}
\end{equation}

The antisymmetric tensor 
${\mathcal V}^{(h)}{\;}_{a_1a_2}(\alpha)$ can be written as a bivector
of the $b^{a}_{\alpha\beta}(\alpha)$

\begin{equation}
{\mathcal V}^{(h)}{\;}^{ab}(\alpha)=
 {1\over n!(n-2)!V(\alpha)}\left(b_{\alpha\alpha+1}^{a}(\alpha)
b_{\alpha-1\alpha}^{b}(\alpha)
-b_{\alpha\alpha+1}^{b}(\alpha)b_{\alpha-1\alpha}^{b}(\alpha)\right)\;\;\;\;,
\label{susi}
\end{equation}

\noindent where $\alpha-1$ and $\alpha+1$ indicate the two 
$n$-simplices which have a face  and the hinge $h$ in common
with $\alpha$. The volume $V(\alpha)$ of the simplex $\alpha$ 
can be written as function of the $b_{\alpha\beta}$ in the 
following way

\begin{equation}
(V(\alpha))^{n-1}={1\over {n!}^{n}}
\epsilon_{a_1...a_{n}}\epsilon^{j_1...j_{n} j}
b^{a_1}_{j_1}(\beta)...b^{a_{n}}_{j_{n}}(\beta)\;\;\;\; ,
\label{vol}
\end{equation}

\noindent where the indices $(j_1,...,j_n)$ run over all possible values of the
dual Vorono{\"\i}
edges $(\alpha,\beta)\;\;,\beta=1,...,n+1$ and $j$ is a fixed index. The
equation \ref{vol} does not
depend from the choice 
of the index $j$ in the $\epsilon$-indices, or, in other words, 
it does not depend from the $b$ that is left
out.

As said in the previous section we can put the origin of the reference frame of
$\alpha$ in the
circumcenter. If $z^{a}_{i}(\alpha)\;\;, a=1,...,n,\;\;i=1,...,n+1$ are the
circumcentric 
coordinates \cite{sorkin2} of the $n+1$ vertices of $\alpha$ we have

\begin{equation}
\sum_{i=1}^{n+1}z^{a}_{i}(\alpha)=0\;\;\;\; .
\label{bari}
\end{equation}

\noindent In these coordinates we can express, as it is easy to see, the vector
$b^{a}_{i}(\alpha),\;i=1,...,n$
in the following way

\begin{equation}
 b^{a}_{i}(\alpha)={1\over n!}
 \sum_{k\neq i}\epsilon^{a}_{a_1,...,a_n}
 \epsilon_{i}^{k
 j_1...j_n}z^{a_1}_{j_1}(\alpha)...z^{a_{n-1}}_{j_{n-1}}(\alpha)\;\;\;\; .
 \label{bubu}
 \end{equation}

\noindent This equation can be inverted, so we have

\begin{equation}
z^{a}_{i}(\alpha)={1\over (n+1)!(n!V(\alpha))^{n-2}}
\sum_{k\neq i}\epsilon^{a}_{a_1...a_{n-1}}\epsilon_{i\;k}^{j_1...j_{n-1}}
b^{a_1}_{j_1}(\alpha)...b^{a_{n-1}}_{j_{n-1}}(\alpha)\;\;\;\;.
\label{oppi}
\end{equation}

\noindent These two equations \ref{bubu} \ref{oppi} show that there is a one to
one correspondence 
between the circumcentric coordinates of the vertices of a simplex $\alpha$ and
the 
$b_i(\alpha)$.

The gravitational action \ref{haction} can be written, by the equation
\ref{susi}, as a function of
the $b^{a}_{i}(\alpha)$. Anyway we are still dealing with a second order
formalism, since the 
connection matrices $U^{a}_{b}(\alpha,\beta)$ and the $b_{\alpha \beta}(\alpha)$
are both
functions of the coordinates of the edge of the simplicial complex. Now we
propose a first
order formalism in which $U^{a}_{b}(\alpha,\beta)$ and $b_{\alpha\beta}(\alpha)$
are independent variables defined on the 
Vorono{\"\i} edge $(\alpha,\beta)$. We consider equation \ref{relatto} as a
constraint of the 
theory. This constraint fix only $n$ degree of freedom for each Vorono{\"\i}
edge
$(\alpha,\beta)$, 
and not the ${n(n-1) \over 2}$ 
conditions that are necessary to determine $U^{a}_{b}(\alpha\beta)$. Pairwise
\ref{chiusura} is
the second constraint of the theory for each Vorono{\"\i} vertex. 

In the first order formalism we consider connection matrices more general
than the Levi-Civita matrices defined by equation \ref{connec} . 

\begin{figure}
\begin{center}
\epsfig{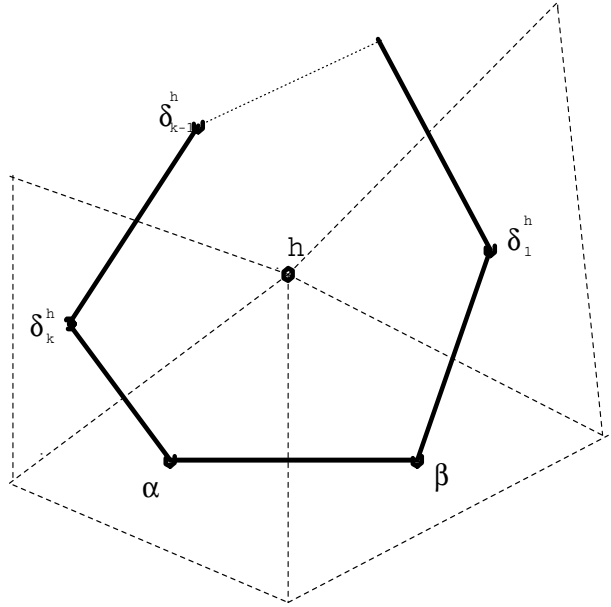}
\caption{Dual Vorono{\"\i} Plaquette}
\end{center}
\end{figure}

This implies that in these cases the gravitational action 
in the form (\ref{haction}) could be
dependent from the reference frame in which it is written since it
is no longer true an equality like \ref{eguaglio} . So if we define the
following 
antisymmetric tensor on the plaquette $h$ as 

\begin{eqnarray}
W^{(h)}_{c_1c_2}(\alpha) &\equiv & 
{1\over k_{h}+2} \Big({\mathcal{V}}^{(h)}(\alpha)+
U_{\alpha\beta}{\mathcal{V}}^{(h)}(\beta)U_{\beta\alpha}+... \nonumber\\
&+&U_{\alpha\beta}...
U_{\delta^{h}_{k-1}\delta^{h}_{k}}
{\mathcal{V}}^{(h)}(\delta^{h}_{k})U_{\delta^{h}_{k}\delta^{h}_{k-1}}...
U_{\beta\alpha} \Big)_{c_1c_2}
\label{media}\;\;\;\;.
\end{eqnarray}

the action can be written in the form

\begin{equation}
S\equiv -{1\over 2}\sum_{h}{\rm Tr}
\left(U^{h}_{\alpha\alpha}W^{h}(\alpha)\right)
\label{caction}
\end{equation}

The above action coincide with the previous action in second order formalism and

since (in matrix notation)

\begin{equation}
W^{(h)}(\alpha)=U_{\alpha\beta}...U_{\delta_{i-1}\delta_{i}}
W^{(h)}(\delta_{i})U_{\delta_{i}\delta_{i-1}}...U_{\beta\alpha}
\label{1lastessa}
\end{equation}

\noindent the action (\ref{caction}) result to be independent form the frame in
which it is written
essentially for the same reasonings of the lemma (\ref{invectiva}).

\noindent Moreover the action (\ref{caction}) is invariant under the following
set of 
transformations

\begin{eqnarray}
U_{\alpha\beta}&\mapsto&
{U}'_{\alpha\beta}=
O(\alpha)U_{\alpha\beta}O^{-1}(\beta)\nonumber\\
b_{\alpha\beta}(\alpha)&\mapsto&
b'_{\alpha\beta}(\alpha)=O(\alpha)b_{\alpha\beta}(\alpha)
\label{gauge}
\end{eqnarray}

where $O(\alpha)$ and $O(\beta)$ are two element of $SO(n)$. 

We want to close this section by adding some consideration on the nature of the
$b_{\alpha\beta}$.
In the interior of every simplex $\sigma$, as we have remarked above, the space 
is flat. So in every point of the cotangent space to $\sigma$ the metric 
is flat

\begin{equation}
{\bf g}(\sigma)=\delta_{ab}{\bf e}^{a}(\sigma)\otimes{\bf e}^{b}(\sigma)
\label{metric}
\end{equation}

\noindent where ${\bf e}^{a}(\sigma), a=1,...,n$ form an orthonormal dual base
in the cotangent 
space of $\sigma$. Of course it is evident that in every point of the cotangent 
space of $\sigma$ we can choose the same base ${\bf e}^{a}(\sigma)$. In the form
(\ref{metric}) the metric coefficients are automatically diagonal and so we can 
say that trivially a $n$-dimensional simplex is a local inertial reference
frame. 

\noindent It seems more natural to choose as base the edges of the simplex
$\sigma$ 
because this makes contact with the geometry of the simplex $\sigma$. Let 

\begin{equation}
{\bf E}_{\mu}(\sigma)\; {\mu=1,...,n}
\label{edge}
\end{equation} 

\noindent $n$ linear independent edge vectors. The other edge vectors can be
expressed 
as linear combination of the $n$ vectors above. Associated to these $n$-vectors
we can
fix (uniquely) $n$ covectors of the dual base such that, by the definition
(\ref{1tetra}), 

\begin{equation}
{\bf \widehat{b}}^{\mu}(\sigma)({\bf E}_{\nu}(\sigma))=
\delta^{\mu}_{\nu}\;\;\;\; .
\label{magno}
\end{equation}

\noindent where 
\begin{equation}
{\bf \widehat{b}}^{\mu}(\sigma)\equiv {{\bf b}^{\mu}(\sigma) \over n!
V(\sigma)}.
\label{1dual}
\end{equation}

\noindent The transformation law between the couple of the orthonormal base     
${\bf e}_{a}(\sigma)$ and their respective duals ${\bf e}^{a}(\sigma)$ and the 
 couple of ${\bf E}_{\nu}(\sigma)$ and ${\bf \widehat {b}}^{\mu}(\sigma)$ are
 given once we fix the 
 component of ${\bf E}_{\nu}(\sigma)$ and ${\bf \widehat{b}}^{\mu}(\sigma)$ in
 the base 
 ${\bf e}_{a}(\sigma)$ and ${\bf e}^{a}(\sigma)$, more precisely
 
\begin{eqnarray}
{\bf e}^{a}(\sigma)=E^{a}_{\mu}(\sigma){\bf
\widehat{b}}^{\mu}(\sigma)\nonumber\\
{\bf e}_{a}(\sigma)={\widehat{b}}_{a}^{\mu}(\sigma){\bf E}_{\mu}(\sigma)\;\;\;\;
.
\label{cambio}
\end{eqnarray}

In these basis we can write the metric tensor as

\begin{eqnarray}
{\bf
g}(\sigma)=\delta^{ab}{\widehat{b}}^{\mu}_{a}(\sigma){\widehat{b}}^{\nu}_{b}(\sigma)
{\bf E}_{\mu}(\sigma)\otimes {\bf E}_{\nu}(\sigma)\nonumber\\
{\bf g}(\sigma)=\delta_{ab}E_{\mu}^{a}(\sigma)E_{\nu}^{b}(\sigma)
{\bf \widehat{b}}^{\mu}(\sigma)\otimes {\bf \widehat{b}}^{\nu}(\sigma)
\label{diago}
\end{eqnarray}
 
\noindent It is now clear that we can do the following identification respect to
the 
continuum theory

\begin{eqnarray}
e^{a}_{\mu}(x) \longmapsto E^{a}_{\mu}(\sigma)\nonumber\\
e^{\mu}_{a}(x) \longmapsto b^{\mu}_{a}(\sigma) \;\;\;\; ,
\label{identifico}
\end{eqnarray}

\noindent where with $e^{a}_{\mu}(x)$ and $e^{\mu}_{a}(x)$ we indicate the
n-bein of the 
continuum theory.
The discrete n-bein, in analogy to the continuum case, 
are determined modulo the action of the orthogonal group
because if we perform an orthogonal transformation on 
$E^{a}_{\mu}(\sigma)$ and $b^{\mu}_{a}(\sigma)$, the equations (\ref{diago})
remain form invariant.

\noindent Moreover the second equation of (\ref{diago}) says that the  
coefficient of the metric tensor in the base of the 
${\bf \widehat{b}}^{\mu}(\sigma)$ are the usual metric coefficients written in
Regge calculus

\begin{eqnarray}
g_{\mu\mu}(\sigma)=l^{2}_{\mu}(\sigma)\nonumber\\
l^{2}_{\mu\nu}(\sigma)={1\over 2}
(g_{\mu\mu}(\sigma)+g_{\nu\nu}(\sigma)-g_{\mu\nu}(\sigma))
\label{zuppa}
\end{eqnarray}

\noindent where, in the standard notation, $l^{2}_{\mu}\equiv |{\bf
E_{\mu}}|^{2}$ and 
with ${\bf E}_{\mu\nu}={\bf E_{\mu}}-{\bf E_{\nu}}$, 
$l^{2}_{\mu\nu}=|{\bf E}_{\mu\nu}|^{2}$.

\section{Remark on the Orientation}
\label{ore}

In the definitions (\ref{susi}), (\ref{media}), (\ref{ruota})  
we have tacitly assumed that
the boundary of the plaquettes is oriented. As a consequence, for example, 
we can establish that the Vorono{\"\i}-edge 
$(\alpha,\beta)$, in the direction from $\alpha$ to $\beta$, is positive
oriented,
 while the opposite Vorono{\"\i}-edge $\beta\alpha$ is negative oriented.
So this naturally fix a rule to
  define the bivectors ${\mathcal{V}}^{(h)}{\;}^{c_1c_2}(\alpha)$,
  $W^{h}(\alpha)$ 
and the holonomy matrix $U^{h}_{\alpha\alpha}$ along the positive direction. 
In particular we use the notation
${\mathcal{V}}^{(h)}{\;}^{c_1c_2}(\alpha)\equiv 
{\mathcal{V}}^{(h)}{\;}^{c_1c_2}_{\alpha\beta}(\alpha)$ to stress that in the 
definition \ref{susi} the positive direction is from $\alpha$ to $\beta$ 
in order that in the bivector we have to use first the vector
$b_{\alpha\beta}(\alpha)$
and after $b_{\alpha\delta_{k}}(\alpha)$. Pairwise we indicate as ${W^{+}}^{h}$
the 
antisymmetric tensor \ref{media} that in these new notations is

\begin{eqnarray}
{W^{+}}^{(h)}_{c_1c_2}(\alpha) &\equiv& 
{1\over k_{h}+2} \Big({\mathcal{V}}^{(h)}_{\alpha\beta}(\alpha)+
U_{\alpha\beta}{\mathcal{V}}^{(h)}_{\beta\delta_1}(\beta)U_{\beta\alpha}+...\nonumber\\
&+&  
U_{\alpha\beta}...U_{\delta^{h}_{k-1}\delta^{h}_{k}}
{\mathcal{V}}^{(h)}_{\delta^{h}_{k}\alpha}(\delta^{h}_{k})U_{\delta^{h}_{k}\delta^{h}_{k-1}}...
U_{\beta\alpha} \Big)_{c_1c_2}\;\;\;\; ,
\label{positiva}
\end{eqnarray} 

\noindent and  ${U^{+}}^{h}_{\alpha\alpha}\equiv U^{h}_{\alpha\alpha}$, that is
the
holonomy 
matrix in \ref{ruota} is defined along the positive direction $(\alpha,\beta)$.

\noindent The negative direction is from $\alpha$ to $\delta_{k}^{h}$, and as in
the 
definitions above we have

\begin{eqnarray}
 {W^{-}}^{(h)}_{c_1c_2}(\alpha) &\equiv& 
{1\over k_{h}+2} \Big({\mathcal{V}}^{(h)}_{\alpha\delta^{h}_{k}}(\alpha)+
U_{\alpha\delta^{h}_{k}}{\mathcal{V}}^{(h)}_{\alpha\delta^{h}_{k}}(\delta^{h}_{k})
U_{\delta^{h}_{k}\alpha}+...\nonumber \\
&+&  
U_{\alpha\delta^{h}_{k}}...U_{\delta^{h}_{1}\beta}
{\mathcal{V}}^{(h)}_{\beta\alpha}(\beta)U_{\beta\delta^{h}_{1}}...
U_{\delta^{h}_{k}\alpha} \Big)_{c_1c_2}\;\;\;\; ,
\label{negativa}
\end{eqnarray}

\noindent and 

\begin{equation}
{U^{-}}^{h}_{\alpha\alpha}\equiv 
U_{\alpha\delta^{h}_{k}}U_{\delta^{h}_{k}\delta^{h}_{k-1}}
...U_{\beta\alpha}\;\;\;\; .
\label{inversa}
\end{equation}

Now we are going to prove the following lemma:

\begin{lemma} 
The action \ref{caction} is independent from the chosen orientation for 
the boundary of the plaquettes $h$ in the sense that

\begin{equation}
S\equiv -{1\over 2}\sum_{h}{\rm Tr}
\left({U^{+}}^{h}_{\alpha\alpha}{W^{+}}^{h}(\alpha)\right)
=-{1\over 2}\sum_{h}{\rm Tr}
\left({U^{-}}^{h}_{\alpha\alpha}{W^{-}}^{h}(\alpha)\right)\;\;\;\;.
\label{tesi}
\end{equation}
\end{lemma}

\noindent {\bf Dim}: since the trace of a matrix and its transposed are the same
we can write

\begin{equation}
-{1 \over 2}{\rm Tr}\left({U^{+}}^{h}_{\alpha\alpha}{W^{+}}^{h}(\alpha)\right)= 
-{1 \over 2}{\rm
Tr}\left({U^{+}}^{h}_{\alpha\alpha}{W^{+}}^{h}(\alpha)\right)^{T}=
-{1 \over 2}{\rm
Tr}\left({{U^{+}}^{h}_{\alpha\alpha}}^{T}{{W^{+}}^{h}(\alpha)}^{T}\right)\;\;\;\;
.
\label{sp2}
\end{equation}

\noindent Now we have that

\begin{equation}
{{U^{+}}^{h}_{\alpha\alpha}}^{T}=U_{\alpha\delta^{h}_{k}}U_{\delta^{h}_{k}\delta^{h}_{k-1}}
...U_{\beta\alpha}
\label{UT}
\end{equation}

\noindent and 

\begin{eqnarray}
{{W^{+}}^{h}(\alpha)}^{T}&=& 
{1\over k_{h}+2} \Big({\mathcal{V}}^{(h)}_{\alpha\delta^{h}_{k}}(\alpha)+
U_{\alpha\beta}{\mathcal{V}}^{(h)}_{\beta\alpha}(\beta)
U_{\beta\alpha}+...\nonumber\\
&+&  
U_{\alpha\beta}...U_{\delta^{h}_{k-1}\delta^{h}_{k}}
{\mathcal{V}}^{(h)}_{\delta^{h}_{k}\delta^{h}_{k-1}}(\delta^{h}_{k})
U_{\delta^{h}_{k}\delta^{h}_{k-1}}...
U_{\beta\alpha} \Big)\;\;\;\; .
\label{WT}
\end{eqnarray}

\noindent Consider the orthogonal matrix 

\begin{equation}
\Gamma=U_{\alpha\beta}U_{\beta\delta^{h}_1}...
U_{\delta^{h}_{k-1}\delta^{h}_{k}}
\label{chiave}
\end{equation}

\noindent and manipulate the last trace in \ref{sp2} in the following way

\begin{equation}
-{1 \over 2}{\rm
Tr}\left({{U^{+}}^{h}_{\alpha\alpha}}^{T}{{W^{+}}^{h}(\alpha)}^{T}\right)=
-{1 \over 2}{\rm Tr}\left(\Gamma^{T}{{U^{+}}^{h}_{\alpha\alpha}}^{T}\Gamma
\Gamma^{T}{{W^{+}}^{h}(\alpha)}^{T}\Gamma\right)\;\;\;\; .
\label{modeo}
\end{equation}

\noindent A straightforward calculation shows that

\begin{eqnarray}
\Gamma^{T}{{W^{+}}^{h}(\alpha)}^{T}\Gamma&=&
{1\over k_{h}+2}
\Big({\mathcal{V}}^{(h)}_{\delta^{h}_{k}\delta^{h}_{k-1}}(\delta^{h}_{k})+
U_{\delta^{h}_{k}\delta^{h}_{k-1}}
{\mathcal{V}}^{(h)}_{\delta^{h}_{k-1}\delta^{h}_{k-2}}(\delta^{h}_{k-1})
U_{\delta^{h}_{k-1}\delta^{h}_{k}}+...\nonumber\\
&+& 
U_{\delta^{h}_{k}\delta^{h}_{k-1}}...U_{\beta\alpha}
{\mathcal{V}}^{(h)}_{\alpha\delta^{h}_{k}}(\alpha)
U_{\alpha\beta}...
U_{\delta^{h}_{k-1}\alpha^{h}_{k}} \Big)\;\;\;\; , 
\label{cisiamo}
\end{eqnarray}

\noindent and

\begin{equation}
\Gamma^{T}{{U^{+}}^{h}_{\alpha\alpha}}^{T}\Gamma=U_{\delta^{h}_{k}\delta^{h}_{k-1}}
U_{\delta^{h}_{k-1}\delta^{h}_{k-2}}...U_{\alpha\delta^{h}_{k}}
\label{alma}\;\;\;\; .
\end{equation}

\noindent These last two equations imply that

\begin{equation}
-{1 \over 2}{\rm
Tr}\left({{U^{+}}^{h}_{\alpha\alpha}}^{T}{{W^{+}}^{h}(\alpha)}^{T}\right)=
 -{1 \over 2}{\rm Tr}\left({U^{-}}^{h}_{\delta^{h}_{k}\delta^{h}_{k}}
 {W^{-}}^{h}(\delta^{h}_{k})\right)
\label{+-}
\end{equation}

\noindent Since the trace \ref{+-} is independent from the reference frame where
it is written, we have  
the equality \ref{tesi}.

This property of the action is necessary not to have ambiguity in defining it.
As we have seen  
on each hinge there
is no preferred orientation (c.f. section \ref{ciube}) and, as is easy to see,
no
orientation on the
corresponding Vorono{\"\i}-dual plaquettes even if the manifold is oriented. 

\section{First Order Field Equations for Small Deficit Angles}
\label{fieno}

As remarked in \cite{Ale}, in the second order 
formalism the action (\ref{haction}) is equivalent to the 
Regge action for small deficit angles $\theta_{h}$ 
(since in this case $sin\theta_{h}\approx \theta_{h}$). In the first 
order formalism we don't have angles $\theta_{h}$, but the only variables
related to the deficit angles are the connection 
matrices $U_{\alpha\beta}$. Then we assume, by definition, 
that the {\it small deficit
angles approximation} in the first order formalism is the passage
from the group variables $U_{\alpha\beta}$ of 
SO(n) to the algebra variables $\phi_{\alpha\beta}$ of so(n). 
As a consequence the 
connection matrices can be written in the form

\begin{equation}
U_{\alpha\beta}=I+\epsilon \; \phi_{\alpha\beta} + o(\epsilon)
\label{approx}
\end{equation}

This implies, as it is easy to verify, 
that the only gauge transformations (\ref{gauge})
which can be compatible with the equation (\ref{approx}) 
are of the type $O(\alpha)=I$, or, in other words, the approximation in the form
(\ref{approx}) is also a gauge fixing. 

\noindent In order to avoid technical complication, that we shall discuss
in the next section, we now suppose to substitute
the constraint
$b_{\alpha\beta}(\alpha)=U_{\alpha\beta}b_{\beta\alpha}(\beta)$ in the  
action for each Vorono{\"\i} edge $(\alpha, \beta)$.

A straightforward expansion of the action up to the first order shows that

\begin{equation}
S=-{1\over 2}\epsilon\;
\sum_{h}Tr\left((\phi_{\alpha\beta}+\phi_{\beta\delta^{h}_{1}}...
+\phi_{\delta^{h}_{k}\alpha}){\;}^{0}W^{h}(\alpha)\right)+ o(\epsilon)
\label{expansion}
\end{equation}

\noindent where ${\;}^{0}W^{h}{\;}^{c_1c_2}$ is the bivector
$W^{h}{\;}^{c_1c_2}$ to the zero order in which we have done for each matrix 
the approximation (\ref{approx})and for each Vorono{\"\i}-edge 
we have solved the constraint (\ref{1tetra}) 
 up to the
first order, that is to say

\begin{equation}
b_{\alpha\beta}(\alpha)=
(I+\epsilon\;\phi_{\alpha\beta})b_{\beta\alpha}(\beta) + o(\epsilon) 
\label{first}
\end{equation}

\noindent Demanding that the action  be stationary under the variation respect 
to $(\phi_{\alpha\beta})$ we obtain

\begin{equation}
{\delta S \over \delta \phi_{\alpha\beta}{\;}_{c_1c_2}}=
\epsilon \sum_{h \in (\alpha\beta)}{\;}^{0}W^{h}{\;}^{c_1c_2}
+o(\epsilon)=0
\label{1ord}
\end{equation}

\noindent where $h \in (\alpha\beta)$ means the sum over all plaquettes $h$ 
to which the Vorono{\"\i}-edge
$\alpha\beta$ belongs. The tensor ${\;}^{0}W^{h}{\;}^{c_1c_2}$ is an
antisymmetric tensor of type $(2,0)$ in $n$-dimensions, 
so has ${n(n-1)\over 2}$ independent components, the same independent
components of $\phi_{\alpha\beta}$ that have to be determined.   
  
Let's consider the face between the simplices $\alpha$ and $\beta$, dual to the
Vorono{\"\i}-edge
$\alpha \beta$. The equations (\cite{Ale}) that determine the Levi-Civita
connection
$U_{\alpha\beta}$ between the reference frames of the two simplices 
, to the first order (\ref{approx}) 
 are:

\begin{eqnarray}
(z^{a}_{1}(\alpha)-z^{a}_{n}(\alpha))&=&
(I+\epsilon\;\phi_{\alpha\beta})^{a}_{b}
(z^{b}_{1}(\beta)-z^{b}_{n}(\beta))
\nonumber\\
...................&..&.................... \nonumber\\
(z^{a}_{n-1}(\alpha)-z^{a}_{n}(\alpha))&=&
(I+\epsilon\;\phi_{\alpha\beta})^{a}_{b}
(z^{b}_{n-1}(\beta)-z^{b}_{n}(\beta))
\label{Levi}\;\;\;\;.
\end{eqnarray}

The $z^{a}_{i}(\alpha)$, $a=1,...,n$ are the coordinates of the vertices
$i=1,...,n$ of the face in the reference frame of the circumcenter in $\alpha$.
The circumcentric coordinates \ref{bari} of $n+1$  vertices of a
n-dimensional simplex are in one to one correspondence with the $n+1$
$b_{i}^{a}(\alpha),\; \{i=1,...,n+1\}$ 
. Equations
(\ref{Levi}) can be seen as an equation 
for determining $\phi_{\alpha\beta}$ as 
function of the $z^{a}_{i}(\alpha)$ (and then of the $b_{i}^{a}(\alpha)$). 
Each $\phi_{\alpha\beta}$, being an antisymmetric matrix in n-dimension,
has ${n(n-1)\over 2}$ degree of freedom.
The number of independent equations are $n(n-1)$. 
Anyway there is the following identity among the edge components 
of the $n-1$-dimensional face $(\alpha,\beta)$ 
 
\begin{equation} 
\sum_{h\in (\alpha\beta)} 
{\mathcal{V}}^{h}{\;}^{c_1c_2}(\alpha)
=0
\label{identity}\;\;\;\;,
\end{equation}
 
\noindent The (\ref{identity}) fixes 
${n(n-1)\over 2}$ degrees of freedom, so that the (\ref{Levi}) together
(\ref{identity}) 
has ${n(n-1)\over 2}$ linear independent equations. By linearity there is for 
$\phi_{\alpha\beta}$ only one solution. Anyway this is the connection that we
adopt in the
second order formalism. 
We want to see if this connection, that is the Levi-Civita or Regge
connection, satisfies equation (\ref{1ord}). If the connection is Regge 
we have (in matrix notation)

\begin{equation}
{\mathcal{V}}^{(h)}(\alpha)
=U_{\alpha\beta}{\mathcal{V}}^{(h)}(\beta)U_{\beta\alpha}
=...=U_{\alpha\beta}...U_{\delta^{h}_{k-1}\delta^{h}_{k}}
{\mathcal{V}}^{(h)}(\delta^{h}_{k})U_{\delta^{h}_{k}\delta^{h}_{k-1}}...
U_{\beta\alpha} 
\label{equo0}
\end{equation}

so that to the zero order 

\begin{equation}
{\mathcal{V}}^{(h)}(\alpha)={\mathcal{V}}^{(h)}(\beta) +O(\epsilon)=...=
{\mathcal{V}}^{(h)}(\delta_{k}^{h})+O(\epsilon)
\label{requo}\;\;\;\;.
\end{equation}

These facts imply that

\begin{equation}
\epsilon\sum_{h \in (\alpha\beta)}{\;}^{0}W^{h}{\;}^{c_1c_2}
=\sum_{h \in (\alpha\beta)}
\left(\epsilon{\mathcal{V}}^{h}{\;}^{c_1c_2}(\alpha) + \epsilon
O(\epsilon)\right)      
=0+O(\epsilon^{2})
\label{soddisf}
\end{equation}

that is to say that Regge connection is solution of our first order 
equations for the connection matrices in the limit of small deficit angles.

\section{First Order Field Equations: the General Case}
\label{pupu}

In the previous section we have seen that in the case of small deficit angles  
the Regge calculus is solution of
the first order field equations. 

We are now going to deal with the general problem. We want to derive 
the equation of motion by varying the action respect to $U_{\alpha\beta}$
and $b_{\alpha\beta}$. This formulation of gravity
on lattice is close to the popular Ashtekar's variables \cite{ash} in
non perturbative canonical continuum quantum gravity (see also
\cite{Ga} for a covariant version). Of course we have to take in account the
constraints
(\ref{relatto}) and (\ref{chiusura}), so in order to perform independent
variation of $U_{\alpha\beta}$ and $b_{\alpha\beta}$ it is necessary to
introduce
the constraints in the action by using Lagrange multipliers. Then the first
order action will be 

\begin{eqnarray}
S&\equiv& -{1\over 2}\sum_{h}Tr\left(U^{h}_{\alpha\alpha}W^{h}(\alpha)\right)+
\sum_{(\alpha\beta)}\lambda_{\alpha\beta}\left(b_{\alpha\beta}(\alpha) -
U_{\alpha\beta}b_{\beta\alpha}(\beta)\right) \nonumber \\
&+& \sum_{(\alpha\beta)}Tr\left({\tilde{\lambda}}^{(\alpha\beta)}
\bigl(U_{\alpha\beta}U^{T}_{\alpha\beta}-I\bigr)\right)
+\sum_{\alpha}\mu(\alpha)\left(\sum_{\beta =1}^{n+1}
b_{\alpha\beta}(\alpha)\right) 
\label{1action}
\end{eqnarray}

where $\lambda^{(\alpha\beta)}$ and $\mu(\alpha)$ are $n$-dimensional vectors 
and are  Lagrange
multipliers. ${\tilde{\lambda}}^{(\alpha\beta)}$ is an $n \times n$ matrix. The
constraint $U_{\alpha\beta}U^{T}_{\alpha\beta}-I$ is introduced to
restrict the variation of $U_{\alpha\beta}$ on the group $SO(n)$. 

\noindent Let's introduce the following lemma:

\noindent {\bf Lemma}: {\it The action, in matrix notation, 

\begin{equation}
S'\equiv Tr(\Lambda A) + Tr\left(\lambda(\Lambda\Lambda^{T}-I)\right)
\label{lemma}
\end{equation}

gives the following equation of motion if we assume that the variation 
respect to $\Lambda$ be stationary}

\begin{equation}
(\Lambda A)= (\Lambda A)^{T}
\label{reslem}\;\;\;\; .
\end{equation}

\noindent {\bf Dim}: If we consider the variation 
of the action respect to $\Lambda$ and
$\Lambda^{T}$, using the property that $Tr(M)=Tr(M^{T})$, we have

\begin{equation}
\delta S'={1\over 2}Tr(\delta\Lambda A+A^{T}\delta\Lambda^{T})
+{1\over 2}Tr\lambda(\delta\Lambda\Lambda^{T}+\Lambda\delta\Lambda^{T})
+{1\over 2}Tr(\delta\Lambda\Lambda^{T}+\Lambda\delta\Lambda^{T})\lambda^{T}
\label{vario}
\end{equation}

\noindent that implies  

\begin{eqnarray}
{\delta S'\over(\delta\Lambda)\Lambda^{T}}&=&\Lambda A+
\lambda+\lambda^{T}=0\nonumber\\
{\delta S'\over\Lambda(\delta\Lambda^{T})}&=&(\Lambda A)^{T}+
\lambda^{T}+\lambda=0
\label{equo}\;\;\;\;,
\end{eqnarray}

\noindent from which, subtracting term by term these two 
equations we have  (\ref{reslem}).

\noindent We can apply the lemma to the action (\ref{1action}) for the 
variation respect to $U_{\alpha\beta}$, we obtain the following
field equations

\begin{equation}
\sum_{h\in (\alpha\beta)}(U^{h}_{\alpha\alpha}W^{h}(\alpha))_{ij}
-\lambda_{\alpha\beta}{\;}_{i}b_{\alpha\beta}(\alpha){\;}_{j}=
\sum_{h\in (\alpha\beta)}(U^{h}_{\alpha\alpha}W^{h}(\alpha))^{T}_{ij}
-\lambda_{\alpha\beta}{\;}_{j}b_{\alpha\beta}(\alpha){\;}_{i}
\label{variola}
\end{equation}

The next step will consist of determining the field equations for the variations
of $b_{\alpha\beta}$. For this aim it is necessary to determine the 
quantity ${\partial V(\alpha) \over \partial b^{a}_{\alpha\beta}}$.
We recall that in the circumcentric coordinates, the coordinates of the 
$i$-th vertex $z^{a}_{i}(\beta)$ is given by formula \ref{oppo}

 We can write the formula \ref{vol} in a way not dependent
from the chosen index $j$. If we call $(V^{(j)}(\alpha))^{n-1}$ 
equation \ref{vol}, we have that it can be rewritten as 

\begin{equation}
(V^{*}(\alpha))^{n-1}= \root{n+1} \of {\left((V^{(1)}(\alpha))^{n-1}\right)...
\left((V^{(n+1)}(\alpha))^{n-1}\right)}
\label{radix}\;\;\;\; ,
\end{equation}

\noindent in such a way that it does not depend from any index. Evaluating 
the derivative of this last expression we get,

\begin{equation}
{\partial \left(1\over V(\alpha))\right)\over \partial b^{a}_{i}(\alpha)}=
-{1\over {n!(n-1)V^{2}(\alpha)}}z^{i}_{a}(\alpha)
\label{derivo}\;\;\;\; .
\end{equation}

We are now ready to derive the field equations for the variations of the 
$b_{\alpha\beta}$. The starting action is  (\ref{1action}). 
Furthermore we have remarked that the action 
is invariant under the orientation of the boundary of each plaquette, so we 
can write it in an orientation independent way

\begin{eqnarray}
S&\equiv& -{1\over 4}\sum_{h}Tr\left(U^{h}_{\alpha\alpha}W^{h}(\alpha)
+{U}^{T}{\;}^{h}_{\alpha\alpha}{W}^{T}{\;}^{h}(\alpha)\right)+
\sum_{(\alpha\beta)}\lambda^{(\alpha\beta)}\left(b_{\alpha\beta}(\alpha) -
U_{\alpha\beta}b_{\beta\alpha}(\beta)\right) \nonumber \\
&+&\sum_{\alpha}\mu(\alpha)\left(\sum_{\beta =1}^{n+1}
b_{\alpha\beta}(\alpha)\right) 
\label{2action}\;\;\;\; .
\end{eqnarray}

A straightforward calculation show that:

with the notation

\begin{equation}
U^{h}_{\alpha\alpha}{\;}^{ij}-U^{h}_{\alpha\alpha}{\;}^{ji}\equiv
\Omega^{h}_{\alpha\alpha}{\;}^{ij}
\end{equation}

\begin{eqnarray}
{\partial S \over \partial b^{i}_{\alpha\beta}(\alpha)}&\equiv&
\sum_{h\in (\alpha\beta)}{1\over 4V(\alpha)}
\left(\Omega^{h}_{\alpha\alpha}{\;}_{ij}b_{\alpha\delta_{k}^{h}}^{j}
-{1\over n!(n-1)}Tr(\Omega^{h}_{\alpha\alpha}{\mathcal{V}}^{h}(\alpha))
  z^{\alpha\beta}_{i}(\alpha)\right)\nonumber\\
  &+&\lambda_{i}{\;}_{\alpha\beta}-\mu_{i}(\alpha)=0
\label{moto}\;\;\;\;.
\end{eqnarray}

In the field equations (\ref{moto}) and (\ref{variola}) we have to determine
also
the Lagrange multipliers $\lambda_{\alpha\beta}^{i}$ and $\mu^{i}(\alpha)$. We 
now propose a method for determining them. In each simplex $\alpha$, or,
equivalently, in each point of the dual complex, we have $n+1$ vectors
$b_{\alpha\beta}^{i}(\alpha), b_{\alpha\delta^{1}}^{i}(\alpha),...,
b_{\alpha\delta^{n}}^{i}(\alpha)$ 
and among them there is the constraint (\ref{chiusura}). Let us choose 
n of this vectors for example 
$b^{i}_{\alpha\beta}(\alpha),...,b^{i}_{\alpha\delta_{n-1}}(\alpha)$, 
in other words we are solving the constraint (\ref{chiusura}). This is a 
base in the tangent space of the $n$-simplex 
$\alpha$ so that we can write each Lagrange multiplier, such that  
$\lambda^{h}_{\alpha\beta}$, considered as a vector in 
tangent space of $\alpha$ as 

\begin{equation}
\lambda_{\alpha\beta}^{h}=c_{\alpha\beta}^{1}b^{h}_{\alpha\beta}(\alpha)...
c_{\alpha\beta}^{n}b^{h}_{\alpha\delta_{n-1}}(\alpha)
\label{expansion}
\end{equation}
 
The set of $n$ vectors
$b^{i}_{\alpha\beta}(\alpha),...,b^{i}_{\alpha\delta_{n-1}}(\alpha)$
are in correspondence with the circumcentric coordinate

\begin{equation}
z^{a}_{n+1}(\alpha)={1\over (n+1)!(n!V(\alpha))^{n-2}}
\sum_{k\neq n+1}\epsilon^{a}_{a_1...a_{n-1}}\epsilon_{n+1\;k}^{j_1...j_{n-1}}
b^{a_1}_{j_1}(\alpha)...b^{a_{n-1}}_{j_{n-1}}(\alpha)
\label{oppo}
\end{equation}

\noindent where the indices $(j_1,...,j_{n-1},k)$ run over 
$\alpha\beta,\alpha\delta_1,...,\alpha\delta_{n-1}$ and in this 
case $n+1=\alpha\delta_{n}$.

\noindent From a geometrical point of view elementary arguments show that
$z^{a}_{n+1}(\alpha)$
is the circumcentric coordinate of the vertex of $\alpha$ which is the
intersection of  the $n$
$n-1$-dimensional faces whose normals are  
$b^{i}_{\alpha\beta}(\alpha),...,b^{i}_{\alpha\delta_{n-1}}(\alpha)$.

\noindent Pairwise to each $b_{\alpha\delta_{i}}(\alpha)$ we can associate a 
circumcentric coordinate $z^{a}_{\alpha\delta_{i}}(\alpha)$ obtained as 
in the definition \ref{oppo} in which  $n+1 \mapsto i=\alpha\delta_i$ and 
$(j_1,...,j_{n-1},k)$ now run over $\alpha\beta, \alpha\delta_1,...,
\alpha\delta_n$ 
in which, of course, $\alpha\delta_i$ is omitted. In this way we can make 
a one to one correspondence within each $b_{\alpha\delta_i}(\alpha)$ and 
the edge of the simplex $\alpha$ $E_{\alpha\delta_i}(\alpha)$ defined in the
circumcentric 
coordinate

\begin{equation}
E^{a}_{\alpha\delta_i}(\alpha)
\equiv z^{a}_{\alpha\delta_i}(\alpha)- z^{a}_{n+1}(\alpha)
\label{lato}
\end{equation}

\noindent It is straightforward to see that

\begin{equation}
\left(b_{\alpha\delta_{i}}(\alpha)\right)^{a}
\left(E_{\alpha\delta_{j}}(\alpha)\right)_{a}=n!V(\alpha)\delta_{ij}
\label{duale}\;\;\;\; .
\end{equation}

Consider the equation (\ref{expansion}), project it along the edge 
$E_{\alpha\beta}^{h}$ and repeat the procedure for each 
$E_{\alpha\delta_{i}}^{h}, {i=1,...,n-1}$. We will obtain

\begin{eqnarray}
c_{\alpha\beta}^{1}&=&{1\over
n!V(\alpha)}(\lambda^{h}_{\alpha\beta}E_{\alpha\beta}{\;}_{h})\nonumber\\
c_{\alpha\beta}^{i+1}&=&{1\over
n!V(\alpha)}(\lambda^{h}_{\alpha\beta}E_{\alpha\delta_{i}}{\;}_{h})\;\;
i=1,...,n-1
\label{proietto}
\end{eqnarray}

In other words for determining $\lambda_{\alpha\beta}$ we have to determine
its projections along the above edges. For this purpose we project equation
(\ref{variola}) along $E^{i}_{\alpha\delta_{i}}(\alpha)$ and
$E^{j}_{\alpha\beta}(\alpha)$ 
and we find that

\begin{equation}
\lambda_{\alpha\beta}^{h}E_{\alpha\delta_{i}}{\;}_{h}(\alpha)=
-{1\over V(\alpha)}\sum_{h\in
(\alpha\beta)}\left((U^{h}_{\alpha\alpha}W^{h}(\alpha))_{pq}
-(U^{h}_{\alpha\alpha}W^{h}(\alpha))_{qp}\right)
E^{p}_{\alpha\delta_{i}}(\alpha)E^{q}_{\alpha\beta}(\alpha)\;\;\;\; ,
\label{propino}
\end{equation}

which fixes the coefficients $c^{i}_{\alpha\beta},\;\; i=2,...,n$. 
A problem arises with the 
coefficient $c^{1}_{\alpha\beta}$, in fact 
equation (\ref{variola}) does not determine the 
component of $\lambda_{\alpha\beta}^{i}$ in 
the direction of $b^{i}_{\alpha\beta}(\alpha)$, 
because the following  transformation 
$\lambda_{\alpha\beta}^{i} \mapsto \lambda_{\alpha\beta}^{i} 
+\alpha b_{\alpha\beta}^{i}$ does not affect the equation. 
It seems necessary to use 
the field equation for $b_{\alpha\beta}$ for evaluating
$c^{1}_{\alpha\beta}$.

\noindent Anyway if we project (\ref{variola}) only along
$E^{j}_{\alpha\beta}(\alpha)$
we obtain that

\begin{equation}
\lambda_{\alpha\beta}^{i}={1\over V(\alpha)}
\left(\lambda_{\alpha\beta}^{h}E_{\alpha\beta}{\;}_{h}(\alpha)\right)b^{i}_{\alpha\beta}(\alpha)
+\sum_{h\in (\alpha\beta)}\left((U^{h}_{\alpha\alpha}W^{h}(\alpha))_{ij}
-(U^{h}_{\alpha\alpha}W^{h}(\alpha))_{ji}\right)E^{j}_{\alpha\beta}(\alpha)
\label{xfisso}
\end{equation}

which, substituted in the starting equation, gives the field equation for
$U_{\alpha\beta}$
without Lagrange multipliers

\begin{eqnarray}
&&\sum_{h\in (\alpha\beta)}\left((U^{h}_{\alpha\alpha}W^{h}(\alpha))_{ij}
-(U^{h}_{\alpha\alpha}W^{h}(\alpha))_{ji}\right)=
\sum_{h\in
(\alpha\beta)}\left((U^{h}_{\alpha\alpha}W^{h}(\alpha))_{ik}
-(U^{h}_{\alpha\alpha}W^{h}(\alpha))_{ki}\right)\nonumber \\
&&\times E^{k}_{\alpha\beta}(\alpha)
b^{j}_{\alpha\beta}(\alpha)
-\sum_{h\in (\alpha\beta)}\left((U^{h}_{\alpha\alpha}W^{h}(\alpha))_{jk}
-(U^{h}_{\alpha\alpha}W^{h}(\alpha))_{kj}\right)E^{k}_{\alpha\beta}(\alpha)
b^{i}_{\alpha\beta}(\alpha)
\label{lagrano}
\end{eqnarray}  

Of course the reasoning for the Vorono{\"\i}-edge $\alpha\beta$ can be applied
to
other Voronoi edges
$\alpha\delta_{i}$. The unknown 
coefficients will be $c_{\alpha\delta_{i}}^{i+1},\;i=2,...,n-1$. 

\noindent In the reference frame $\alpha$ we write the equation for the
variation 
respect to $b_{\alpha\beta}$ and $b_{\alpha\delta_{i}}$ in the following manner

\begin{eqnarray}
B_{\alpha\beta}^{a}+\lambda_{\alpha\beta}^{a} +\mu^{a}(\alpha)&=& 0\nonumber \\
B_{\alpha\delta_{i}}^{a}+\lambda_{\alpha\delta_{i}}^{a} +\mu^{a}(\alpha)&=& 0
\label{tradux}
\end{eqnarray}

\noindent where in the symbols $B$ we have summarized all the terms in the
equations (\ref{2action})  
which don't contain the Lagrange multipliers. We recall that the expansion 
of $\lambda_{\alpha\beta}$ in the base of $b$ is given by (\ref{expansion}), 
pairwise

\begin{equation}
\lambda^{a}_{\alpha\delta_{i}}=c_{\alpha\delta_{i}}^{1}b_{\alpha\beta}^{a}+...+
c^{i+1}_{\alpha\delta_{i}}b_{\alpha\delta_{i}}^{a}+...
+c^{n}_{\alpha\delta_{n-1}}b^{a}_{\alpha\delta_{n-1}}
\label{expantion2}
\end{equation}

Subtracting the two equations (\ref{tradux}) between them, we have

\begin{equation}
(B^{a}_{\alpha\beta}-B^{a}_{\alpha\delta_{i}})+
(c_{\alpha\beta}^{1}-c_{\alpha\delta_{i}}^{1})b^{a}_{\alpha\beta}
+...+(c_{\alpha\beta}^{n}-c_{\alpha\delta_{n-1}})b^{a}_{\alpha\delta_{n-1}}=0
\label{differe}\;\;\;\; .
\end{equation}

\noindent Projecting along $(E_{\alpha\beta}(\alpha))_{a}$ we get 

\begin{equation}
c_{\alpha\beta}^{1}=c_{\alpha\delta_{i}}^{1}-{1\over
n!V(\alpha)}(B^{a}_{\alpha\beta}
-B^{a}_{\alpha\delta_{i}})(E_{\alpha\beta}(\alpha))_{a}
\label{1c}
\end{equation}

in this way we determine $c^{1}_{\alpha\beta}$, and a similar procedure gives

\begin{equation}
c_{\alpha\delta_{i}}^{i+1}=c_{\alpha\beta}^{i+1}-{1\over
n!V(\alpha)}(B^{a}_{\alpha\delta_{i}}
-B^{a}_{\alpha\beta})(E_{\alpha\delta_{i}}(\alpha))_{a}
\label{2c}
\end{equation}

Once we have determined $\lambda_{\alpha\beta}^{a}$ and
$\lambda_{\alpha\delta_{i}}^{a}$
we can determine $\mu^{a}(\alpha)$ by summing the two equations

\begin{equation}
\mu^{a}(\alpha)=-{1\over
2}(\lambda_{\alpha\beta}^{a}+\lambda_{\alpha\delta_{i}}^{a})-
(B^{a}_{\alpha\beta}
+B^{a}_{\alpha\delta_{i}})
\label{mu}
\end{equation}

We have used the equations for $b_{\alpha\beta}$ and $b_{\alpha\delta_{i}}$ to
determine
$\lambda^{a}_{\alpha\beta}$, $\lambda_{\alpha\delta_{i}}$, and
$\mu^{a}_{\alpha}$. To determine
$b_{\alpha\beta}$ and $b_{\alpha\delta_{i}}$ we can solve the equations for
$b_{\beta\alpha}(\beta)$ 
and $b_{\delta_{i}\alpha}(\delta_{i})$ in which the Lagrange multipliers 
$\mu^{a}(\beta),\;\mu^{a}(\delta_{i})$ have been 
determined by two couple of different equations. Then we will use the constraint
$b_{\alpha\beta}(\alpha)=U_{\alpha\beta}b_{\beta\alpha}(\beta)$ and 
$b_{\alpha{\delta}_{i}}(\alpha)=
U_{\alpha\delta_i}b_{{\delta}_{i}\alpha}(\delta_{i})$ and so on.
In this method, at the end, the only unknown quantity is the Lagrange multiplier
$\lambda^{a}_{\alpha\delta_{n}}$ corresponding 
to the Vorono{\"\i}-edge $\alpha\delta_{n}$ and to $b_{\alpha\delta_{n}}$.
Anyway 
we have that $b_{\alpha\delta_{n}}$ is determined once we know all the other $b$
and 
we can use the equation for $b_{\alpha\delta_{n}}$ to determine
$\lambda_{\alpha\delta_{n}}$ 
as function of the other $b$. 

Let do an example that show how this method work in practice. Consider the
boundary 
of a tetrahedron as a two dimensional triangulation of the sphere

Let call $\alpha$ the triangle ABC, $\beta$ the triangle BCD, $\gamma$ the
triangle ACD
and $\delta$ the triangle ABD.

\begin{figure}
\begin{center}
\epsfig{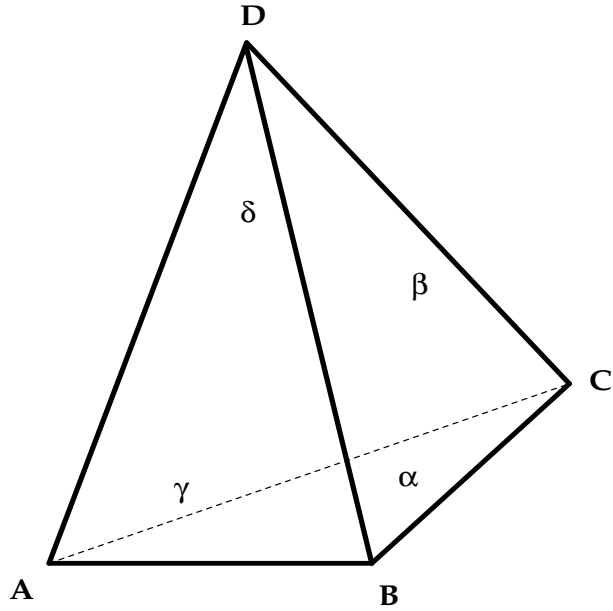}
\caption{The Most Elementary Triangulation of the 2-Sfere}
\label{fig:tetra}
\end{center}
\end{figure}

\noindent The edge BD is the face (one-dimensional) between $\alpha$ and
$\beta$, 
$b_{\alpha\beta}(\alpha)$ will be the vector in $\alpha$ perpendicular  to this 
face. DC is the face between $\beta$ and $\gamma$, and $b_{\beta\gamma}(\beta)$ 
the vector perpendicular to it. AD is the face between $\gamma$ and $\alpha$,
and 
$b_{\gamma\alpha}(\gamma)$ is the vector perpendicular to it, and so on. AC is 
perpendicular to the Vorono{\"\i}-edge $\alpha\gamma$, BC to ($\beta\delta$), AB
to
$\alpha\delta$.

Let's write the equation of motion for the variation of the $b$ for this system:

\begin{eqnarray}
B^{a}_{\alpha\beta}+\lambda^{a}_{\alpha\beta} + \mu^{a}(\alpha)&=&0\;\;\;\;{\rm
edge}\;(\alpha\beta)\nonumber\\
B^{a}_{\alpha\gamma}+\lambda^{a}_{\alpha\gamma} +
\mu^{a}(\alpha)&=&0\;\;\;\;{\rm
edge}\;(\alpha\gamma)\nonumber\\
B^{a}_{\alpha\delta}+\lambda^{a}_{\alpha\delta} +
\mu^{a}(\alpha)&=&0\;\;\;\;{\rm
edge}\;(\alpha\delta)\nonumber\\
B^{a}_{\beta\alpha}-(\lambda_{\alpha\beta}U_{\alpha\beta})^{a} 
+ \mu^{a}(\beta)&=&0\;\;\;\;{\rm edge}\;(\beta\alpha)\nonumber\\
B^{a}_{\beta\gamma}+\lambda^{a}_{\beta\gamma} + \mu^{a}(\beta)&=&0\;\;\;\;{\rm
edge}\;(\beta\gamma)\nonumber\\
B^{a}_{\beta\delta}+\lambda^{a}_{\beta\delta} + \mu^{a}(\beta)&=&0\;\;\;\;{\rm
edge}\;(\beta\delta)\nonumber\\
B^{a}_{\gamma\beta}-(\lambda_{\beta\gamma}U_{\beta\gamma})^{a} 
+ \mu^{a}(\gamma)&=&0\;\;\;\;{\rm edge}\;(\gamma\beta)\nonumber\\
B^{a}_{\gamma\alpha}-(\lambda_{\alpha\gamma}U_{\alpha\gamma})^{a} 
+ \mu^{a}(\gamma)&=&0\;\;\;\;{\rm edge}\;(\gamma\alpha)\nonumber\\
B^{a}_{\gamma\delta}+\lambda^{a}_{\gamma\delta} +
\mu^{a}(\gamma)&=&0\;\;\;\;{\rm
edge}\;(\gamma\delta)\nonumber\\
B^{a}_{\delta\alpha}-(\lambda_{\alpha\delta}U_{\alpha\delta})^{a} 
+ \mu^{a}(\delta)&=&0\;\;\;\;{\rm edge}\;(\delta\alpha)\nonumber\\
B^{a}_{\delta\beta}-(\lambda_{\beta\delta}U_{\beta\delta})^{a} 
+ \mu^{a}(\delta)&=&0\;\;\;\;{\rm edge}\;(\delta\beta)\nonumber\\
B^{a}_{\delta\gamma}-(\lambda_{\gamma\delta}U_{\gamma\delta})^{a} 
+ \mu^{a}(\delta)&=&0\;\;\;\;{\rm edge}\;(\delta\gamma)
\label{tutto}
\end{eqnarray}

\noindent From the first two equations we choose to determine
$\lambda^{a}_{\alpha\beta}$,
 $\lambda^{a}_{\alpha\gamma}$ and $\mu^{a}(\alpha)$ by the above method. The
 equation for
 the Vorono{\"\i}-edge $\alpha\delta$ will be used to determine
 $\lambda^{a}_{\alpha\delta}$. The
 equations for the Vorono{\"\i}-edges $\beta\gamma$ and $\beta\delta$ will be
 used to
 determine 
 $\lambda^{a}_{\beta\gamma}$, $\lambda_{\beta\delta}$ and $\mu^{a}(\beta)$. The
 equation
 for the Vorono{\"\i}-edge $\beta\alpha$ will be used to determine
 $b_{\beta\alpha}$.
 From the
 constraint $b_{\alpha\beta}=U_{\alpha\beta}b_{\beta\alpha}$ we determine
 $b_{\alpha\beta}$ too. The equation for the Vorono{\"\i}-edge $\gamma\beta$ may
 be used
 to determine 
 $\mu^{a}({\gamma})$. The equation for $\gamma\alpha$ may be seen as an equation
 for
 determining $b_{\gamma\alpha}$, and so we determine $b_{\alpha\gamma}$ too.
 The equation 
 for $\gamma\delta$ may be used to determine $\lambda_{\gamma\delta}$. The
 equation
 for 
 $\delta\alpha$ determine $\mu^{a}(\delta)$, those for $\delta\beta$ and
 $\delta\gamma$ determine $b_{\delta\beta}$ and $b_{\delta\gamma}$ and so also
 $b_{\beta\delta}$ and $b_{\gamma\delta}$. 
 The other $b$ are determined from the constraint
 on the $b$ (\ref{chiusura}).
 
 In this way we determine all the Lagrange multiplier as functions of the
 $U_{\alpha\beta}$ and $b_{\alpha\beta}$. Using also the constraint 
 we have that all the field equations fix a number of degrees of freedom 
 that is equal to the number of degrees of freedom of  $U_{\alpha\beta}$ 
 and $b_{\alpha\beta}$ .

\section{Quantization of the Model}
\label{quanta}

In this section we shall discuss the quantum measure to associate to the
previous classical action. The action \ref{caction} is invariant under the 
action of the group $SO(n)$, the gauge group (\ref{gauge}).

\noindent Let us fix the following notation: we define

\begin{equation}
\mu(b_{\alpha\beta}(\alpha))\equiv 
db^{1}_{\alpha\beta}(\alpha)...db^{n}_{\alpha\beta}(\alpha)
\label{bmes}
\end{equation}

\noindent and with 

\begin{equation}
\mu(U_{\alpha\beta})
\label{Haar}
\end{equation}

\noindent the Haar measure of $SO(n)$.  The partition function for this theory
will
be:

\begin{eqnarray}
Z&=&\int e^{{1\over 2}\sum_{h}Tr\left(U^{h}_{\alpha\alpha}W^{h}(\alpha)\right)}
\prod_{\alpha}\delta(\sum_{\beta=1}^{n+1}b_{\alpha\beta}(\alpha))\\ \nonumber
&&\prod_{\alpha\beta}
\delta\left(b_{\alpha\beta}(\alpha)-U_{\alpha\beta}b_{\beta\alpha}(\beta)\right)
\mu(U_{\alpha\beta})\mu(b_{\alpha\beta})
\label{part}
\end{eqnarray}

\noindent in which the first product $\prod_{\alpha}$ is a product over all
vertices 
of the dual complex, and the second product $\prod_{\alpha\beta}$ is the 
product over all the edges starting from the vertex $\alpha$. The second 
delta function $\delta\left(b_{\alpha\beta}(\alpha)-
U_{\alpha\beta}b_{\beta\alpha}(\beta)\right)$ has to be introduced one time
for each edge, so that if we have introduced it for the edge $\alpha\beta$,
we have not to introduce it for the edge $\beta\alpha$ in the reference frame
$\beta$. It is straightforward to see that the measure of the partition function
is invariant under the gauge  transformation (\ref{gauge}). 
In fact if we perform a gauge transformation of the type
(\ref{gauge}), the modified terms of the measure  are:

\begin{equation}
\delta\left(\sum_{\beta=1}^{n+1}b'_{\alpha\beta}(\alpha)\right)
\delta\left(b'_{\alpha\beta}(\alpha)-{U}'_{\alpha\beta}b'_{\beta\alpha}(\beta)\right)
\mu({U}'_{\alpha\beta})\mu(b'_{\alpha\beta}) 
\label{modifico}\;\;\;\; .
\end{equation}

Now $\mu(b'_{\alpha\beta})$ is equal to 
$det\left(O(\alpha)\right)\mu(b_{\alpha\beta})$ and so is equal to
$\mu(b_{\alpha\beta})$. The Haar measure of $SO(n)$ is right and left 
invariant so $\mu({U}'_{\alpha\beta})=\mu({U}_{\alpha\beta})$.
By the equations (\ref{gauge}), we have that the two delta of
(\ref{modifico}) can be written in the following form

\begin{eqnarray}
\delta\left(O(\alpha)\Bigl(\sum_{\beta=1}^{n+1}b_{\alpha\beta}(\alpha)\Bigr)\right)\nonumber\\
\delta\left(O(\alpha)
\Bigl(b_{\alpha\beta}(\alpha)-{U}_{\alpha\beta}b_{\beta\alpha}(\beta)\Bigr)\right)
\label{soso}
\end{eqnarray}

\noindent that, from the properties of the delta function, 
will reduce to 1 over the determinant
of $O(\alpha)$ for the deltas of the quantities in  parenthesis that
multiply $O(\alpha)$. Then we have that equation (\ref{modifico})
can be written as 

\begin{equation}
\delta\left(\sum_{\beta=1}^{n+1}b_{\alpha\beta}(\alpha)\right)
\delta\left(b_{\alpha\beta}(\alpha)-{U}_{\alpha\beta}b_{\beta\alpha}(\beta)\right)
\mu({U}_{\alpha\beta})\mu(b_{\alpha\beta}) 
\label{rivedo}\;\;\;\; ,
\end{equation}
 
and then the invariance of the measure under gauge transformations. 

\section{Coupling with Matter}
\label{copy}

In the continuum theory on Riemannian manifolds the coupling with fermionic
matter is given 
by the following 
term of the Lagrangian density

\begin{equation}
{\mathcal L}\equiv \left({\bar \psi}e^{\mu}_{a}\gamma^{a}{\nabla}_{\mu}\psi
- {\nabla}_{\mu}e^{\mu}_{a}{\bar \psi}\gamma^{a}\psi\right)
+m\psi {\bar \psi}
\label{lagra}
\end{equation}

where $\gamma^{a},\; a=1,...n$ are the Dirac-matrices satisfying the Clifford
algebra 

\begin{equation}
\gamma^{a}\gamma^{b}+\gamma^{b}\gamma^{a}=2\delta^{a\;b}
\label{gamma}
\end{equation}

$\psi$ the n-dimensional Dirac spinor field 
(${\bar \psi}\equiv {\psi}^{\dagger}{\gamma}^{1}$),${\nabla}_{\mu}$ the
covariant derivative, $e_{a}^{\mu}$ the n-beins
that is, on the tangent space of the Riemannian manifolds $(M,g)$ 
where the Lagrangian density (\ref{lagra}) is 
defined, the vector fields such that

\begin{equation}
g^{\mu\nu}(x)={e^{\mu}_{a}}(x){e^{\nu}_{b}}(x)\delta^{ab}
\label{tetrad}\;\;\;\; .
\end{equation}
 
\noindent We are assuming that the Riemannian manifolds $(M,g)$ in question have
a
spin structure, that is the second Stiefel-Whitney class is zero.

We have now all the ingredients to define the coupling of gravity with 
fermionic matter on the lattice in analogy with continuum case
(for related works see \cite{frohlich} \cite{drummond} \cite{ren}
\cite{jourjine}). 

\noindent Let $\nu=2n$ or $\nu=2n+1$ (depending if 
$n$ is even or odd) and consider the $2^{\nu}$  dimensional 
two-fold covering group of $SO(n)$. So instead to consider the
connection matrices $U_{\alpha\beta}$, we have to consider the
$2^{\nu}\times 2^{\nu}$ connection matrices $D_{\alpha\beta}$ such that

\begin{equation}
D_{\alpha\beta}{\gamma}^{a}D_{\alpha\beta}^{-1}
=(U_{\alpha\beta})^{a}_{b}{\gamma}^{b}
\label{2-fold}
\end{equation}

\noindent where in the left hand part of (\ref{2-fold}) we have 
used matrix notation. Given $D_{\alpha\beta}$ we determine
$U_{\alpha\beta}$, but if we know $U_{\alpha\beta}$ we determine
$D_{\alpha\beta}$ up to a sign. In particular, as it is easy to derive from 
(\ref{2-fold}), we can write $U_{\alpha\beta}$ as (\cite{frohlich})

\begin{equation}
(U_{\alpha\beta})_{ab}={1\over
2^{\nu}}Tr(\gamma_{a}D_{\alpha\beta}\gamma_{b})
\label{questa}
\end{equation}

In the discrete theory we assume that the spinor field is a $2^{\nu}$ complex 
vector defined in each vertices of the dual simplicial complex, that is to say 
a map that to each vertex $\alpha$ associates the $2^{\nu}$ complex 
vector $\psi(\alpha)$.

\noindent In order to define the covariant derivative on lattice we have to 
derive the distance $|\alpha\beta|$ between the two neighborly circumcenters
in $\alpha$ and $\beta$. The distance $\triangle h_{1}$ of the circumcenter
in $\alpha$ from the face $\alpha\beta$ can be calculated by deriving 
the volume of the $n$-dimensional simplex obtained by joining the circumcenter
with the $n$-vertices of the face $\alpha\beta$ and dividing it by $n$ and 
the volume of the face itself. So we have that

\begin{equation}
\triangle h_{1}={1\over
n^{2}}{\sum_{i=1}^{n}b^{a}_{\alpha\beta}(\alpha)z_{a}^{i}(\alpha) \over
|b_{\alpha\beta}(\alpha)|}
\label{dist}
\end{equation}

\noindent in which $z^{a}_{i}$ are the circumcentric coordinates of the vertices
of the
face $\alpha\beta$ and with $|b_{\alpha\beta}(\alpha)|$ the module of the 
vector that divided by $(n-1)!$ is equal to the volume of the face.

\noindent In the same manner we have

\begin{equation}
\triangle h_{2}={1\over
n^{2}}{\sum_{i=1}^{n}b^{a}_{\beta\alpha}(\beta)z_{a}^{i}(\beta) \over
|b_{\beta\alpha}(\beta)|}
\label{dist2}
\end{equation}

\noindent At the end we have $|\alpha\beta|=\triangle h_{1} + \triangle h_{2}$.

At this point we are ready to define the covariant derivative
$(\nabla_{\mu}\psi)(\alpha)$ on lattice

\begin{equation}
(\nabla_{\mu}\psi)(\alpha)\equiv
{ D_{\alpha\beta}\psi(\beta) - \psi(\alpha)\over |\alpha\beta|}
\label{covariant}\;\;\;\;.
\end{equation}

So far the discrete version of the the action for the coupling between gravity
and fermionic matter can be written in the form

\begin{eqnarray}
S_{F} &\equiv& \sum_{\alpha}\Bigl(\sum_{(\alpha\beta), \beta=1,...,n+1}
{1\over |\alpha\beta|}\big({\bar \psi}(\alpha)b^{a}_{\alpha\beta}\gamma_{a}
D_{\alpha\beta}\psi(\beta) \nonumber \\
&-& 
D_{\alpha\beta}{\bar \psi}(\beta)
b^{a}_{\alpha\beta}\gamma_{a}\psi(\alpha)\bigl)
+m{\bar\psi}(\alpha)\psi(\alpha)\Bigr)
\label{fermion} 
\end{eqnarray}

so that the quantum measure that includes also fermionic matter can be written 
as

\begin{eqnarray}
Z&=&\int e^{-(S+S_{F})}
\prod_{\alpha}\mu\Bigl(\psi(\alpha)\Bigr)
\mu\Bigl({\bar\psi}(\alpha)\Bigr)
\delta\left(\sum_{\beta=1}^{n+1}b_{\alpha\beta}(\alpha)\right)\nonumber \\ 
&&\prod_{\alpha\beta}
\delta\left(b_{\alpha\beta}(\alpha)-U_{\alpha\beta}b_{\beta\alpha}(\beta)\right)
\mu(D_{\alpha\beta})\mu(b_{\alpha\beta})
\label{mine}
\end{eqnarray}

\noindent where $S$ is the action for pure gravity (\ref{caction}), and 
$\mu(D_{\alpha\beta})$ the Haar measure on the two-fold covering group 
of $SO(n)$ while $\mu(\psi(\alpha))=d\psi(\alpha)$ the standard measure 
on $C^{2^{\nu}}$. 

The coupling with Fermionic matter on lattice presented 
in this paragraph is quite similar to the 
one introduced in reference \cite{ren}. The differences 
consist of the fact that 
the action is here group theoretical and not Regge, 
and this formalism implies that  
from the beginnings
the coupling is written on the dual Vorono{\"\i} complex.

\appendix

\chapter{Branched Polymers}
\label{abranch}

The notion of Branched Polymers is crucial in the classification of the 
so called "Elongated Phase" of 4-D simplicial quantum gravity. We are now 
going to review \cite{frohlich2} some basic concepts 
of statistical mechanics on graphs.

An abstract {\it graph} is a set of points called "vertices" connected 
by lines. These lines introduce a natural concept of connectedness. A graph
can be seen as a one complex. From now on all our graphs will be abstract, 
not embedded in some affine space so that we may omit the adjective abstract. A
 graph is {\it simple} if two distinct vertices are connected at most  
by only one line and the vertices at the end of each line are always distinct.
A graph that is connected and simple is called a {\it tree}. The number of 
lines $l(i)$ at the vertex $i$ is called "the coordination number" of the 
vertex $i$. In a tree with $n$ vertices we have that $\sum_{i=1}^{n}l(i)=
2(n-1)$. A tree is {\it rooted} if we label one vertex which is called
the {\it root}.

\noindent Two trees $T$ and $T'$ are combinatorial equivalent if there is 
a one to one  map $\phi$ between them such that the vertices of $T$ are 
mapped uniquely into the vertices of $T'$ so that a line joining two 
vertices of $T'$, $v'_{i}$ and $v'_{j}$, is a line of $T'$ if and only if 
it is the image of a line of $T$ by $\phi$, in the sense that there exist two
vertices 
of $T$ $v_{i}$ and $v_{j}$ joined by a line of $T$ 
whose images by $\phi$ are $v'_{i}$ and $v'_{j}$ and the line joining them
is mapped into the line joining $v'_{i}$ and $v'_{j}$. 
If the two trees are 
rooted they are combinatorially equivalent if there is a one to one 
map such that together with the properties described before it 
maps the root of $T$ into the root of $T'$ and viceversa. 
In statistical mechanics tree graphs are called 
{\it Branched Polymers}. The statistical mechanics of these abstract tree
graphs is also known as {\it mean field theory} of Branched Polymers to 
distinguish it from the case of embedded Branched Polymers. 

Let's define as $r^{\alpha}(n)$ the number of combinatorial inequivalent 
rooted tree graphs with $n$ vertices in which the root has coordination 
number equal to one and all others vertices have coordination 
number less than or equal $\alpha$. By Caley's formula we have

\begin{equation}
r^{\alpha}(n)=\sum^{\alpha}_{l_{1}...l_{n}=1 
\atop \sum_{i=2}^{n}l_{i}=2(n-1)-1} 
{1 \over (n-1)} \prod_{i=2}^{n}
{1 \over (l_{i}-1)!}
\mathop{\longrightarrow}_{\alpha \mapsto \infty}
{(n-1)^{n-2} \over (n-1)!}
\label{root}\;\;\;\; .
\end{equation}                              

Let $\xi^{\alpha}(n)$ indicates the number of combinatorial inequivalent 
tree graphs with $n$ vertices whose coordination number is less than or 
equal to $\alpha$. We have that

\begin{equation}
\xi^{\alpha}(n)=\sum^{\alpha}_{l_{1}...l_{n}=1 
\atop \sum_{i=1}^{n}l_{i}=2(n-1)} 
{1 \over n(n-1)} \prod_{i=1}^{n}
{1 \over (l_{i}-1)!}
\mathop{\longrightarrow}_{\alpha \mapsto \infty}
{n^{n-2} \over n!}
\label{unroot}\;\;\;\; .
\end{equation}                              

From the above formula or from geometrical arguments it is easy to see 
that

\begin{equation}
\xi^{\alpha}(n)={1 \over n}r^{\alpha}(n+1)\;\;\;\; .
\label{iden1}
\end{equation}

The partition function $R^{\alpha}(\beta)$ over the ensemble of rooted tree
graphs, whose vertices have coordination number less than or equal $\alpha$,  
is defined in the following way

\begin{equation}
R^{\alpha}(\beta)\equiv \sum_{n=2}^{\infty}
\beta^{n-1}r^{\alpha}(n)\;\;\;\; ,
\label{ropart}
\end{equation}

\noindent and the analogous $Z^{\alpha}(\beta)$ unrooted 

\begin{equation}
Z^{\alpha}(\beta)\equiv \sum_{n=1}^{\infty}
\beta^{n}r^{\alpha}(n)\;\;\;\; .
\label{ropart}
\end{equation}

\noindent From these definitions and the identity \ref{iden1} 
we have that

\begin{equation}
{d \over d\beta}Z^{\alpha}(\beta)=
{1 \over \beta}R^{\alpha}(\beta)\;\;\;\; .
\label{iden2}
\end{equation}

As regard the definition of the Green function for tree graphs with 
coordination numbers up to $\alpha$, the strategy 
is completely similar to the same case for dynamical triangulations (c.f.
section \ref{sgreen}). A path between two vertices of a tree graph is given by a

sequence $\{i\}_{i=1}^{i=l}$ of vertices such that the vertices 
$j$ and $j+1$ ($j+1 \neq j$ and $j=1,...,l-1$)
are joined by a line and $i_{1}$ coincides with the first vertex and 
$i_{l}$ with the second vertex. The length of the path is the number of its 
line, $l-1$ in this specific example. The distance between two vertices of a 
tree graph is the length of the minimal path that joins them.

\noindent As usual the micro-canonical Green function is the sum over all 
combinatorial inequivalent tree graphs $T^{\alpha}(r,n)$ whose vertices have 
coordination numbers up to $\alpha$  with $n$ vertices and with 
two labelled vertex at distance $r$

\begin{equation}
G^{\alpha}(r,n)\equiv \sum_{T^{\alpha}(r,n)}
\label{microtubo}
\end{equation}

\noindent and the relative grand-canonical partition function is

\begin{equation}
G^{\alpha}(r,\beta)\equiv 
\sum_{n=1}^{\infty}\beta^{n}G^{\alpha}(r,n)\;\;\;\; .
\label{grantubo}
\end{equation}

\noindent The relative susceptibility function is defined as follows

\begin{equation}
\chi^{\alpha}(\beta)\equiv \sum_{r=0}^{\infty}G^{\alpha}(r,\beta)
\label{squalo}
\end{equation}

As for the dynamical triangulations in the definition above we can exchange
the sum over $r$ with the sum over $n$ so that

\begin{equation}
\sum_{r}\sum_{T^{\alpha}(r,n)}
\label{sempre}
\end{equation}

\noindent is the sum of all combinatorial inequivalent tree graphs with 
coordination number up to $\alpha$ and with two labelled vertices. Since
for large $n$ asymmetric tree graphs prevails this sum is asymptotical 
equal to $n^{2}G^{\alpha}(r,n)$, so that as in \ref{cuccu} 
the susceptibility function is asymptotically equal to

\begin{equation}
\chi^{\alpha}(\beta) \asymp {d^{2} \over d\beta^{2}}Z^{\alpha}(\beta)\;\;\;\;.
\label{rivoto}
\end{equation} 

Now we are going to use the following recursion relation

\begin{lemma}
The function $r^{\alpha}(n)$ can be written as

\begin{equation}
r^{\alpha}(n)=\sum_{\gamma=0}^{\alpha-1}
{1 \over \gamma !}
\sum_{n_1,...,n_{\gamma}\geq 2 \atop \sum_{i=1}^{\gamma}n_{i}=n+\gamma -2}
\prod_{i=1}^{\gamma} r^{\alpha}(n_i)\;\;\;\; . 
\label{recurro}
\end{equation}
\end{lemma}

\noindent This recursion formula can be proved straightforwardly or by 
using formula \ref{root} or by geometric considerations. Anyway from this
lemma we have the following evident corollary

\begin{corollary}

\begin{equation}
R^{\alpha}(\beta)=\beta \sum_{\gamma=0}^{\alpha-1}
{1 \over \gamma!}[R^{\alpha}(\beta)]^{\gamma}
\label{pote}
\end{equation}

\end{corollary}

At this point  \ref{pote} can be used to derive the differential 
equation for $R^{\alpha}(\beta)$,

\begin{equation}
{d \over d\beta}R^{\alpha}(\beta)=
{R^{\alpha}(\beta) \over \beta}
\left[1 +\beta{(R^{\alpha}(\beta))^{\alpha-1} \over (\alpha-1)!}
-R^{\alpha}(\beta)\right]^{-1}
\label{calda}
\end{equation}

\noindent This equation is divergent at the point $\beta^{c}$ 
where the denominator 
vanish, that is to say

\begin{equation}
R^{\alpha}(\beta^{c})=1 + \beta^{c}
{\left(R^{\alpha}(\beta)\right)^{\alpha-1} \over (\alpha-1)!}
\label{critico}
\end{equation}

\noindent This equation has only one solution for each value of $\alpha$. 
Since ${d \over d\beta}R^{\alpha}(\beta)$ is divergent in $\beta^{c}$ 
we have that

\begin{equation}
\left. {d\beta(R^{\alpha})  
\over dR^{\alpha}}\right|_{R^{\alpha}(\beta^{c})}=0\;\;\;\; .
\label{inveo}
\end{equation}

\noindent By the implicit function theorem $\beta(R^{\alpha})$ result to be 
analytic around $R^{\alpha}(\beta^c)$ so that we can expand it in Taylor
series

\begin{equation}
\beta(R^{\alpha})-\beta^{c}={1 \over 2} 
{d^{2}\beta(R^{\alpha}) \over {dR^{\alpha}}^{2}}
(R^{\alpha}-R^{\alpha}(\beta^c))^{2} +
o\left((R^{\alpha}-R^{\alpha}(\beta^c))^{2}\right)\;\;\;\; ,
\label{expo}
\end{equation}

\noindent and for $\beta$ near $\beta^{c}$ we get

\begin{equation}
R^{\alpha}(\beta)\asymp R^{\alpha}(\beta^{c})+ 
C(\beta^{c}-\beta)^{1\over 2}
\label{arrogo}
\end{equation}

These considerations together with equations \ref{iden2} and \ref{rivoto} imply
that
for $\beta$ near $\beta^{c}$ 

\begin{equation}
\chi^{\alpha}(\beta) \asymp (\beta^{c}-\beta)^{-{1\over 2}}
\label{exporde}
\end{equation}

\noindent so that the susceptibility of Branched Polymers has the same
exponent $\gamma_{s} ={1 \over 2}$ of the mean field.

\chapter{Baby Universes}
\label{ababy}

The concept of Baby Universe in dynamical triangulations 
has been introduced for the first time by S. Jain and S.D.Mathur \cite{jain}.
The original idea of these authors has been to introduce a way for describing
the roughness of triangulated surfaces. Successively the average number of Baby
Universes has been used as a sort of order parameter to classify the two phases,
{\it crumpled} and {\it branched polymer} phases which have been seen for the
first
time in the Monte Carlo simulations of 4-D dynamical triangulations. In fact in
the 
articles \cite{baby4} and \cite{4dynamical} numerical simulations have shown
very few 
baby universes in the crumpled phase and a cascade of baby universes in the
branched
polymer phase. Motivated by these data it has been 
tried to put in correspondence the baby universes
with the branched polymers. The deep reason of this correlation is
discussed in chapter \ref{geo}. Anyway seminal arguments in favor of the
connection between branched
polymers 
and baby universes of minimal size can be found in the article \cite{semi}. 
    
Let's give the main definitions regarding  
Baby Universes in the spirit of the article \cite{jain}. We shall restrict, for 
convenience, to the four dimensions and to the case in which the baby universe
has 
the neck of minimum size (Minbu). A Minbu can be considered as a triangulation 
of a $PL$ manifold in which we can distinguish two parts: the Mother universe
that is 
the part that has the majority of the volume of the triangulation, and the baby
universe 
which is the part with the minority of the volume. The two parts are glued
together by the
boundary of a four-simplex that is called the {\it neck} of the baby universe. 

\noindent Now we are going to introduce, following  \cite{jain}, a way to
enumerate 
the average number of baby universes with given volume $V$ in the canonical
ensemble
of all triangulations of a PL-manifold with $N_{4}$ four-dimensional simplices. 
Anyway we want to stress that this enumeration is
not rigorous but it is in good agreement with the numerical results
\cite{baby4}, \cite{4dynamical}, \cite{catteral}. Suppose that the topology is
that one of 
the sphere $S^{4}$ since our future discussions will be concentrated on this
case. Imagine 
to separate the mother universe and the baby. We will obtain two triangulations
of 
the sphere $S^{4}$
with a boundary made by the boundary of a four-simplex. The mother has $N_{4}-V$
four dimensional
simplices and $N'_{2}$ two-dimensional simplices, the baby $V$ four-dimensional 
simplices and $N''_{2}$ two-dimensional simplices. Obviously the sum
$N'_{2}+N''_{2}$ 
will be equal to $N_{2}+10$, where $N_{2}$ is the number of two-dimensional
simplices 
of the original triangulation, because in separating the mother and the baby we
are 
counting the bone of the neck twice. Suppose now to fill the hole of the
boundaries 
by gluing a four simplex to each of them. So that the number of combinatorial
inequivalent triangulations of the mother universe is now 
$\rho_{a}(S^{4},N_{4}-V+1,N'_{2})$  (c.f. \ref{micocanonica}) and that one of
the baby 
is $\rho_{a}(S^{4},V+1,N''_{2})$. The {\it antsaz} is that all the combinatorial
inequivalent configuration with a Minbu are obtained by marking a four simplex 
in the mother universe and one in the baby, and by gluing them through
identifying
the two marked simplices.
We suppose, as usual, that we are in the regime of large $N_{4}$ so that
asymmetric
triangulations prevail. This implies that the marking process will produce
$(N_{4}-V+1)\rho_{a}(S^{4},N_{4}-V+1,N'_{2})$ combinatorial inequivalent
triangulations
for the mother universe and $(V+1)\rho_{a}(S^{4},V+1,N''_{2})$ for the baby
universes.
So that, at the end, the number of the combinatorial inequivalent Minbu 
$M(S^{4},N_{4},N_{2},V)$ is

\begin{eqnarray}
M(S^{4},N_{4},N_{2},V)=&&5!(N_{4}-V+1)(V+1)\\ \nonumber
&&\times\rho_{a}(S^{4},N_{4}-V+1,N'_{2})
\rho_{a}(S^{4},V+1,N''_{2})\;\;\;\; . 
\label{prole}
\end{eqnarray}
 
\noindent where $5!$ comes from the fact that in the case of asymmetrical
triangulations
there are $5!$ combinatorial inequivalent ways of joining two four-simplices. 
Any Minbu of the characteristics defined above can be obtained in this way, so
that 
\ref{prole} really can be considered as a formula that gives an estimate of
the 
number of all Minbus of baby-size
$V$ and number of bone $N''_{2}$ in the micro-canonical ensemble of all
triangulations of $S^{4}$ with 
$N_{4}$ four-simplices and $N_{2}$
bones. Consider a triangulation $T$ of $S^{4}$ with $N_{4}$ four-simplices and
$N_{2}$ 
bones,  which has a Minbu of size $V$ and $N_{2}''$ bones. 
Let us indicate by $n_{T}(S^{4},N_{4},N_{2},V)$ the number of combinatorial
inequivalent
triangulations that can be obtained from a given triangulation $T$ by cutting a
baby universe and
attach it 
on another  four simplex of $T$, previously marked, as described above.  
 It is clear that
$M(S^{4},N_{4},N_{2},V)=\sum_{T}n_{T}(S^{4},N_{4},N_{2},V)$, so that the average
number
of combinatorial inequivalent Minbu in the micro-canonical ensamble can be
approximated by 
dividing $M(S^{4},N_{4},N_{2},V)$ by the micro-canonical partition function
$\rho_{a}(S^{4},N_{4},N_{2})$

\begin{eqnarray}
{\mathcal N}(S^{4},N_{4},N_{2},V)&\approx& 
5!(N_{4}-V+1)(V+1)\\ \nonumber
&&\times{\rho_{a}(S^{4},N_{4}-V+1,N'_{2})\rho_{a}(S^{4},V+1,N''_{2}) 
\over \rho_{a}(S^{4},N_{4},N_{2})} \;\;\;\;. 
\label{miprole}
\end{eqnarray}

\noindent In the canonical ensemble the average number of Minbu with volume $V$
can be calculated
in the following way

\begin{eqnarray}  
&&{\mathcal N}(S^{4},N_{4},k_{2},V) \approx 
5!(N_{4}-V+1)(V+1)\\ \nonumber
&&\times{\sum_{N_{2}}e^{k_{2}N_{2}}\rho_{a}(S^{4},N_{4}-V+1,N'_{2})
 \rho_{a}(S^{4},V+1,N''_{2})  
\over Z(S^{4},N_{2},k_{2})}\;\;\;\; ,
\label{canu}
\end{eqnarray}

\noindent which using the relation $N'_{2}+N''_{2} =N_{2}+10$ is equal to

\begin{eqnarray}  
&&{\mathcal N}(S^{4},N_{4},k_{2},V)\approx 
5!(N_{4}-V+1)(V+1)e^{-10k_2}\\ \nonumber
&&\times{\sum_{N'_{2}}e^{k_2N'_2}\rho_{a}(S^{4},N_{4}-V+1,N'_{2})
\sum_{ N''_2}e^{k_2N''_2} \rho_{a}(S^{4},V+1,N''_{2})   
\over Z(S^{4},N_{2},k_{2})}\;\;\;\; .
\label{canu}
\end{eqnarray}

\noindent Finally

\begin{eqnarray}
{\mathcal N}(S^{4},N_{4},k_{2},V)&\approx&
5!(N_{4}-V+1)(V+1)e^{-10k_{2}}\\ \nonumber
&&\times{Z(S^{4},N_{4}-V+1,k_{2})
 Z(S^{4},V+1,k_{2})  
\over Z(S^{4},N_{2},k_{2})}\;\;\;\; .
\label{canuta}
\end{eqnarray}

\noindent For large value of $N_{4}$ and $V$ the previous formula will reduce to
the 
standard formula for the average number of baby universe in the canonical
ensemble

\begin{equation}
{\mathcal N}(S^{4},N_{4},k_{2},V)\approx
{5!(N_{4}-V+1)(V+1)Z(S^{4},N_{4}-V,k_{2})
 Z(S^{4},V,k_{2})  
\over Z(S^{4},N_{2},k_{2})}\;\;\;\; .
\label{rimorchio}
\end{equation}

We remember that in each phase the canonical partition function of $A$
four-simplices 
has the behavior $Z(S^{4},A,k_{2})\asymp A^{\gamma_{str}-3}e^{Ak_{4}(k_{2})}$,
which 
substituted in \ref{rimorchio} gives

\begin{equation}
{\mathcal N}(S^{4},N_{4},k_{2},V)\approx
N_{4}\left(V \Big(1-{V\over N_{4}}\Big)\right)^{\gamma_{str}-2}\;\;\;\; .
\label{computo}
\end{equation}

\noindent In this way by computing the average number of baby universes with
volume $V$
in a canonical ensamble of $N_{4}$ simplices we can obtain informations on the
susceptibility
exponent $\gamma_{str}$. This is the reason too for which the argument of the
baby universes is 
so popular among people doing numerical simulations.

\chapter{Vorono{\"\i}  Cells}
\label{voronoi}

In this section we shall introduce the definition of Vorono{\"\i} Cells 
since we are going to use it for the definition of the "metric" dual 
of a triangulation of a PL-manifold. 

Consider a point set $\Xi$ of a $n$-dimensional Euclidean space $E^{n}$. We say 
that $E^{n}$ is discrete if there is a positive real number $r_{0}$ such that  
$\forall x,y \in E^{n}$  $d(x,y)\geq 0$, where $d$ is the distance function 
between two points of $E^{n}$ induced by the Euclidean metric \cite{senechal}.

\noindent It follows immediately from the above definition this 
theorem:

\begin{theorem} 
If $\Xi$ is discrete it has the local finiteness property: for every closed
ball ${\overline{B}}_{r}(x)$ with center in $x\in E^{n}$ and radius $r$, 
${\overline{B}}_{r}(x)\cap \Xi$ is a finite set.
\end{theorem}

\noindent A point set $\Xi$ is relative dense in $E^{n}$ if there is a real
positive number $R_{0}$ such that all the spheres of $E^{n}$ of radius 
greater than $R_{0}$ have at least one point of $\Xi$ in their interior. 
$R_{0}$ is
called the covering radius of $\Xi$.

A set point $\Xi$ which is discrete and relative dense is a Delaunay or Delone
set \cite{senechal} \cite{okabe}, and is also indicated as $(r,R)$ from 
its discrete radius $r$ and covering radius $R$.

\noindent In a Delone set $\Xi$ the $r$-star, $r>0$, of a point $x \in \Xi$ 
is the finite point set ${\overline{B}}_{r}(x)\cap \Xi$.

Consider now a Delone set $\Xi \in E^{n}$ (or, equivalently, 
a set with a finite number of points). Let $x \in \Xi$, the Vorono{\"\i} 
cell of $x$ is, by definition, the convex region of points of $E^{n}$ 
which are more close to $x$ than to any other point of $\Xi$. 

\begin{figure}
\begin{center}
\epsfig{file=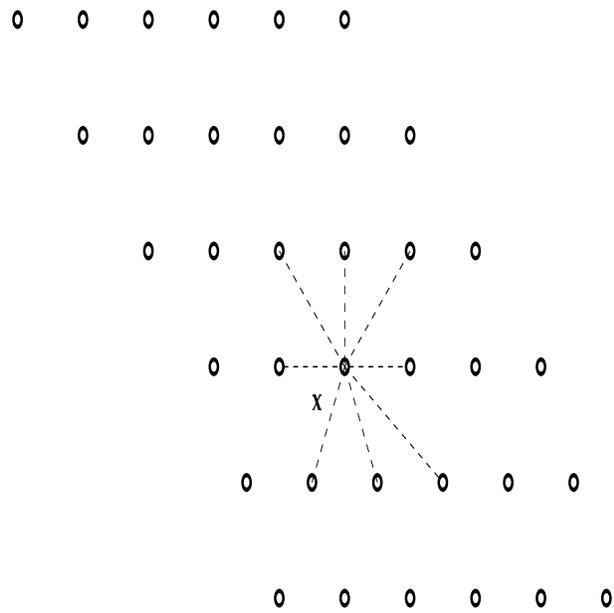,height=8truecm,width=8truecm}
\caption{In this regular crystal the points to be considered for constructing
the Vorono{\"\i}  cell $V(x)$ of $x$ are those ones which belong to the star of
$x$
made by the points closer to $x$. These points are also the points which form an
elementary cell of this crystal}
\label{fig:reti}
\end{center}
\end{figure}

\noindent In practice a Vorono{\"\i} cell of a point $x \in \Xi$ is constructed 
by considering the segments that join $x$ to each point 
of $\Xi$ in the star of $x$. Anyway for the moment we are considering
the case in which $\Xi$ is a sort of periodic crystal so that the 
radius of the star has to be chosen in such a way that it contains $x$ and 
the points of its elementary cell (see figure \ref{fig:reti}).

\begin{figure}
\begin{center}
\epsfig{file=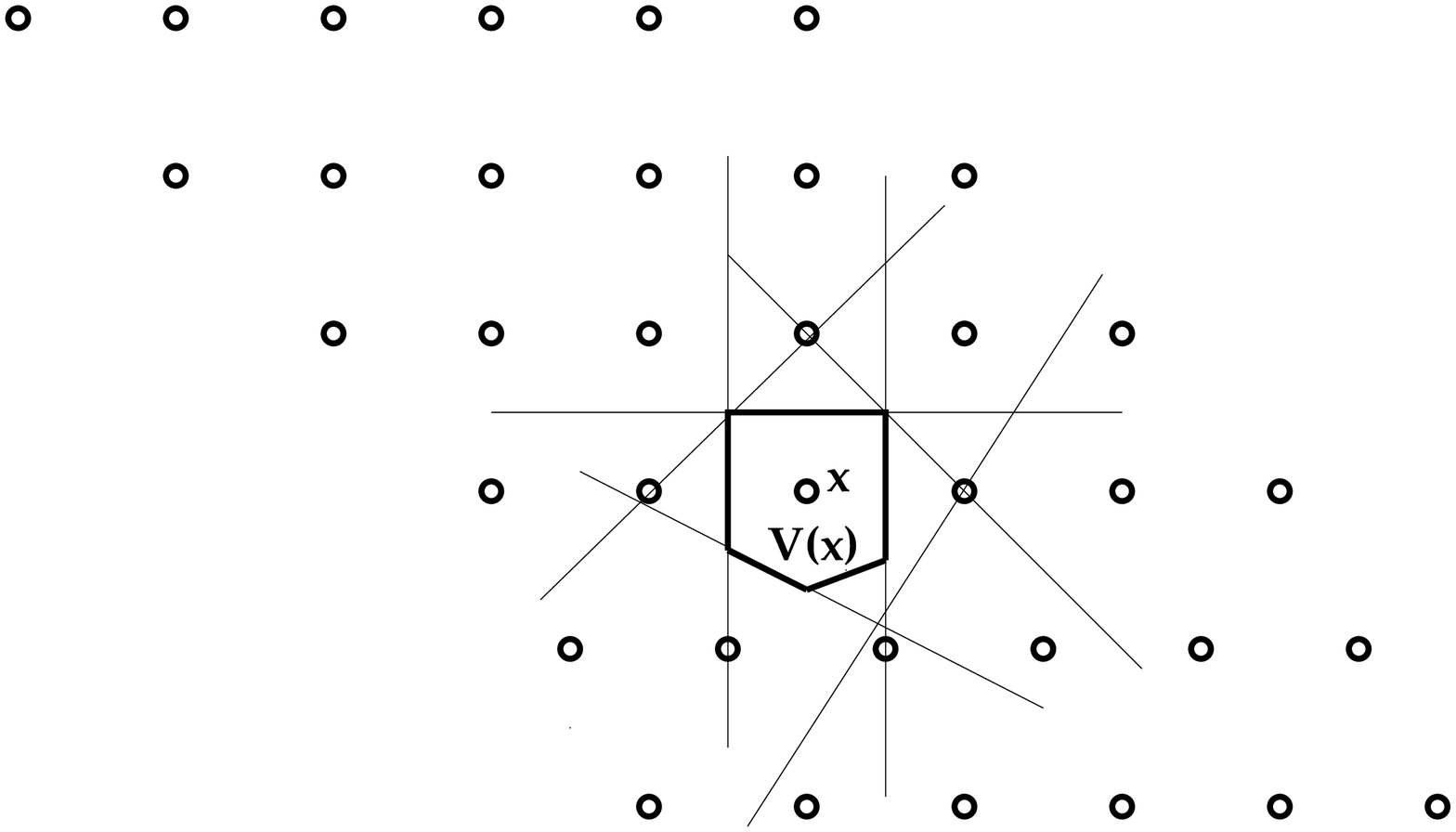,height=10truecm,width=10truecm}
\caption{Vorono{\"\i} cell $V(x)$ of the point $x$ belonging to the crystal 
of the figure \ref{fig:reti}}
\label{fig:roro}
\end{center}
\end{figure}

\noindent Then the
$n-1$-dimensional faces of the Vorono{\"\i} cell of $x$ $V(x)$, called facets,
are obtained by considering the $n-1$-dimensional 
orthogonal bisectors to the segments. The intersection of the facets $V(x)$ 
are the $n-2$ dimensional face of the Vorono{\"\i} cell and so on (see figure
\ref{fig:roro}).

\noindent If we carry on this construction for each point of $\Xi$ we obtain 
a partition of $E^{n}$ into cells that is called 
{\it Vorono{\"\i} tessellation}. This tessellation has the following properties:

\noindent 1)The Vorono{\"\i} cells are convex region of $E^{n}$ 
and two cells can have in common 
only one $n-1$ dimensional facets.

\noindent 2)The points of $\Xi$ whose Vorono{\"\i} cells share a vertex $\nu$
lie on a sphere with center in $\nu$ which has no points of $\Xi$ in its
interior. 
Furthermore it is easy to see that the sphere centered in $\nu$ pass among at
least 
$n+1$ points of $\Xi$.

If $\Xi$ has an infinite number of points 
and it doesn't have any periodic structure, 
for constructing a Vorono\"{i}  
cell of a point $x \in \Xi$ in the general 
case we need an infinite number of operations. 
But we are going to show that really they are 
still finite, thanks to the following 
theorem

\begin{theorem}
Let $\Xi$ a Delone set in $E^{n}$ of type $(r,R)$ in the previous notations. 
To construct a
Vorono\"{i} cell in $E^{n}$ of the point 
$x\in \Xi$ we have to join $x$ to all point of $\Xi$
in the ball ${\overline{B}}_{2R}(x)$.
\end{theorem}

\noindent {\bf Dim}:

Consider the Vorono{\"\i} tessellation of $E^{n}$ based on the Delone set $\Xi$.
Let $x$ a point
of $\Xi$ and $V(x)$ the facets of the Vorono{\"\i} cell relative to $x$. By
definition the $V(x)$ lie
on the $n-1$ iperplains that are bisectors of the segment joining $x$ to its
neighbor points of 
$\Xi$. Let $\nu$ any vertex of $V(x)$. We know that $\nu$ is the center of a
sphere on which there 
are at least $n$ point of $\Xi$ whose relative Vorono{\"\i}  cells  intersect at
$\nu$.
In the interior of this sphere there are no points of $\Xi$ so that
$d(x,\nu)\leq R$. So that if we
consider the distance between $x$ and any other point $y$ among all other at
least $n$ points above, 
we have $d(x,y)\leq d(x,\nu)+d(y,\nu)$ since the reasonings for $x$ 
can be applied equivalently to any other point $y$, we have $d(y,\nu)\leq R$ and
then $d(x,y)\leq 2R$. 
This implies that
to construct a Vorono{\"\i} cell in $E^{n}$ of a point $x$ belonging to a Delone

set it is sufficient to consider
the points of $\Xi$ which are in the ball ${\overline{B}}_{2R}(x)$.

The triangulations $\Sigma$ of a PL-manifold, which we are considering, have
always a finite number
of $n$ simplices so that they have a finite number of vertices. On
$\Sigma$ it is defined a metric tensor such that in the interior of each
$n$-dimensional simplex
it coincides with the Euclidean metric and near the hinges it is like the metric
tensor of a cone.
So it does make sense to define on the simplicial manifold the Vorono{\"\i} cell

(see reference \cite{ruth} p. 395)
of a vertex as the convex 
region of points of the simplicial complex that are closer to the vertex than to
any other vertex of the 
simplicial complex. To construct this Vorono{\"\i} cell relative to a vertex 
we consider on the simplicial complex all the 
vertices that lie in the star of this vertex and the edges joining them with the
vertex. The facets of 
this Vorono{\"\i} cell lie on the $n-1$ iperplain orthogonal to the edges in
their middle points.
Standard facts of elementary Euclidean geometry tell us that the vertices of the
Vorono{\"\i} cells coincides 
with the circumcenters of the simplices, that is the centers of the spheres
circumscribed to the simplices.
In this way we have constructed the dual of the original triangulation $\Sigma$.
In fact 
to each vertex uniquely it corresponds a $n$-dimensional Vorono{\"\i} cell, to
an
edge a $n-1$ dimensional
facet of the Vorono{\"\i} cell and in general to a $k$ simplex a
$n-k$-Vorono{\"\i} polyhedron orthogonal
to it. In particular this dual application maps a $n$-simplex to its
circumcenter.

\end{document}